\documentclass[aps,prd,reprint,superscriptaddress,floatfix,nofootinbib]{revtex4-1}

\usepackage{microtype}
\usepackage{comment}
\usepackage{bbm}

\usepackage{graphicx}
\usepackage[dvipsnames]{xcolor}
\usepackage{rotating}
\usepackage[caption=false,singlelinecheck=false]{subfig}
\usepackage{amsmath}
\usepackage{amssymb}
\usepackage{multirow}
\usepackage{amsfonts,dsfont}
\usepackage{bm}
\usepackage{physics}
\usepackage{mathtools}
\usepackage{slashed}

\usepackage{booktabs}
\usepackage{array}
\usepackage{enumitem}
\usepackage[normalem]{ulem}

\usepackage{url}
\usepackage{xspace}
\usepackage{siunitx}
\usepackage{xfrac}
\usepackage{hyperref}
\usepackage[nameinlink]{cleveref}
\usepackage{appendix}

\usepackage{xifthen}
\usepackage{xcolor}
\hypersetup{colorlinks,	linkcolor={green!47!black},	citecolor={green!60!black},	urlcolor={green!55!black}}

\usepackage{lmodern}     
\usepackage[T1]{fontenc}

\DeclareSymbolFont{myletters}{OML}{ztmcm}{m}{it}
\DeclareMathSymbol{\uplambda}{\mathord}{myletters}{"15}

\newcommand{\imag}{\text{i}}

\newcommand{\orcid}[1]{\href{https://orcid.org/#1}{\includegraphics[height=1.7ex,width=1.7ex]{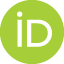}}}

\newcolumntype{x}[1]{>{\centering\arraybackslash\hspace{0.cm}}p{#1}}

\DeclareSymbolFont{symbolsC}{U}{pxsyc}{m}{n}

\newcommand{\kSSB}{k_{\chi{\rm SB}}}
\newcommand{\kconf}{k_{\rm conf}}
\newcommand{\SSB}{\chi{\rm SB}}
\newcommand{\dSSB}{{\rm d}\chi{\rm SB}}
\newcommand{\acrit}{\alpha^{\rm crit}_{\SSB}}

\graphicspath{{figures/}}

\begin{document}
	

	\title{Gauge-Fermion Cartography:\\[.75ex] from confinement and chiral symmetry breaking to conformality}

    \author{Florian~Goertz~\orcid{0000-0001-8880-0157}}
	\affiliation{Max-Planck-Institut f\"ur Kernphysik P.O. Box 103980, D 69029, Heidelberg, Germany}
	\author{\'Alvaro~Pastor-Guti\'errez~\orcid{0000-0001-5152-3678}}
	\affiliation{Max-Planck-Institut f\"ur Kernphysik P.O. Box 103980, D 69029, Heidelberg, Germany}
	\affiliation{Institut für Theoretische Physik, Universität Heidelberg, Philosophenweg 16, 69120 Heidelberg, Germany}
	\author{Jan M. Pawlowski~\orcid{0000-0003-0003-7180}\,}
	\affiliation{Institut für Theoretische Physik, Universität Heidelberg, Philosophenweg 16, 69120 Heidelberg, Germany}
	\affiliation{ExtreMe Matter Institute EMMI, GSI Helmholtzzentrum für Schwerionenforschung mbH, Planckstr.\ 1, 64291 Darmstadt, Germany}

\begin{abstract}
We study, for the first time, the interplay between colour-confining and chiral symmetry-breaking dynamics in gauge-fermion systems with a general number of flavours and colours.  Specifically, we work out the flavour dependence of the confinement and chiral symmetry breaking scales.  We connect the QCD-like regime, in quantitative agreement with lattice data, with the perturbative conformal limit, thereby exploring uncharted region of theory space. This analysis is done within the first-principles functional renormalisation group approach to gauge-fermion systems and is facilitated by a novel approximation scheme introduced here. This novel scheme enables a relatively simple access to the confining dynamics. This allows us to investigate the whole landscape of many-flavour theories and to provide a cartography of their phase  structure. In particular, we uncover a novel phase with the locking of confining and chiral dynamics at intermediate flavour numbers. We also  explore the close-conformal region that displays a walking behaviour. Finally, we provide a quantitative estimate for the lower boundary of the conformal Caswell-Banks-Zaks window, with a $N^{\rm crit}_f(N_c=3)= 9.60^{+0.55}_{-0.53}$.  This work offers a self-consistent framework for charting the landscape of strongly interacting gauge-fermion theories necessary to reliably study strongly coupled extensions of the Standard Model of particle physics.
\end{abstract}

\maketitle

\quad\newpage\quad\newpage
\tableofcontents 
\newpage

\section{Introduction}
\label{sec:Introduction}

The Standard Model (SM) of particle physics provides a comprehensive description of high-energy phenomena. It has been impressively confirmed by the measurement of scattering events at the highest collision energies currently accessible, with the Higgs discovery being a prominent example \cite{Aad:2012tfa,Chatrchyan:2012xdj}. It has also been impressively confirmed in experiments (collider and fixed target) probing the strongly correlated regime of the strong interactions at sufficiently small collision energies.

The low-energy dynamics of the SM are governed by two inherently non-perturbative phenomena: chiral symmetry breaking and confinement. While confinement is exclusive to the Quantum Chromodynamics (QCD) sector of the theory, spontaneous chiral symmetry breaking occurs at multiple scales. Chiral symmetry breaking at the electroweak scale is driven by the Higgs mechanism. At around 1 GeV, dynamical strong chiral symmetry breaking ($\dSSB$) is driven by composite quark bilinears. There, the pions take the rôle of pseudo-Goldstone bosons, while the $\sigma$-mode takes that of the radial excitation, similar to the Higgs. In the absence of electroweak symmetry breaking, a massless spectrum of pions corresponding to exact Goldstone bosons would result in a very different realization of nature.

In this context, understanding the landscape of gauge-fermion quantum field theories (QFTs), where QCD is found, is of high relevance. Exploring the phase structure of these theories provides a better insight into the interplay between dynamical phenomena and can reveal new, interesting phenomena and spectra. Additionally, these theories could underlie extensions of the Standard Model, needed to resolve open problems and puzzles in fundamental physics. For example, nature may exhibit additional fundamental forces, potentially including a strongly coupled dark sector containing fermions, which could be responsible for the existence of dark matter \cite{Renner:2018fhh,Bai:2013xga,Cline:2013zca,Cirelli:2024ssz}. Furthermore, the Higgs boson could be realized as a bound state of fermions charged under a more fundamental strong force. This is the case in Composite Higgs~\cite{Kaplan:1983fs,Kaplan:1983sm,Georgi:1984ef,Dugan:1984hq,Contino:2003ve,Agashe:2004rs} or Technicolour-like~\cite{Susskind:1978ms,Weinberg:1975gm,Chivukula:1998if,Susskind:1982mw,Hill:2002ap,Yamawaki:1985zg,Galloway:2010bp,Appelquist:2010gy,Belyaev:2013ida,Arbey:2015exa} models, where the Higgs is realized as a pseudo Goldstone boson of a spontaneously broken global flavour symmetry or as a radial mode. To judge the viability of such models, it is important to study how successful low-energy scenarios can emerge from fundamental theories. This includes the appearance of suitable bound-state spectra compatible with experimental measurements and of  almost conformal \textit{walking} regimes spanning over several orders of magnitude above the confinement scale, see e.g.~\cite{Cacciapaglia:2020kgq,Cacciapaglia2022a}. Here, we present a framework that can be used to examine these strongly coupled setups in a self-consistent manner, with functional implementations of the SM~\cite{Pastor-Gutierrez:2022nki,Goertz:2023pvn}.

In particular, we present a combined study of the emergence and interplay of chiral symmetry breaking and confinement in gauge-fermion systems. Both phenomena are signalled by the dynamical emergence of a mass gap, which is reflected in the hadron and glueball spectrum of QCD. $\dSSB$ is governed by its order parameter, the chiral condensate, which vanishes in the chiral symmetric phase and sets the characteristic mass scale (cubed) in the chirally broken phase. We connect the well-known QCD limit with the conformal limit, where infrared (IR) Caswell-Banks-Zaks (CBZ) fixed points are found. Thereby, we approach interesting close-conformal theories that display a walking regime from first principles. 

For our analysis, we use the functional renormalisation group (fRG) approach to gauge-fermion systems 
for general $N_f$. As a diagrammatic approach, the fRG requires gauge fixing and the study is performed in the Landau gauge. This choice is the most common one which has both conceptual as well as computational reasons. Moreover, the fRG has been used for the quantitative study of QCD in the vacuum and at finite temperature and density. For reviews of results see \cite{Dupuis:2020fhh, Fu:2022gou}, as well as \cite{Ihssen:2024miv} for a recent discussion of the systematic error control. The striking potential of the fRG for qualitative as well as quantitative studies of gauge-fermion systems is well-proven by now. Specifically, chiral symmetry breaking is covered and unravelled very directly and concisely within the fRG approach with emergent composites or dynamical hadronisation \cite{Gies:2001nw, Pawlowski:2005xe, Floerchinger:2009uf}, while the confining mass gap in the physical spectrum is reflected in a mass gap in the gluon propagator in the Landau gauge, see \cite{Kugo:1979gm, Cornwall:1981zr}. In short, both phenomena are readily accessed in the fRG approach, as they are entailed in the IR behaviour of low order quark and gluon correlation functions, whose stable and reliable computation is a specific virtue of the fRG approach.  

In this work we present the first comprehensive analysis of confinement and chiral symmetry breaking and their interrelation for a general number of flavours $N_f$. In particular this includes predictions for the flavour dependence of the confinement and chiral symmetry breaking scales as well as the emergent low lying hadron spectra. This analysis includes the walking regime and the CBZ fixed point, the former being specifically interesting for beyond Standard Model (BSM) theories, as these gauge-fermion systems provide a natural way of separating scales as well as prolonging the onset of instabilities. This analysis is made possible due to technical advances in the description of confinement in the fRG approach to Landau gauge QCD, presented below, that efficiently and qualitatively simplify the respective numerical computations. 

We close the introduction with a bird's-eye view on the work. In \Cref{sec:ChiSBandConf} we provide an introduction to colour confinement, ${\rm d}\chi{\rm SB}$ and conformality in gauge-fermion theories, including the manifestation of these phenomena in Landau gauge QCD. In \Cref{sec:truncation} we introduce the functional renormalisation group (fRG) approach to gauge-fermion theories, including a detailed discussion of the truncation used in the present work. In particular, we present the fRG-methodology to accommodate confinement, and the emergent description of composite degrees of freedom in \Cref{sec:econf,sec:emergentcomposites}. In \Cref{sec:InterplayLowNf} we study confining and non-confining solutions for a QCD-like theory with three colours and two and three flavours in the chiral limit. This Section provides benchmark tests of the truncation, but also contains new results concerning the interplay of the dynamics. 
In~\Cref{sec:phasesYM+chiral} we chart the landscape of gauge-fermion theories from QCD-like scenarios to the conformal limit. In this regime we use an improved approximation, which accommodates the non-perturbative IR regime with confinement and chiral symmetry breaking as well as higher orders of perturbation theory relevant in the approach to conformality. Some important technical details can be found in~\Cref{app:largeNfapproximation}. We also analyse the close-conformal properties of walking theories and derive unique quantities, such as the size of the walking regime. Finally, in \Cref{sec:BoundaryCBZ} we address the presence of $\dSSB$ in close-conformal theories and provide a quantitative estimate of the boundary of the CBZ conformal window including a systematic error analysis. The reader familiarised with the fRG approach to gauge-fermion QFTs may directly proceed to \Cref{sec:phasesYM+chiral,sec:BoundaryCBZ} where the main results of this work are presented, see \Cref{fig:SummaryBZ} for the cartography of the phase structure.

\section{Gauge-fermion theories}
\label{sec:ChiSBandConf}

The results on the confining and chiral dynamics of gauge-fermion systems in the present work are obtained in a gauge-fixed functional approach to multi-flavour QCD, based on the effective action $\Gamma[A_\mu,c,\bar c, \psi, \bar \psi]$, where $A_\mu$ is the SU($N_c$) gauge (mean) field, $c,\bar c$ are the ghost fields introduced in the Faddeev-Popov gauge fixing, and $\psi,\bar \psi$ are the fermions and anti-fermions in the fundamental representation. The one-particle irreducible part of the gauge-fixed correlation functions are obtained as field-derivatives of the effective action and are the building blocks of the present approach, in particular the full two-point correlation functions or propagators.  In this Section we discuss, how the confining and chiral dynamics of multi-flavour QCD, including its conformal regime for a large number of flavours, manifests itself in such a gauge-fixed setting, in particular in the correlation functions.  

In \Cref{sec:confinement} we discuss confinement and its signatures in the gauge field propagator, while in \Cref{sec:dSSB} we treat $\dSSB$ and its manifestation in the full four-fermion scattering vertex. Last, in \Cref{sec:CBZ} we introduce the conformal limit of gauge-fermion QFTs.

\subsection{Confinement} 
\label{sec:confinement}

Confinement is the absence of coloured asymptotic states and the existence of a mass gap for the (colourless) asymptotic states. 
For example, in physical QCD, the lowest lying state is the scalar glueball with a mass of about 1800\,MeV, see eg.~\cite{Pawlowski:2022zhh,Huber:2023mls,Huber:2022dsn,Morningstar:1999rf,Athenodorou:2020ani}, which is directly related to $\Lambda_\textrm{QCD}$. 

\begin{figure*}[t!]
	\centering
	\includegraphics[width=\columnwidth]{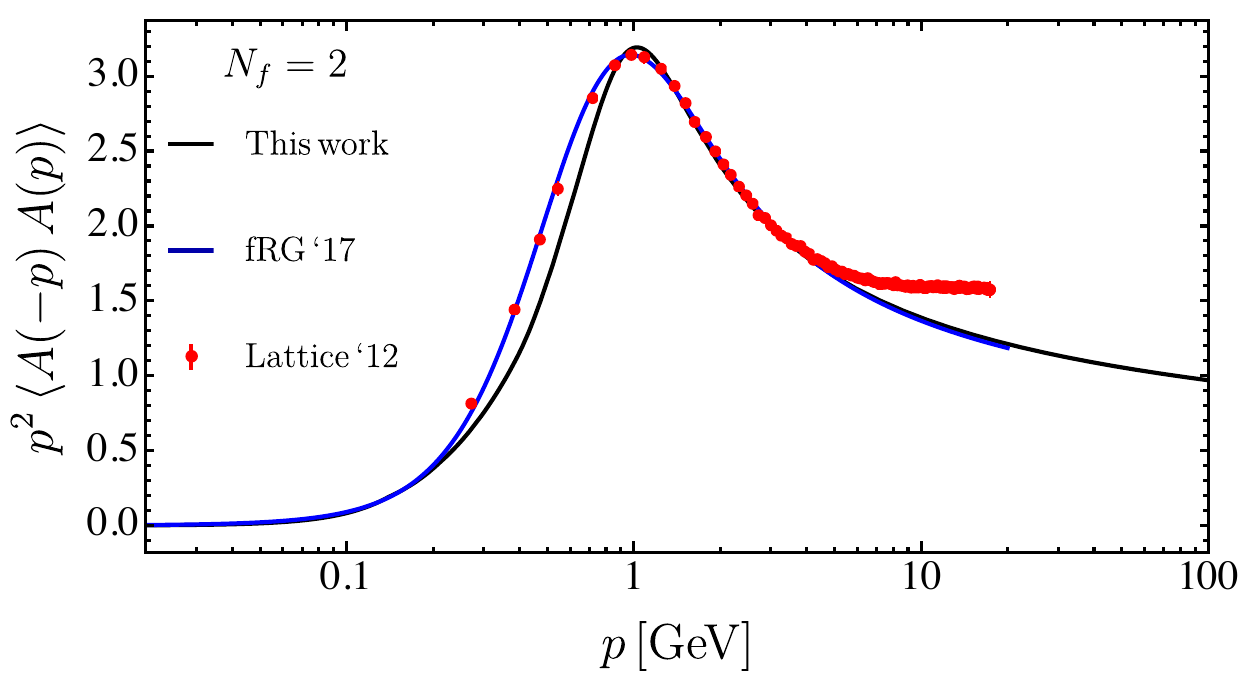}\hspace{.5cm}
	\includegraphics[width=\columnwidth]{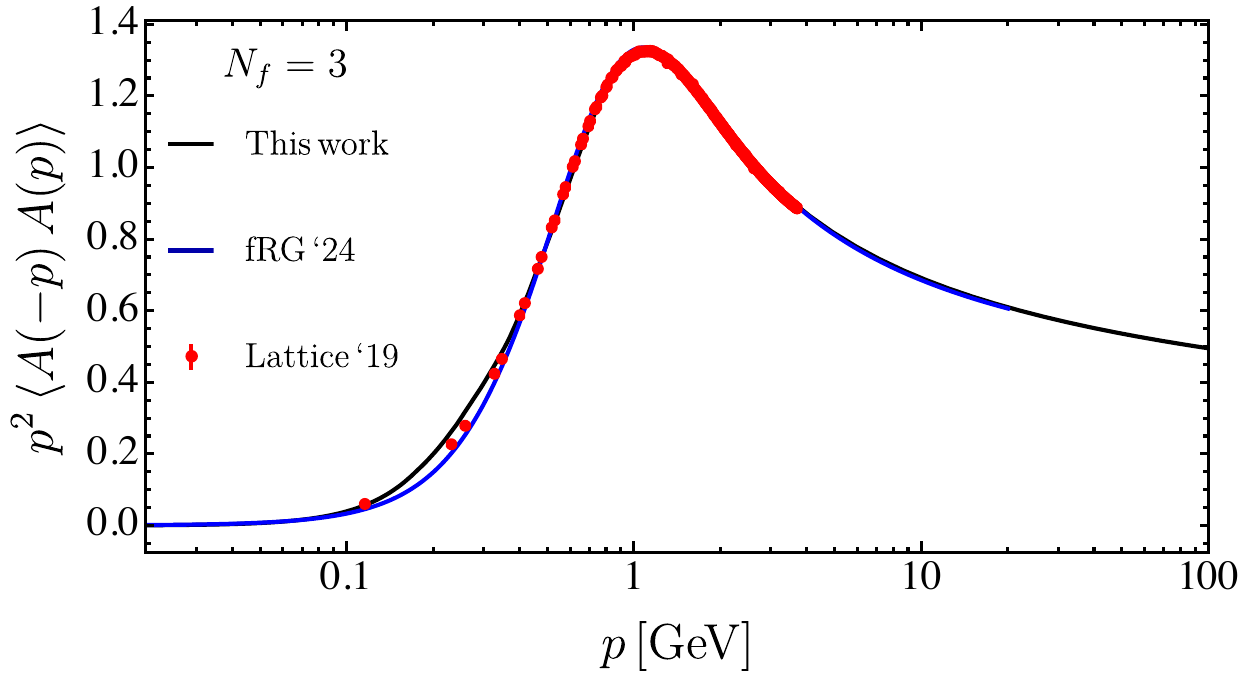}
	\caption{Dressing function $p^2 \,G_A(p)$ of the scalar part $G_A(p)$ of the gluon propagator  \labelcref{eq:ZA(p)}. We compare results for a SU(3) gauge theory with $N_f=2$ and 3 in the approximation used in the present work (black lines), results from quantitative fRG truncations \cite{Cyrol:2017ewj} ($N_f=2$) and \cite{Ihssen:2024miv} ($N_f=2+1$) (blue lines) and lattice simulations, \cite{Sternbeck:2012qs} ($N_f=2$) and \cite{Zafeiropoulos:2019flq} ($N_f=2+1$) (red data points).}
	\label{fig:ZAcomparisonQCD}
\end{figure*}
%

\subsubsection{Linear quark-antiquark potential and the Wilson loop}
\label{sec:LinPot+Wilson}

The mass gap goes hand in hand with a linear potential between colour charges and the respective flux tubes connecting these charges. For a quark--antiquark pair this linear potential is derived from the expectation value of the closed Wilson loop operator, that follows from the operator of a gauge invariant quark-antiquark state 
\begin{align} 
e^{F_{\bar q q}({{\cal C}_{x,y}})}\propto 	\langle \bar q(y) {\cal W}^{\ }_{{\cal C}_{x,y}}q(x)\rangle\,,
	\label{eq:qbarq}
\end{align}
with the Wilson line 
\begin{align}
 {\cal W}^{\ }_{{\cal C}_{x,y}} ={\cal P} \exp \left\{\imag g \int\limits_{{\cal C}_{x,y}} d z_\mu\,A_\mu  \right\}\,,
\label{eq:WilsonLine}
\end{align}
where $\cal P$ stands for path ordering and ${\cal W}$ is the path-dependent Wilson line from $x$ to $y$.
If we consider a quark-antiquark pair, created at a time $t_0$, pulled apart and kept at a spatial distance $r=\|\boldsymbol{x}-\boldsymbol{y}\|$ and annihilated at some later time $t_1$, the path ${\cal C}_{x,y}$ simply is the world line of this process. Hence, it is a closed rectangular curve enclosing an area $\cal A$, and confinement is tantamount to an area law for \labelcref{eq:qbarq} for large areas ${\cal A}\to \infty$:  
\begin{align} 
e^{F_{\bar q q}}({\sigma \cal A}\to \infty) \propto e^{-\sigma {\cal A}}\,, 
	\label{eq:AreaLaw}
\end{align}
with the string tension $\sigma$, which is related to $\Lambda_\textrm{QCD}$ and the mass gap of QCD. The area law entails the linear potential if $t_0,t_1$ is kept fixed and the distance $r$ is increased.

\subsubsection{Confinement and the gluon mass gap}
\label{sec:LinPot+Wilson2}

In the present work we use a gauge-fixed functional approach to many flavour QCD, built upon correlation functions of gluons, ghosts and quarks. How confinement or rather the linear potential and the area law \labelcref{eq:AreaLaw} can be extracted from gluon, ghost and quark correlation functions in such a gauge-fixed functional approach has been studied for decades both in the vacuum and at finite temperature, for a recent review see \cite{Dupuis:2020fhh}. At the center of these studies are the gluon, ghost, and quark propagators in the background of non-trivial gauge field configurations and the respective field strengths. For example, confinement leads to non-vanishing expectation values of the field strength squared (Savvidy vacuum) as well as its chromo-electric and chromo-magnetic components, see \cite{Eichhorn:2010zc, Horak:2022aqx, Pelaez:2024mtq}. In these works, the existence and the quantitative values of the non-trivial condensates are computed from the propagators. Most importantly, it turns out that both, the underlying mechanisms and the existence of the mass gap and the linear potential/area law, as well as the quantitative values of the respective observables are directly deduced from the mass gap $m_{\textrm{gap}}$ in the gluon propagator, the lack of a mass gap in the ghost propagator and the chiral dynamics of the quark propagator. 

This mass gap is well-captured by the dressing function $p^2 G_A(p)$ of the gluon propagator in the Landau gauge, see \Cref{fig:ZAcomparisonQCD} for a comparison of lattice and functional renormalisation group data in $N_f=2$ and $N_f=2+1$ flavour QCD. The function $G_A$(p) is the scalar part of the gluon propagator 
\begin{align}\nonumber 
	G_A{}^{ab}_{\mu\nu}(p,q)=&\, \langle A^a_\mu(p) A^b_\nu (q) \rangle\\[1ex]
	= &\, (2\pi)^4\delta(p+q) \, \delta^{ab} \,\Pi^\bot_{\mu\nu}(p)\,G_A(p) \,,
	\label{eq:GA}
\end{align}
with the transverse projection operator  
\begin{align}
	\Pi^\bot_{\mu\nu}(p)=\left( \delta_{\mu\nu}-\frac{p_\mu p_\mu}{p^2}\right) \,. 
\label{eq:Pibot}
\end{align}
In the Landau gauge here used, the gluon propagator is transverse which is one of the many technical simplifications gained in this gauge. The gluon mas gap is conveniently defined as the inverse correlation length or screening mass of the gluon two-point function. This information is stored in the dressing $1/Z_A(p)$ of the gluon propagator with 
\begin{align}
	G_A(p) = \frac{1}{Z_A(p) } \frac{1}{p^2}\,.
\label{eq:ZA(p)}
\end{align}
While the propagator is not renormalisation group invariant, the respective correlation length is. Its presence in the gluon propagator leads to the behaviour 
\begin{align}
	\lim_{{p^2}/{\Lambda^2_\textrm{QCD}}\to 0} \frac{1}{Z_A(p)}= 0\,, 
	\label{eq:GAGap}
\end{align}
for the dressing $p^2 G_A(p)=1/Z_A(p)$ of the gluon propagator. We emphasise that the value of the gluon propagator at $p=0$ is not the gluon mass gap, which is rather determined by its singularity structure in the complex plane, see e.g.~\cite{Cyrol:2018xeq, Horak:2022myj, FP2W2024, HPW}: it is the distance of the closed singularity of the gluon propagator to the Euclidean frequency axis (not counting possible cuts at $p=0$) and is of the order $1$\,GeV in physical QCD. Moreover, it can be shown that it is gauge-independent, even though it is extracted from a gauge-variant correlation function. Indeed, it is this mass gap that is directly related to the glueball mass, see \cite{Pawlowski:2022zhh}. 

For a more detailed analysis as well as the numerical extraction of the gluon mass gap we refer to the works~\cite{Cyrol:2017qkl, Cyrol:2018xeq, Horak:2022myj, HPW} and in particular \cite{FP2W2024}. Here, we shall use an easy-access proxy for the gluon mass gap: \Cref{eq:GAGap} implies a peak of the dressing as it also decays (logarithmically) for $p^2/\Lambda^2_\textrm{QCD}\to \infty$. As observables are computed from the dimensionless parts of the correlation functions, it is precisely the dressing $p^2 G_A(p)$ that enters the computation and its peak position is a good approximation for its mass gap and hence a signature for confinement and the strength of the linear potential. This is nicely illustrated in \Cref{fig:ZAcomparisonQCD} for $N_c=3$ and $N_f=2,3$ with the peak position at approximately 1\,GeV.

\subsubsection{Confinement-deconfinement phase transtion}
\label{sec:Conf-deconf}

A variant of \labelcref{eq:AreaLaw} also serves as an order parameter for the confinement-deconfinement phase transition at finite temperature or at a finite spatial distance. We discuss this here for a twofold purpose. Firstly, in a follow-up work we shall extend the present set-up to finite temperature, aiming at a study of signatures of phase transitions in the early universe within the present class of BSM models. Secondly, the study of the confinement-deconfinement phase transition illustrates very nicely the relation between confinement and the gluon mass gap present in the gluon propagator in covariant gauges: while the propagator is a gauge-variant quantity, as stressed before, its mass gap carries a gauge invariant information and it is directly related to the physical mass gap in QCD  measured with the scalar glueball mass and the confinement-deconfinement phase transition temperature~$T_\textrm{conf}$. 

At finite temperature we consider the free energy of a static quark at the location $\vec x$ with a static antiquark  at the location $\vec y$. This is related to the correlation  function of the Polyakov loop (Wilson loop in the time direction) and its adjoint, 
\begin{align} 
\langle L(\vec x)\, L^\dagger (\vec y)\rangle \,,\qquad L(\vec x)= \frac{1}{N_c} \textrm{tr}\, {\cal P} e^{ \imag g \int_0^\beta A_0(\tau , \vec x) }\,, 
\label{eq:LbarL}
\end{align}
with $\beta=1/T$. Roughly speaking, \labelcref{eq:LbarL} is a Wilson loop that winds around the full time direction at finite temperature with $\tau \in [0,\beta]$. For asymptotically large distances $r =\| \vec x - \vec y\|$ and small temperatures \labelcref{eq:LbarL} shows the area law with ${\cal A} = \beta \, r$ which implies a linear confinement potential. This entails  
\begin{align}
	\langle L(\vec x)\rangle = \left\{ 
	\begin{array}{rcl} 
		0 & \qquad & T<T_\textrm{conf} \\[1ex] 
        \neq 0 & & T>   T_\textrm{conf} 
        \end{array}                     
\right.\,.
\end{align}
For asymptotically large temperatures the Polyakov loop tends towards unity, which signals the deconfined phase. Here, the  dynamics of the $A_0$-component is perturbative and the temporal background gauge field $\bar A_0$ takes a vanishing value, $\bar A_0=\langle A_0\rangle = 0$. In turn, $\langle L(\vec x)\rangle=0$ implies a non-trivial $\bar A_0$, its value depending on the gauge group. 

The respective order parameter potential is denoted by $V_\textrm{eff}(A_0)$, and $\bar  A_0$ is the solution of the equations of motion (EoM). In the fRG, this effective potential can be computed from the gluon, ghost and quark propagators alone. It can be shown that the gluon contribution leads to a deconfining potential with a minimum at $\bar A_0=0$. At large temperatures, one even has an analytic access to $V_\textrm{eff}(A_0)$ within thermal perturbation theory, see \cite{Gross:1980br, Weiss:1980rj}. In turn, the ghost loop, including the minus sign, generates a confining potential, which also can be computed analytically at large temperatures. There, the gluon contribution  dominates, leading to deconfinement. Note that both, the chromo-electric (parallel to the heat bath) and chromo-magnetic (perpendicular to the heat bath) gluon propagators, are gapped at large temperatures  as the gluonic mass gap receives a thermal contribution, similar to the standard Debye or thermal screening mass, see \cite{Cyrol:2017qkl}. 

At sufficiently low temperatures, the gluonic contribution is effectively switched off due to the gluon mass gap present in the vacuum, reflected in \labelcref{eq:GAGap}. This has first been discussed in \cite{Braun:2007bx}, for further works see \cite{Braun:2010cy, Fister:2013bh} and the review \cite{Dupuis:2020fhh}. This entails that a necessary and sufficient condition for confinement is the relative suppression of the gluon loop in comparison to the ghost loop.

\subsubsection{Signatures of confinement in Landau gauge QCD}
\label{sec:ConfWrapup}

To wrap up, in Landau gauge QCD, confinement at vanishing temperature is in one-to-one correspondence to the existence of a mass gap in the gluon propagator signalled by a vanishing dressing at vanishing momentum, see \labelcref{eq:GAGap}. Its size is given by the distance of the closed singularity in the complex plane to the Euclidean frequency axis (apart from the cut at $p=0$). This distance is related to the location of the maximum of the dressing and we shall use the value
\begin{align}
p_\textrm{peak}:\  p_\textrm{peak}^2 G_A(p_\textrm{peak}) \geq p^2 G_A(p)\,,\qquad \forall p\in \mathbbm{R}\,
\label{eq:ConfScale}
\end{align}
for the mass gap. This concludes our discussion of the signatures of confinement in Landau gauge QCD.

\subsection{Dynamical chiral symmetry breaking }
\label{sec:dSSB}

For a sufficiently small number of flavours the (perturbative) $\beta$-function of many-flavour QCD is negative and entails asymptotic freedom with $\alpha_g(p\to \infty) \to 0$, where 
\begin{align}
	\alpha_g(p) =\frac{1}{4 \pi} g^2(p)\,, 
\end{align}
with the running gauge coupling $g(p)$. This entails that qauge-fermion systems with a sufficiently small number of flavours exist in the ultraviolet (UV) and are asymptotically free. In turn, towards the IR the gauge coupling rises and finally triggers dynamical (strong) chiral symmetry breaking. In the chiral limit the occurrence of massless pseudoscalar composite Goldstone bosons, the pions, is reflected in a $p=0$ singularity of the dressing $ \lambda_\textrm{\tiny{SP}}$ of the pseudoscalar channel of the four-quark interaction, defined by   
\begin{align}\nonumber 
	 \lambda_\textrm{\tiny{SP}}(p_1,...,p_3)\,(2\pi)^4 \delta(p_1+\cdots+p_4) &\\[1ex] 
	 &\hspace{-3cm} \propto \left. \langle  \psi(p_1)  \psi(p_3) \bar\psi(p_2)\bar \psi(p_4)\rangle_\textrm{1PI}  \right|_{\rm (S-P)}\,,
\end{align}   
where the subscript ${}_\textrm{(S-P)}$ indicates the projection on the scalar-pseudoscalar channel. The four-fermion scattering vertex admits an expansion in a rather large complete basis of tensor structures, that carry colour, Dirac and flavour indices. For example, restricting ourselves to the case of two-flavour physical QCD, there are already ten momentum-independent tensor structures, and the complete basis has hundreds of elements. These typically include a scalar-pseudoscalar element whose precise definition depends on the basis chosen: it is not unique due to the Fierz ambiguity, see e.g.~the fRG review \cite{Braun:2011pp}. 

The vanishing momentum limit of the dressing $ \lambda_\textrm{\tiny{SP}}(p_1,p_2,p_3)$ at the symmetric point $p$ behaves like 
\begin{align}
	 \lambda_\textrm{\tiny{SP}}(p\to 0) \propto \frac{1}{p^2} \,.  
\end{align}
Strictly speaking, it is the $t$-momentum channel that carries the pion singularity. In the fRG  approach this phenomenon is captured in a rather simple form in terms of the RG flow of the four-quark scattering vertex at vanishing momentum, where the fRG cut-off scale $k$ takes over the rôle of an average symmetric point momentum scale. This will be discussed in detail in \Cref{sec:truncation}. The respective order parameter for dynamical chiral symmetry breaking is the chiral condensate, 
\begin{align}
\Delta= \int_x \langle \bar\psi(x) \psi(x)\rangle\,, 
\label{eq:OrderChiral}
\end{align} 
and its third root sets its scale.

\subsection{Conformality and Caswell-Banks-Zaks fixed~point}
\label{sec:CBZ}

The dynamical phenomena of confinement and $\chi{\rm SB}$ with their emergent mass scales constitute  the non-perturbative character of gauge-fermion theories. In the scaling regime of these theories both phenomena are absent by definition. 
This regime is approached by softening the growth of the gauge coupling towards the IR via the enhancement of anti-screening fermionic corrections by incrementing $N_f$. 
This is well-captured by the beta function $\beta_g$ of the strong coupling. As we aim at a description of physics in the conformal regime with $\beta_g\to 0$ with a relatively small fixed point coupling, we approach this problem within a perturbative expansion. In the chiral limit and hence in the absence of any explicit mass scales we find 
\begin{align}
	\label{eq:betafunctiong_perturbation}
	\beta(g,N_c,N_f)= &-g\,\sum_{n=1}^{\infty}  \left(\frac{g^2}{16 \pi^2 }\right)^{n}\,\beta^{(n)}(N_c,N_f)
\end{align}
with 
\begin{align} \nonumber 
	\beta^{(1)}(N_c,N_f)&=\left(\frac{11}{3}C_{\rm A }-\frac{4}{3}T_{\rm F} N_f\right) 
	\,,\\[1ex]
	\beta^{(2)}(N_c,N_f)&=\bigg(\frac{34}{3}C_{\rm A }- \left[4 C_{\rm F} +\frac{20}{3}C_{\rm A}\right]T_{\rm F} N_f \bigg)\,,
	\label{eq:one-twoloop}
\end{align}
being the one and two loop terms in the expansion. Here, $C_{\rm A }$ and $C_{\rm F}$  are the Casimir operators of the colour group and  of the representation under which the fermions transform, respectively, and $T_{\rm F}$ is the Dynkin index of the latter. 

Roots of the beta function correspond to fixed point solutions $g^*$,
\begin{align}
 \beta(g^* ,N_c,N_f)=0\,.
\label{eq:CBZ fixed point}
\end{align}
For sufficiently small number of flavours, $N_f < N_f^* <  \left(11 C_{\rm A }\right)/\left(4 \,T_{\rm F}\right)$, the existence of an UV Gau\ss ian fixed point $g^*$ in \labelcref{eq:CBZ fixed point} is ensured by the one loop term in \labelcref{eq:one-twoloop}, leading to asymptotically free theories as in the physical QCD case. In this regime with sufficiently small $N_f$ the theory also features $\dSSB$ and confinement.  In turn, for sufficiently large number of flavours, $N_f >  \left(11 C_{\rm A }\right)/\left(4 \,T_{\rm F}\right)$, the UV Gau\ss ian fixed point turns into a Gau\ss ian IR fixed point with a strongly correlated UV regime, similarly to the situation in quantum electrodynamics with its  UV Landau pole in the chiral limit. 

Finally, in an intermediate regime with $N^*_f < N_f <  \left(11 C_{\rm A }\right)/\left(4 \,T_{\rm F}\right)$, non-trivial roots with $g_*\neq 0$ of the beta function exist. For example, considering up to two loops, one can find $(N_c,N_f)$-pairs for which \labelcref{eq:CBZ fixed point} is satisfied for a non-vanishing value $g^*$ of the gauge coupling, which signals an interacting IR fixed point known as Caswell-Banks-Zaks (CBZ) fixed point \cite{Caswell:1974gg,Banks:1981nn}. In this regime,  correlation functions are invariant under scale transformations at the fixed point and enjoy quantum scale invariance, elevating  that of the classical gauge-fermion action to the quantum effective action. 

In the absence of threshold effects and for mass-independent RG schemes, the beta functions of marginal couplings are known to be universal up to two loops \cite{Weinberg:1996kr}, meaning that their form is scheme independent. The existence of the perturbative CBZ fixed point is present at the two-loop level and beyond, consequently expected to be universal in the weak coupling limit. 

Perturbative approaches have proven to be very useful in the determination of this fixed point as they account in a systematic manner for the higher-order contributions in powers of the gauge coupling. This allows to properly account for colour and flavour factors which cancellation between loop orders leads to the CBZ fixed point. Currently, $\overline{\rm MS}$ results are available up to five-loop orders \cite{Herzog:2017ohr} and re-summation techniques have been developed to improve the highest order results \cite{DiPietro:2020jne}. 

While the upper boundary of the CBZ conformal window in the $N_f-N_c$ plane of theories is well-known, the lower boundary towards the QCD-like regime is given by the  loss of quantum scale invariance by the dynamical emergence of a scale. Due to the non-perturbative nature of these phenomena, the location of the lower boundary cannot be exactly computed and remains a topic of research. Generally, the disappearance of the CBZ fixed point \labelcref{eq:CBZ fixed point} can occur in three ways \cite{Kaplan:2009kr}:
\begin{enumerate}
	\item \underline{\it Fixed point goes to 0:} This is the case of the upper boundary of the conformal window where asymptotic freedom is lost into the non-Abelian-QED regime. This is caused by a change of sign in the one-loop beta function term \labelcref{eq:one-twoloop}.
	\item \underline{\it Fixed point goes to $\infty$:} The gauge fixed point coupling disappears into the strong limit where the perturbative expansion breaks down before the fixed-point solution disappears. This is the scenario found in the lower boundary of the CBZ window for two-, three- and four-loop $\overline{\rm MS}$ results. Perturbative approaches fail to properly pin-point the precise value given that they cannot describe the dynamical appearance of scales. We will add on this point for the remaining of this Section and \Cref{app:gammam}. 
	\item \underline{\it Fixed-point merger into the complex plane:} This scenario is given as the CBZ and QCD$^\star$ fixed points merge, disappearing in the complex plane \cite{Kusafuka:2011fd,Antipin:2012kc,Gukov:2016tnp}. This is found only for five-loop $\overline{\rm MS}$ results where the fixed point disappears at weak coupling values. 
\end{enumerate}

Despite their high success and practical use, perturbative approaches lack the capability to track the emergence of dynamical phenomena such as $\chi{\rm SB}$. In other words, such approaches can provide strong coupling fixed-point solutions which are unviable given the prior appearance of dynamics.

A common approach employed to estimate whether ${\rm d}\chi{\rm SB}$ occurs at perturbatively computed fixed points is to analyse the magnitude of the fermion mass operator anomalous dimension~\cite{Yamawaki1996,Miransky1989,Miransky1989a,Miransky1989b,Miransky1997,Miransky:1998dh,Appelquist1988,Appelquist:1996dq,Dietrich:2006cm,Appelquist:1998rb},
\begin{align}\label{eq:gammampert}
	\gamma_m &=\gamma^{(1)}_m+\gamma^{(>1)}_m= -\frac{3\, \alpha_{g} \, C_{\rm F} }{2 \pi}+ \ldots
\end{align}
where we have expanded in powers of $\alpha_g$ singling out the one-loop contribution $\gamma^{(1)}_m$.  Commonly, ${\rm d}\chi{\rm SB}$ is said to occur in theories where $\left|\gamma^{(1)}_m\right|_{g^*}\gtrsim 1$ providing an approximate methodology to determine the boundary of the conformal window \cite{Ryttov2011,Ryttov2016,Ryttov2016a,Ryttov2017,Ryttov2018,Lee2021}. 

On the other hand, non-perturbative approaches have investigated self-consistently the close-conformal regime of theory space. Fully self-consistent computations employing the Dyson-Schwinger equation (DSE) in the many flavour limit have been performed \cite{Hopfer:2012qr,Zierler2023} and the boundary of the conformal window is found at a rather low number of flavours $N_f^{\rm crit}=4.67\pm0.02$ for $N_c=3$~\cite{Zierler2023}. This is known to be caused by too weak gauge-fermion dynamics.  Additionally, see \cite{Aoki:2012ve} for a DSE scaling analysis at large $N_f$ and \cite{Miura:2015mna} for computations with an IR cut-off.

Monte Carlo lattice simulations have also investigated the many-flavour region, see eg. \cite{Appelquist:2007hu,DeGrand:2009mt,Iwasaki:2003de,Appelquist:2011dp} and  \cite{Hasenfratz:2023sqa,Ayyar:2024dmt,LSD:2014obp,LatticeStrongDynamicsLSD:2013elk,DeGrand:2015zxa,DelDebbio:2010hx,LSD:2009yru} for studies in the context of BSM physics. However, even contemporary state-of-the-art lattice simulations  only consider gauge-fermion systems significantly away from the chiral limit (hence with sizeable fermion masses). This certainly complicates the interpretation of lattice results in the walking regime and conformal window. 

There have been various efforts studying  $N_f=12$ \cite{Peterson:2024pfn,Hasenfratz:2016dou,Aoki:2012eq}, showing strong numerical evidence for the existence of an IR fixed point. Additionally, there are studies which suggest $N_f=10$ is inside the conformal window \cite{LatticeStrongDynamics:2020uwo,Hasenfratz:2023wbr,Chiu:2016uui,Hayakawa:2010yn,Hasenfratz:2020ess,Hasenfratz:2017qyr}. For lower number of flavours well within the non-conformal regime, the gauge beta functions have been determined for $N_f=6,\,4$ \cite{Hasenfratz:2022yws} and  $N_f=8$ \cite{LatKMI:2016xxi, Hasenfratz:2022zsa,LatticeStrongDynamics:2018hun}.  

Within the fRG approach, the appearance of $\dSSB$ can be clearly traced, as will be demonstrated in \Cref{sec:emergentcomposites,sec:ScalechiralSB,sec:BoundaryCBZ}. This phenomenon can be diagnosed using the four-Fermi framework through a singular coupling \cite{Braun:2010qs,Braun:2009ns,Gies:2005as,Gies:2003dp}, or within the bosonized formulation \cite{Gies:2001nw,Pawlowski:2005xe} by the emergence of a non-trivial minimum in the chiral potential. Consequently, the fRG approach offers distinct advantages in determining the onset of a dynamical scale and, thus, the boundaries of the conformal window. In fact, this approach has been employed for this purpose in \cite{Gies:2005as}. In \Cref{sec:BoundaryCBZ}, we address this task in more detail and present quantitative and qualitative improvements in estimating the boundary of the CBZ window.

In the vicinity of the conformal window, but below the boundary curve $N_f^{\rm crit}(N_c)$, gauge-fermion theories are expected to exhibit close conformal scaling where $\dSSB$ and confinement are still present. Specifically, they will show a very slowly growing coupling along a near conformal regime, commonly known as {\it walking regime}. This phenomenon has been widely employed in BSM phenomenology, for example Composite Higgs realisations require of such walking  to generate flavour hierarchies in the fermionic sector, see eg. \cite{Ferretti:2013kya,Kaplan:1991dc,Goertz:2023nii,Sannino:2016sfx,Evans:2005pu,DelDebbio:2010hx,DelDebbio:2010hu,Foadi:2007ue,Gudnason:2006ug}. 
Walking dynamics have been studied with conformal perturbation theory  \cite{Gorbenko:2018ncu}, effective field theories \cite{Benini:2019dfy} and Monte Carlo lattice simulation \cite{LatKMI:2016xxi}. We will study in detail such particular scenarios in \Cref{sec:EmergentWalking}, analysing the interplay of dynamics and the spectrum of the theories from first principles.

\section{Functional RG approach to gauge-fermion theories}
\label{sec:truncation}

In this Section we introduce a unified functional approach to confinement and $\dSSB$, based on the effective action of many-flavour and many-colour QCD, 
\begin{align}
	\Gamma[\Phi]=\Gamma_\textrm{glue}[A,c,\bar c]+\Gamma_\textrm{mat}[\Phi]\,. 
		\label{eq:G-glue-matter}
\end{align}
Here, $\Gamma_\textrm{glue}$ is the pure glue part, that only depends on the gluons $A_\mu^a$ and ghosts $c^a, \bar c^a$, where small Latin letters $a, b,...$ with $a=1,...,N_c^2-1$ label the adjoint representation of the gauge group. Moreover, $\Gamma_\textrm{mat}$ is the gauge-fermion part, which vanishes for vanishing fermion fields $\psi,\bar \psi=0$, where the gauge group indices of the representation under which the fermions transform is kept implicit. The field $\Phi$ in \labelcref{eq:G-glue-matter} is the super mean field that collects all mean-field components, 
\begin{align}
 \Phi = (A_\mu, c, \bar c, \psi,\bar \psi)\,, \qquad \Phi=\langle \hat \Phi\rangle\,,
	\label{eq:Superfield}
\end{align}
where $\hat\Phi$ indicates the field operator. The $n$th field derivatives of \labelcref{eq:G-glue-matter},   
\begin{align}
	\Gamma^{(n)}_{\Phi_{i_1}\cdots \Phi_{i_n}}(p_1,...,p_n) =\frac{\delta}{\delta \Phi_{i_1}(p_1)}\cdots \frac{\delta}{\delta\Phi_{i_n}(p_n)}\Gamma[\Phi]\,, 
	\label{eq:DefGn}
\end{align}
are the one-particle irreducible parts of the respective $n$-point functions $\langle \hat\Phi_{i_1} \cdots \hat\Phi_{i_n} \rangle$. We shall use the notation for functional derivatives in \labelcref{eq:DefGn} not only for the effective action but for general functionals. 

The correlation functions $\Gamma^{(n)}$ carry all physics information of the theory, and one can expand the effective action accordingly, 
\begin{align}
	\Gamma[\Phi]= \sum^{\infty}_{n}  \int_{\boldsymbol{p}}  \Gamma^{(n)}_{\Phi_{i_1}\cdots \Phi_{i_n}}(\boldsymbol{p})\, \Phi_{i_n}(p_n)\cdots  \Phi_{i_1}(p_1)\,,
\label{eq:DefGammaVertexExpansion}
\end{align}
with the multi-momentum variable $\boldsymbol{p}=(p_1,...,p_n)$. In the following we shall access the effective action by computing its expansion coefficients $\Gamma^{(n)}(\boldsymbol{p})$ from their functional renormalisation group flows, which are one-loop exact relations. 

In the following \Cref{sec:fRG} we briefly introduce the functional renormalisation group. In \Cref{sec:econf} we detail the bootstrap implementation of confinement via a mass gap. In \Cref{sec:emergentcomposites} we focus on the fermionic sector and discuss the emergent composites formalism which provides quantitative access to bound states and ${\rm d}\chi{\rm SB}$. Finally, in \Cref{sec:couplings}, we summarise our approximation and list the flows derived for this truncation.

\subsection{The functional renormalisation group}
\label{sec:fRG}

The correlation functions \labelcref{eq:DefGn} can be computed within functional approaches such as the functional renormalisation group (fRG) or Dyson-Schwinger equations (DSEs). In these approaches the correlation functions of gauge-fermion systems obey  coupled sets of diagrammatic loop equations. In the present work we use we the functional RG, formulated for the 1PI effective action~\cite{Wetterich:1991be,Morris:1993qb,Ellwanger:1993mw}. In this approach an IR cut-off term is added to the classical action in the path integral, effectively suppressing the low momentum modes. This cut-off term is quadratic in the fields,  
\begin{align}
	\Delta S_k[\Phi]=\frac{1}{2}  \int_p \, \Phi(-p) \,R_k(p) \,\Phi(p)\,, 
	\label{eq:Regulator}
\end{align} 
with the block diagonal regulator matrix $R_k(p)$ specified in \labelcref{eq:QCD-RegulatorMatrix} in \Cref{app:regs}. 
Its components are momentum dependent mass functions $R_A(p), R_c(p), R_\psi(p)$, that decay rapidly in the UV for momenta $p^2/k^2 \gtrsim 1$, and act as a mass terms for momenta  $p^2/k^2 \lesssim 1$, suppressing the respective momentum modes. 

In the presence of \labelcref{eq:Regulator}, the effective action analogue depends on the cut-off scale $k$, $\Gamma[\Phi]\to \Gamma_k[\Phi]$, and consequently lacks the physics of smaller momenta. By lowering the cut-off scale $k$ infinitesimally, the physics of the momentum shell $p^2 \approx k^2$ are progressively included into the effective action. Hence, the scale-dependent effective action $\Gamma_k[\Phi]$  interpolates between the classical or UV relevant action $S[\Phi]$ and the full effective action at $k=0$, $\Gamma[\Phi]=\Gamma_{k\to 0}[\Phi]$. The evolution of the scale-dependent effective action  is described by the Wetterich or flow equation \cite{Wetterich:1992yh}, 
\begin{align}
	\partial_t \Gamma_k [\Phi ] &= \frac{1}{2} \textrm{Tr}\left[ \frac{1}{\Gamma^{(2)}_k[\Phi] +R_k }\, \partial_t R_k \right]\, ,
	\label{eq:Floweq}
\end{align}
with the logarithmic scale derivative or (negative) RG-time 
\begin{align}
t = \log \frac{k}{k_\textrm{ref}}\,.
\end{align}
Here, $k_\textrm{ref}$ is a suitable reference scale, typically chosen to be the initial UV cut-off scale, $k_\textrm{ref} = \Lambda_{\rm UV
}$. This leads to $t(k=\Lambda_{\rm UV})=0$ and $t(k=0) =-\infty$. 

The fRG has been applied to a broad range of physics areas and phenomena, see \cite{Dupuis:2020fhh} for an overview. Specifically interesting for the present work are its applications to physical QCD, see in particular \cite{Braun:2014ata, Mitter:2014wpa, Cyrol:2017ewj, Cyrol:2017qkl, Corell:2018yil, Fu:2019hdw, Ihssen:2024miv}, and the many flavour limit, see \cite{Braun,Gies:2005as, Braun:2006jd,Braun:2009ns, Braun:2010qs} which provide the technical and conceptual foundations. Investigations of SM-like theories and BSM extensions have also been performed, see eg.~\cite{Gies:2013fua, Gies:2014xha, Eichhorn:2015kea, Pawlowski:2018ixd, Held:2018cxd, Eichhorn:2020upj, Pastor-Gutierrez:2022nki, Goertz:2023nii,Gies:2023jzd}.

\subsection{Gauge dynamics and the mass gap}
\label{sec:econf}

The analysis of the $N_f,N_c$-dependence of gauge-fermion systems requires a flexible fRG setup which allows to scan efficiently and reliably the $N_f-N_c$ theory plane, and in particular the presence or absence of confinement signalled by the gauge field mass gap. In this work we employ the effective average action for the gauge sector \labelcref{eq:Gglue}, discussed in detail in \Cref{app:Effective action}. This truncation has been employed in quantitative QCD studies \cite{Ihssen:2024miv, Cyrol:2017ewj, Cyrol:2016tym, Fu:2019hdw}, where so far the scalar part $G_A(p)$ of the cut-off-dependent gluon propagator \labelcref{eq:GA}, 
\begin{align} 
	G_{A,k}(p) =\frac{1}{Z_{A,k}(p^2) \,p^2 + R_{A,\,k}(p)}\,,
	\label{eq:GAk}
\end{align}
has been computed in pure Yang-Mills (YM) \cite{Cyrol:2016tym,Cyrol:2017qkl,Corell:2018yil}, for $N_f=2$ \cite{Cyrol:2017ewj} and  2+1 flavour QCD \cite{Fu:2019hdw, Ihssen:2024miv}. These quantitative works required of a considerable numerical effort to include momentum-dependent vertices for the sake of quantitative precision. The gluon propagator dressing is obtained from \labelcref{eq:GAk} in the $k\to 0$ limit  and that from \cite{Cyrol:2017ewj} and \cite{Ihssen:2024miv} are shown in the left and right panels of \Cref{fig:ZAcomparisonQCD}, respectively. In these works, the confining solution is obtained within a bootstrap approach provided by the existence of a unique confining solution. In the Kugo-Ojima confinement scenario \cite{Kugo:1979gm} in the Landau gauge, the ghost and gluon propagators display a scaling IR limit with 
\begin{align}
	Z_A(p^2\to 0) \propto (p^2)^{\kappa_A} 
	\label{eq:kappaA}
\end{align}
with a scaling coefficient $\kappa_A \approx -1.2$ in YM. This solution in particular requires the existence of a BRST charge and not only that of infinitesimal BRST transformations, for a detailed discussion see \cite{Fischer:2008uz, Cyrol:2016tym}. In the fRG approach, the precise scaling in \labelcref{eq:kappaA} is arranged by a unique combination of the UV relevant parameters at a large cut-off scale. This is a numerically challenging finite-tuning problem, which comes with high numerical costs.

\subsubsection{Novel expansion scheme capturing confinement}
\label{sec:ExpandConfinement}
 
Below we discuss a novel approximation scheme that reduces the numerical costs significantly and allows for semi-analytical access to the correlation functions describing confinement. It is key to this work as allows to depart towards the uncharted many-flavour limit of QCD-like theories. Importantly, the present novel scheme keeps semi-quantitative reliability even in the strongly correlated regime of physical QCD with $N_f\leq 3$. 

The starting point for the crucial computational reduction is the following observation that also underlies previous expansion schemes used in \cite{Braun:2014ata, Rennecke:2015eba, Fu:2019hdw, Ihssen:2024miv}: In the diagrams of flow equations for correlation functions $\partial_t \Gamma^{(n)}(p_1,...,p_n)$ with $p_i^2\lesssim k^2$, the scalar part of the gluon propagators is well approximated by 
\begin{align}
		G_{A,k}(l) \approx \frac{1}{ Z_{A,k}  \left( l^2 + m_{\textrm{gap},k}^2\right)+ R_{A,k}(l)}\,, 
	\label{eq:GAk2}
\end{align}
where the different propagators in the loop carry the momentum $l=q+p$, where $p$ stands for sums of the external momenta: This includes $p=0$ for the cutoff line proportional to $G_A(q) \partial_t R_A(q) G_A(q)$, where we suppressed the subscript ${}_k$. The above setup implies that $l^2\lesssim k^2$ due to the restricted external momenta and loop momentum $q^2\lesssim k^2$. The latter property follows from the regulator insertion $\partial_t R_k(q)$, that drops off quickly for $q^2 \gtrsim k^2$. 	\Cref{eq:GAk2} leads to a significant simplification of the computations as the right hand side only depends on two momentum-independent parameters, the wave function $Z_{A,k}$ and the mass gap $m_{\textrm{gap},k}$. 

We proceed with a discussion of the semi-quantitative nature of this approximation. To begn with, in \cite{Braun:2014ata, Rennecke:2015eba, Fu:2019hdw, Ihssen:2024miv} this insight has been exploited for the computation of the gluon propagator for $N_f=2$ and $2+1$ flavour QCD in the vacuum and at finite temperature and density. In these works, $m^2_{\textrm{gap},k}=0$ has been used, as only flavour, temperature and density corrections of the gluon propagator have been computed on top of the input of vacuum YM and $N_f=2$ propagators. In particular, the quantitative approximation 
\begin{align}
	Z_{A,\bar k}(p^2)\, p^2 \approx  \left.Z_{A,k}(k^2 +m_{\textrm{gap},k}^2)\right|_{k\approx  p}\,, 
	\label{eq:ZAk2}
\end{align}
is used to determine the two momentum-independent parameters on the right-hand side, from the full momentum dependent wave function $Z_{A,k}(p^2)$. In \labelcref{eq:ZAk2}, $\bar k$ on the left hand side is either used with the same relation as on the right hand side, $\bar k\approx p$, or $\bar k=0$ is chosen. The latter approximation allows one to also use input from other functional approaches or lattice data. Then, the result of the above fit is used on the flow equations of the difference $\Delta Z_{A,k}$ of, e.g., the $2+1$ flavour propagator and the two flavour or YM propagators. The results can then be used to approximate the full two-point functions by using the results on the right-hand side of \labelcref{eq:ZAk2}. 

The quantitative reliability of this approach, even if dropping the mass gap, $m_{\textrm{gap},k}\approx 0$, has been confirmed by the comparison to the full results in two and 2+1 flavour QCD. Respective discussions can be found in \cite{Braun:2014ata, Rennecke:2015eba, Fu:2019hdw, Ihssen:2024miv}, and in conclusion the approach with $m_{\textrm{gap},k}\approx 0$ is tailor made for physical QCD as the input (or rather expansion point) from YM theory and two-flavour QCD already induces confinement with the correct physical scales. Moreover, the additional flavours as well as temperature and density are simply perturbations in the pure glue sector. 

For general gauge-fermion systems, such an expansion about a theory with a small numbers of flavours with $m_{\textrm{gap},k}\approx 0$ may not capture the qualitative change of the interplay of confinement and chiral dynamics, in particular in the approach to the conformal window. For this reason we have developed a novel expansion scheme which accommodates the confinement dynamics for all flavours and does not require of any input while additionally keeps the computational simplicity described above. It comes at the price of a minimally reduced quantitative precision for small flavour numbers, see \Cref{fig:ZAcomparisonQCD} and the discussion in \Cref{sec:InterplayLowNf}.  

In contradistinction to the previous works, in our approach the parametrisation \labelcref{eq:GAk2} still depends on two parameters, $Z_{A,k}, m_{\textrm{gap},k}$, whose flows  have to be determined such that the full gauge field propagator, \labelcref{eq:GAk}, and the approximated one, \labelcref{eq:GAk2}, agree best in the momentum regime $l^2 \lesssim k^2$ relevant in the flow equations. Now, we use that all the coupling parameters computed are extracted from $\partial_t  \Gamma^{(n)}(p_1,...,p_n)$ and $p_i$-derivatives thereof, evaluated at vanishing momenta $p_i=0$ for all $i$. This includes the flows of the parameters $Z_{A,k}$ and $m_{\textrm{gap},k}$, the crucial part of this novel scheme. 

Consequently, all gauge field propagators in the diagrams are given by $G_{A,k}(q)$ in \labelcref{eq:GAk2} and its derivatives. They only depend on the loop momentum $q^2\lesssim k^2$ as we evaluate the diagrams at $p_i=0$ for all $i$. The derivation of the respective flows is detailed in \Cref{app:flowmGap} and are chosen such that they optimise and stabilise the approximation \labelcref{eq:GAk2} for $q^2\lesssim k^2$. For this range of loop-momenta, the regulator adds a mass gap $R_{A,k}(q^2 \lesssim k^2) \approx Z_{A,k} k^2$. Accordingly, the term $Z_{A,k} \,q^2$ is sub-leading. We conclude that the quality of the approximation \labelcref{eq:GAk2} for momenta smaller than the cut-off scale is governed by the accurate determination of the gluon two-point function at vanishing momentum,    $Z_{A,k} m_{\textrm{gap},k}^2$. This product is readily and uniquely determined by its flow. In turn, the determination of $Z_{A,k}$ and hence also of  $m_{\textrm{gap},k}^2$ carries an ambiguity and its value depends on the chosen projection procedure. Importantly, its value does not affect the accuracy of the approximation scheme, and it can be utilised to guarantee the stability of the approximation. A detailed analysis and discussion can be found in \Cref{app:flowmGap}.  

In this novel approximation scheme, the confinement scaling condition following from the Kugo-Ojima criterium \labelcref{eq:kappaA}, translates into 
\begin{align}
	Z_{A,k\to 0} \propto  k^{2+2\kappa_A}\to \infty\,
	\label{eq:kappaAk}
\end{align}
and leads to the unique solution for the mass gap. In terms of the initial condition at the UV cut-off scale $k=\Lambda_{\rm UV}$ this entails 
\begin{align}
m_{\textrm{gap},\Lambda_{\rm UV}} = m_{\textrm{scaling},\Lambda_{\rm UV}}\,, 
	\label{eq:mscaling}
\end{align}
which uniquely fixes the whole trajectory $m_{\textrm{gap},k}$. In short, $m_{\textrm{gap}}$ is not a free parameter but is fixed by the scaling condition for confinement. We note in passing that the full propagator also has a logarithmic UV and IR running where the latter leads to a cut at $p=0$. This causes an additional intricacy for the extraction procedure of the mass gap from the full momentum-dependent propagator, and its resolution has been discussed in \cite{FP2W2024}. In the present approximation this intricacy is resolved approximately by the very set-up of the novel expansion scheme. 

It is worth noting that while it is tempting to identify $m_{\textrm{gap},k=0}$ with the confining physical mass gap, the latter rather is the screening mass of the gluon which is related to the peak of the gluon dressing, see \labelcref{eq:ConfScale} in \Cref{sec:ConfWrapup} and the discussion there. In our approximation this information is carried by the scale-dependence of the combination $Z_{A,k} m_{\textrm{gap},k}$ whose accurate determination governs the quantitative reliability of the approximation.  

We close this Section with a remark on the reliability of the novel scheme. In \Cref{fig:ZAcomparisonQCD} we confront the respective results for the $N_f=2$ and 3 gluon propagators using \labelcref{eq:ZAk2}, with fully quantitative ones from functional approaches and lattice results. This illustrates the impressive semi-quantitative performance of the novel expansion scheme. Indeed, except for the deep IR below the confinement scale, we even find quantitative agreement. Note that this IR regime is efficiently suppressed in the flow diagrams due to the loop integration.

\subsubsection{Bootstrap approach to confinement}
\label{sec:UVSimplification}

A full account of the approximation of the gauge part $\Gamma_\textrm{glue}$ of the effective action is  presented in \Cref{app:Effective action}. Here we concentrate on the terms quadratic in the gauge field, 
	\begin{align}\nonumber
		\hspace{-.3cm}\Gamma_{\textrm{glue}}[A,c,\bar c]  =&\,
		\frac12 \int\limits_p \, A^a_\mu(p) \,  \Biggl[  Z_{A}\,\left( p^2+m^2_{\textrm{gap}} \right)\Pi^\perp_{\mu\nu}(p) \\[1ex]
		&\hspace{-1.4cm} +\frac{ 1}{\xi}Z^\parallel_{A}\,\left( p^2 
		+ m^2_{\textrm{\tiny{mSTI}}}\right)  \frac{p_\mu p_\nu}{p^2} \Biggr] \, A^a_\nu(-p)  +\cdots \,,
		\label{eq:GglueA2}
\end{align}
with the transverse projection operator $\Pi_{\mu\nu}^\bot $ defined in \labelcref{eq:Pibot} and $\cdots$ stands for the rest of the glue action in \labelcref{eq:Gglue}. For the sake of simplicity we have dropped the $k$-argument of dressings and couplings. The longitudinal gauge field mass $m^2_{\textrm{\tiny{mSTI}},k}$ is induced by the modified Slavnov-Taylor identities (mSTIs) in the presence of a momentum cut-off. In the physical limit $k\to 0$ the mSTIs approach the standard STIs of Landau gauge QCD and the longitudinal mass vanishes, $m^2_{\textrm{\tiny{mSTI}},k=0}=0$. Hence, $m^2_{\textrm{\tiny{mSTI}},k}$ does not signal a massive YM theory, but the modification of the Landau gauge STIs in the presence of the IR regulator. For a detailed discussion of this subtlety see e.g.~\cite{Cyrol:2016tym} and the review \cite{Dupuis:2020fhh}. 

\begin{figure}[t]
	\centering
\includegraphics[width=.95\columnwidth]{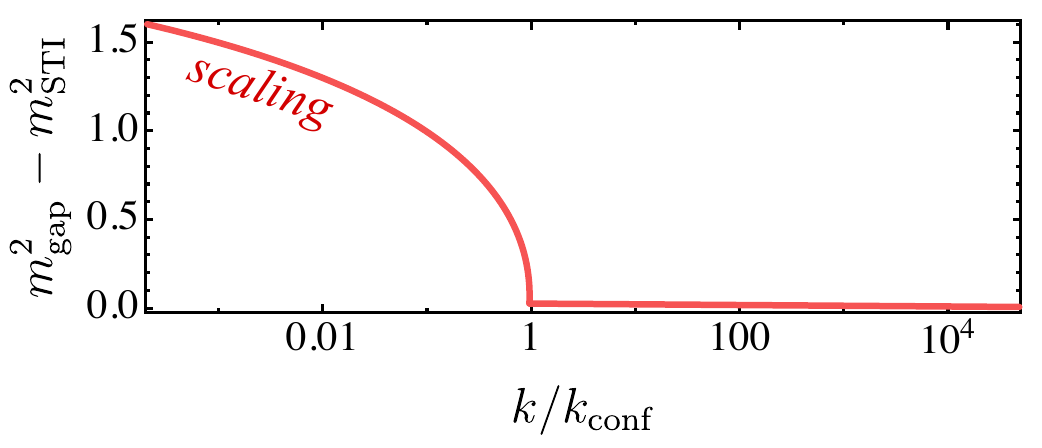}
	\caption{Sketch of the cut-off dependence of the difference between the transverse mass gap $m_\textrm{gap}^2$ and the longitudinal mass $m^2_{\textrm{\tiny{mSTI}}}$. The onset scale of the difference signifies the onset of the dynamical confinement mechanism and defines $k_\textrm{conf}$. In the deep infrared the transverse mass shows scaling.}
	\label{fig:MassDifferenceManyFlavour}
\end{figure}
In turn, the transverse gauge field mass gap is finite, signalled by $m^2_{\textrm{gap},k=0}\neq 0$. Moreover, for cutoff scales above the confinement scale the two masses agree
\begin{align}
m^2_{\textrm{gap},k} = m^2_{\textrm{\tiny{mSTI}},k}  \,, \qquad k \geq  k_\textrm{conf}\,,
\label{eq:mgap=mSTI}
\end{align}
as their difference can only be triggered by irregular vertices. Here, $k_\textrm{conf}$ is the cut-off scale, at which the confining dynamics kicks in: at this scale the longitudinal vertices develop massless modes, either with the Schwinger mechanism \cite{Aguilar:2011xe, Aguilar:2021uwa, Ferreira:2023fva, Aguilar:2022thg} or with the BRST quartet mechanism \cite{Alkofer:2011pe}. In this regime the flow of the transverse mass gap is given by \labelcref{eq:flowmGap} in \Cref{app:flowmGap}. For a discussion in the fRG approach to QCD see \cite{Cyrol:2016tym}.  Note, that $m^2_{\textrm{gap},k\to \infty}$ can be negative or positive, depending on the choice of the regulators $R_A(p), R_c(p), R_\psi(p)$ and the matter content of the theory. For its importance we discuss the key properties of the difference between the transverse mass gap $m^2_{\textrm{gap}}$ and the longitudinal mass $m^2_{\textrm{\tiny{mSTI}}}$ in \Cref{fig:MassDifferenceManyFlavour} as a function of the cut-off scale. 

In \cite{Fischer:2008uz, Cyrol:2016tym} a bootstrap approach was put forward, that circumvents the necessity of dynamically generating the confining mass gap via the Schwinger mechanism or the BRST quartet mechanism. For its importance in the present work we briefly recall its ingredients also in the light of the recent dissection and quantitative evaluation of the Schwinger mechanism in full QCD \cite{Aguilar:2022thg}. 

At the core of the bootstrap approach lies the fact, that it is the dynamically generated massless excitations in the longitudinal part of the ghost-gluon, three-gluon and quark-gluon vertices, that lead to a non-trivial difference between the transverse mass gap and the longitudinal mass. We therefore write all these vertices (and further ones) as a sum of a regular and an irregular part, e.g.~for the three-gluon vertex 
\begin{subequations} 
\label{eq:Reg+irr}	
	\begin{align}
	\Gamma^{(3)}_{\mu\nu\rho}(p,p_1) = \Gamma^{(3),\textrm{reg}}_{\mu\nu\rho} (p,p_1)  +  \Gamma^{(3),\textrm{irr}}_{\mu\nu\rho} (p,p_1) 	\,,
\label{eq:Gamma3Reg+Irr}
\end{align}
with 
\begin{align} 
	\Pi_{\mu\nu}^\bot(p) \Gamma^{(3),\textrm{irr}}_{\mu\nu\rho} (p,p_1)  =0\,.
\label{eq:irrLong}
\end{align}
\end{subequations}
\Cref{eq:irrLong} entails that the required irregularity of the vertices is hosted entirely in their longitudinal part. We emphasise that this structure is hardwired, as the transverse part cannot host massless excitations as they would be physical. Note also that the regular part is uniform in the limit of vanishing momenta: its transverse and longitudinal part have to agree in this limit as otherwise it would contain irregularities arising from the projection operators. More details can be found in \cite{Cyrol:2016tym}. 

The above analysis suggests that the flow of the mass function with regular \textit{uniform} vertices is a good approximation for the flow of the transverse mass gap $m_\textrm{gap}^2$. The second crucial ingredient  of the bootstrap approach is the \textit{assumption} of an IR closure of the gauge fixing which admits BRST charges, an assumption that underlies the Kugo-Ojima confinement scenario. Then, the initial condition for the transverse mass gap simply has to be tuned to infrared scaling \labelcref{eq:mscaling}. 

With this argument, the computation of the gluon propagator in QCD is reduced to solving a quadratic fine-tuning condition for the transverse mass gap parameter $m^2_\textrm{gap}$. This assumption can be checked with a self-consistency evaluation of the results: Varying the regulator matrix $R_k$ triggers vastly different scale dependences of the mass gap parameter and the initial condition \labelcref{eq:mscaling}. These regulator dependences should be absent in the final result for the gluon propagator at $k=0$ only if the above assumption for the fully physical nature of the flow of the mass gap is correct. This analysis has been done in \cite{Cyrol:2016tym} for Yang-Mills theory and in \cite{Cyrol:2017ewj} for two-flavour QCD, and the gluon propagator as well as all other correlation functions showed no sensitivity to sizeable variations of $R_k$. Note however, that this insensitivity was only present for the fully quantitative approximation with full momentum dependences for propagators and vertices. Indeed, these momentum dependences and the respective momentum transfer carry also the cancellation of momentum-dependent cut-off dependences in the flows, triggered by the sizeable mass gap $m_{\textrm{gap},k}^2$ at $k\neq 0$. On the basis of the above arguments we expect that this  bootstrap approach works well for general $N_f$ and provides us with a simple computational procedure to compute glue correlation functions in general gauge-fermion systems. 

Moreover, the above analysis suggests that \labelcref{eq:Reg+irr} can also be used to compute gauge field propagators in simple approximations such as the present one by tuning the initial condition \labelcref{eq:mscaling}. Accordingly, it is the (integrated) flow of $m^2_{\textrm{gap},k}$ that carries the information of confinement and the respective scales, both  $k_\textrm{conf}$ and the physical mass gap. This flow, or rather that of the dimensionless mass gap,  
\begin{align}
	\bar m^2_{\textrm{gap},k}=  m_{\textrm{gap}, k} ^2/ k^2 \,,
	\label{eq:DimlessMassGap}
\end{align} 
is discussed in detail in \Cref{app:flowmGap} and shown in \labelcref{eq:flowmGap}. It consists of two terms: one carries the combination of the canonical and anomalous dimensions and is proportional to $\bar m^2_{\textrm{gap},k}$, and the second term is given by the flow of the gauge boson two-point function evaluated at vanishing momentum, see \labelcref{eq:flowbarm}. Evidently, the total flow of the mass gap is subject to an UV quadratic fine-tuning problem given by the presence of a power-law cut-off dependent term reminiscent of that for a standard mass term. However, in contradistinction to the latter, the initial mass is uniquely fixed by the confinement property in the IR, namely \labelcref{eq:kappaA,eq:kappaAk}. In \cite{Fischer:2008uz, Cyrol:2016tym, Cyrol:2017ewj, Cyrol:2017qkl, Corell:2018yil}, this quadratic fine-tuning problem was simply solved numerically, and the results for the gluon propagator or rather its dressing $p^2 G_A(p)$ match that on the lattice quantitatively. 

In the present work we shall proceed along these lines. Moreover, using the collected experience and understanding of the underlying mechanisms, we shall devise efficient approximate solution of these fine-tuning problems. Its reliability is then checked with the benchmark results of the fully quantitative fRG and lattice works. To begin with, we expect that the lack of full momentum dependences of vertices and propagator in the present approximation limits our choices of $R_k$: the approximation does not allow for momentum transfer in the flow and hence requires rather local correlation functions that admit the interpretation \labelcref{eq:ZAk2}. In short, large values of $m_\textrm{\tiny{mSTI}}(R_k)/k^2$ are bound to require momentum transfers for the cancellation of mSTI effects in $m_\textrm{gap}^2$, and hence these regulator choices require approximations with full momentum dependences. We note in passing that these properties support Curci-Ferrari theories as well-working effective theories of QCD: while they feature a UV gluon mass not present in QCD, the respective renormalisation has no quadratic running. Consequently, in view of the QCD discussion here, they harbour minimal mSTI effects; for a recent review on their use in QCD see \cite{Pelaez:2021tpq}.  

It is in the context of the mSTI effects and the transverse gluon mass gap, where the quantitative dissection of the Schwinger mechanism in QCD in \cite{Aguilar:2022thg} provides an important additional insight. There it has been shown explicitly that the flavour dependence of the confining mass gap is rather small and accounts for less than five percent of the mass gap in two-flavour QCD. This is to be expected from the flavour dependence of the glueball masses in QCD and Yang-Mills theory, and the underlying structure, the Schwinger mechanism, is solely sourced in the glue sector of QCD; it would not be present in the QED-like glue-matter part. Furthermore, the structure of the respective coupled Bethe-Salpeter equations, see \cite{Aguilar:2022thg}, suggests that the importance of the matter part is growing at most linearly also for $N_f>3$ and hence stays small. This estimate worsens successively with growing $N_f$ and for $N_f\gtrsim 8$ we may have to resort to a more complete analysis. However, as will be shown in \Cref{sec:phasesYM+chiral}, in this regime confinement happens at far smaller scales than chiral symmetry breaking, which effectively reduces the analysis to that of Yang-Mills theory. In summary we conclude that the flavour dependence on the confining mass gap is universally small.  

In turn, the power-law flow of $m^2_{\textrm{gap},k}$ with flat regulators \labelcref{eq:Regs} and \labelcref{eq:regxdeff} shows a sizeable dependence on $N_f$, whose effects have to cancel out in the vanishing cut-off limit. Accommodating this property requires either full momentum dependencies that allow for the necessary momentum transfer or a restriction to regulators that maximize momentum locality in the flow. Such regulators naturally lead to near-zero flows for cut-offs $k > k_\textrm{conf}$. Flat regulators do not belong to this class but have near-zero flows in a small $N_f,N_c$ window. For $N_c = 3$ this window is given by a neighbourhood of $N_f = 3$. This analysis is supported by the present results obtained with flat regulators, where the full flow of the mass gap yields quantitative agreement of the gluon propagator with that from fRG computations with full momentum dependencies for $N_f = 3$, as well as with lattice results. However, the results show an increasing deviation for both smaller and larger $N_f$, which enhances the magnitude of the power-law part of the flow. For $N_f = 0, 2$, fRG computations with full momentum dependencies exhibit quantitative agreement with lattice results, underscoring the importance of momentum transfer.

In summary, we may either restrict ourselves to the set of regulators with small mSTI masses $m_\textrm{\tiny{mSTI}}$ stable results (lack of momentum transfer), or we have to improve upon the momentum dependence of the approximation. Both options will be explored in a forthcoming publication, using the computational framework reported on in \cite{Cyrol:2017ewj, Ihssen:2024miv}. Here we choose a more heuristic way and simply use the $N_f=3$ flavour flows for the mass gap with the flat regulators \labelcref{eq:Regs,eq:regxdeff} for all flavours. This approximation is based on its quantitative agreement with the full $N_f=3$ flavour gluon propagator and the small flavour dependence of the mass gap discussed above. Importantly, it allows use to still use the flat regulators, leading to analytic flow in the present approximation. This facilitates the access to the physics mechanisms at work, in particular also in the walking regime. We close this discussion with the remark that this minimisation of $m_\textrm{\tiny{mSTI}}$ can be replicated for different number of colours and flavours.

\subsubsection{Wrap-up of expansion scheme capturing confinement}
\label{sec:ExpandWrapup}

In summary, we have set-up an efficient and semi-quantitative fRG-scheme for the computation of gauge correlation functions in the Landau gauge which includes the confining mass gap. The results obtained are in very good agreement with previous high precision and quantitative fRG studies in pure YM \cite{Cyrol:2016tym} and QCD \cite{Cyrol:2017ewj, Fu:2019hdw, Ihssen:2024miv}. Moreover, the dressings in this approximation are defined as functions of the cut-off scale and we can exploit the direct relation between the cut-off scale and the (average) momentum scale. In short, we identify the momentum-dependent gauge propagator at $k=0$ with its $k$-dependent version at $p=k$, 
\begin{align}
G_A(p)= 	\frac{1}{Z_{A,p} \,p^2 + m^2_{\textrm{gap},p}} \,,
\label{eq:Gluondressingp=k}
\end{align}
following from \labelcref{eq:GAk2}. With \labelcref{eq:Gluondressingp=k} at hand, we readily identify the confinement scale using \labelcref{eq:ConfScale}, to wit, 
\begin{align} 
	k_\textrm{conf} = k_\textrm{peak}\,, 
	\label{eq:kconf}
\end{align}
with $k_\textrm{peak}=p_\textrm{peak}$ defined in \labelcref{eq:ConfScale}. 
Note that $k_\textrm{peak}$ is a proxy for the mass gap itself while $ k_\textrm{conf} $ has been defined as the cut-off scale, at which the confinement dynamics kicks in, see \labelcref{eq:mgap=mSTI}. Clearly, the latter scale is larger than $k_\textrm{peak}$, but YM computations \cite{Cyrol:2016tym} indicate $k_\textrm{conf} \lesssim  2\, k_\textrm{peak}$. In the following we shall use \labelcref{eq:kconf} but the more precise estimate gets important in our technical implementation in \Cref{eq:FineTuningUV}. 

The present approximation to the gauge sector with $\Gamma_\textrm{glue}$ in \labelcref{eq:Gglue} contains the wave functions of ghost and gauge fields, the running mass gap and the avatars of the gauge coupling which are obtained from the three-gauge, four-gauge and ghost-gauge dressings,
\begin{align} \nonumber 
	\alpha_{A^3}&=\frac{\lambda^2_{A^3}}{4 \pi}\frac{1}{\left(1+ \bar m_\textrm{gap}^2\right)^3 }\,, \\[1ex]  \nonumber 
	\alpha_{A^4}&=\frac{\lambda_{A^4}}{4 \pi}\frac{1}{\left(1+ \bar m_\textrm{gap}^2\right)^2 }\,, \\[1ex]    
	\alpha_{c\bar c A}&=\frac{\lambda^2_{c \bar c A}}{4 \pi}\frac{1}{\left(1+ \bar m_\textrm{gap}^2\right) }\,.
	\label{eq:alphasGlue}
\end{align}
Note that both, \labelcref{eq:alphasGlue} and $\lambda_{A^3}, \lambda_{A^3}^{1/2}, \lambda_{c \bar c A}$ can be understood as avatars of $\alpha_g$ or the gauge coupling. The difference is the RG-invariant decoupling factor due to the gauge field mass gap, which is present for exchange couplings that occur in the flow diagrams. In any case, the gauge avatars $\alpha_i$ with $i=A^3, A^4, c\bar c A$ agree for large cut-off scales due to perturbative universality and are genuinely different in the strongly correlated IR regime. This is showcased in the chiral  two-flavour QCD case in \Cref{fig: alpha_s Nf=2 scaling}. 
The parametrisation of the different $\alpha_i$'s in \labelcref{eq:alphasGlue} makes the mass gap factor apparent which induces the low energy decoupling of gauge fluctuations. We will see how, as the perturbative conformal window is approached, the mass gap tends to zero and the IR decoupling disappears.

\subsection{Fermion-gauge dynamics and chiral symmetry breaking}
\label{sec:fermiongaugedynamics}

We proceed with the discussion of the matter part of the effective action.  It contains a running Dirac term with the fundamental fermion-gauge interaction as well as higher order self-interaction terms of the fermions, see the discussion below. We concentrate on the latter terms and explain, how resonant interaction channels as well as multi-scatterings in these channels are efficiently accommodated with the fRG approach with \textit{emergent composites} \cite{Gies:2001nw, Pawlowski:2005xe, Floerchinger:2009uf}. Its application to QCD-like theories is also called \textit{dynamical hadronisation} and has been applied and further developed in \cite{Gies:2002hq, Braun:2008pi, Mitter:2014wpa, Braun:2014ata, Rennecke:2015eba, Cyrol:2017ewj, Fu:2019hdw, Fukushima:2021ctq, Ihssen:2024miv} and the current application is based on \cite{Pawlowski:2005xe, Fu:2019hdw, Fukushima:2021ctq, Ihssen:2024miv}. 

At an asymptotically large cut-off scale the full effective action tends towards the classical action as all quantum fluctuations are suppressed. Note that strictly speaking this is only true in asymptotically free theories. Moreover, we have already discussed in \Cref{sec:econf} that the UV effective action consists of all UV-relevant terms that are compatible with the symmetries of the theory (namely the classical action), including also additional ones that are sourced by the breaking of symmetries due to the regulator insertion. While a momentum space regulator inevitably breaks apparent gauge invariance and leads to the mSTIs, chiral symmetry is preserved in a class of chiral fermion regulators $R_\psi(p)$.  These regulators do not introduce a scalar momentum-dependent mass but a chiral IR suppression that is proportional to $\slashed {p} /\sqrt{p^2}$, see \labelcref{eq:Regs} in \Cref{app:regs}.  

In the chiral limit (absence of explicit fermion mass terms), gauge-fermion systems have a global left-right flavour symmetry, which is spontaneously broken to its vector part, when $\dSSB$ sets in, 
\begin{align}
	& {\rm SU}(N_f)_L \times {\rm SU}(N_f)_R \to {\rm SU}(N_f)_V 
	\label{eq:symmetries}
\end{align}
Accordingly, for chirally invariant regulators, all terms in the effective action $\Gamma_k[\Phi]$ of gauge-fermion theories have the full left-right symmetry, displayed on the left hand side of \labelcref{eq:symmetries}.

\subsubsection{Matter sector of the effective action}
\label{sec:MatterEffAct}

The full matter part $\Gamma_\textrm{mat}[\Phi]$ of the effective action in the approximation used in this work contains the running version of the Dirac term, 
\begin{align}
\Gamma_{{\rm D}} \left[A,\psi,\bar \psi\right]= \int Z_{\psi}\,\bar{\psi}\,\gamma_\mu \left( {\partial}_\mu -  \lambda_{\psi\bar\psi A} \,  Z^{1/2}_{A} A_\mu  \right)\psi\,, 
\label{eq:DiracAction}
\end{align}
introducing the scale-dependent wave function $Z_\psi$ of the fermion as well as the fermion-gauge avatar of the strong coupling $\lambda_{\psi\bar\psi A} $ and the respective exchange coupling  
\begin{align} 
	\alpha_{\psi\bar\psi A} = \frac{\lambda^2_{\psi\bar\psi A}}{4\pi} \frac{1}{1+\bar m^2_\textrm{gap}}\,,
\label{eq:alphapsibarpsiA}
\end{align}
with the decoupling factor $1/(1+ \bar m^2_\textrm{gap})$. \Cref{eq:alphapsibarpsiA} is depicted in \Cref{fig:diagramsexchangecouplings} and complements the avatars of the gauge coupling listed in \labelcref{eq:alphasGlue}. Moreover, the fermion-gauge exchange coupling $\alpha_{\psi\bar\psi A}$ agrees with the other avatars for large cut-off scales up to subleading corrections as a consequence of the mSTIs. Accordingly, no further UV-relevant parameter is introduced by \labelcref{eq:alphapsibarpsiA} and we only have one single UV-marginally-relevant parameter: the gauge coupling. A compilation of all avatars of $\alpha_g$ is shown in \Cref{fig: alpha_s Nf=2 scaling} for $N_c=3$ and $N_f=2$. 

Using \labelcref{eq:alphapsibarpsiA} in the flow equations generates an infinite series of pure fermionic and fermion-gauge interactions that are compatible with the flavour symmetry \labelcref{eq:symmetries}. All of them are UV-irrelevant in the asymptotically free regime of many-flavour QCD. Accordingly, they can be set to zero at the initial cutoff scale, and they are generated by the flow. 

In this work, we start with the four-Fermi operators $(\bar \psi {\cal T}_i \psi)^2$, where ${\cal T}_i$ are dimensionless tensor structures. Additional terms involving higher powers of fermions and/or gauge bosons, or momentum-dependent tensor structures, have higher canonical dimensions and are therefore, in principle, neglected. These terms would require very strong dynamics to become IR relevant and are expected to have a negligible effect. Nevertheless, we will include very high-order operators in the dominant ${\cal T}_i$ channel aiming for quantitative precision. However, we note that this reasoning does not apply to the operator $\bar \psi A^2 \psi$. As shown in \cite{Mitter:2014wpa, Cyrol:2017ewj} for physical QCD, this operator has a subleading significance and is thus omitted from the present analysis.

This leaves us with the following four-Fermi part of $\Gamma_\textrm{mat}$, 
\begin{align}
	 \Gamma_{4 \psi}\left[\bar \psi,\psi\right]=& -\int_x Z_\psi^2\bigg\{\lambda_\textrm{\tiny{SP}} {\cal T}_{\left({\rm S-P}\right)}+\lambda_{+} {\cal T}_{\left({\rm V+A}\right)}\notag\\[1ex]
	&\hspace{.8cm}+ \lambda_{-} {\cal T}_{\left({\rm V-A}\right)}+  \lambda_\textrm{\tiny{VA}}  {\cal T}_{\left({\rm V-A}\right)^{\rm adj}} \bigg\}\,, 
\label{eq:eff4Fermi}
\end{align}
where the global factor $Z_\psi^2$ carries the (inverse) RG-scaling of the fields and hence the couplings $\lambda_\textrm{\tiny{SP}},\lambda_\pm, \lambda_\textrm{\tiny{VA}}$ are RG-invariant.  The tensors ${\cal T}_i$ constitute a Fierz-complete basis compatible with the full flavour symmetry, see \labelcref{eq:4FermiTensors} in \Cref{app:Effective action}. The ${\cal T}_i$ define different channels of the full four-Fermi interaction with given quantum numbers. 

In gauge-fermion systems, the axial U(1)$_{A}$ symmetry is broken anomalously. This anomaly induces fermionic breaking terms, and in the chiral limit the lowest order term generated by the anomaly, the 't Hooft determinant \cite{tHooft:1976snw}, is a $2 N_f$ fermion operator, for a discussion in the fRG approach see \cite{Pawlowski:1996ch}. This operator feeds back into all interaction terms via the flow and in particular it lifts the degeneracy of the scalar--pseudoscalar four-Fermi sector. Then, the size of the Fierz-complete four-Fermi basis increases rapidly with $N_f$. Already in physical QCD with $N_f=2$ the basis contains ten tensor structures, see e.g.~\cite{Mitter:2014wpa, Braun:2020mhk, Ihssen:2024miv}, and for $N_f=2+1$ the basis contains 26 elements, see e.g.~\cite{Braun:2020mhk}. In particular, the scalar-pseudoscalar tensor structure can be written in terms of two different ones,  
\begin{align}
	\lambda_\textrm{\tiny{SP}} {\cal T}_{\left({\rm S-P}\right)}\to \lambda^{(\sigma-\pi)}_\textrm{\tiny{SP}}\, {\cal T}_{\left({\sigma-\pi}\right)}+\lambda^{({\eta-a})}_\textrm{\tiny{SP}}\,{\cal T}_{({\eta-a})}\,, 
\label{eq:SPchannelbroken}
\end{align}
where the subscripts and superscripts ${(\sigma-\pi)}$ and ${(\eta-a)}$ for the tensors and the dressings respectively are taken over from the two-flavour case. There, the 't Hooft determinant is a four-Fermi term and its tensor structure can be used directly in the four-Fermi basis. Moreover, for $N_f=2$ and $N_f=2+1$ the degeneracy of the mass spectrum is lifted considerably by a couple of hundred MeV. In turn, the very nature of the axial U$(1)_{\rm A}$-breaking term as a $2N_f$-fermion term entails that its impact is getting successively irrelevant with larger flavour number. Accordingly, in the large flavour limit we expect the anomalous U$(1)_{\rm A}$-breaking to be absent effectively in the four-Fermi terms, leading to 
\begin{align}
\lambda_{\textrm{\tiny{SP}}}=\lambda^{(\sigma-\pi)}_\textrm{\tiny{SP}}=\lambda^{({\eta-a})}_\textrm{\tiny{SP}}\,.
\label{eq:SPchannelsymmetriccouplings}
\end{align}
In the present work we refrain from computing the flow of the 't Hooft determinant via \cite{Pawlowski:1996ch} and its increasingly irrelevant impact with larger $N_f$ on the four-Fermi terms. Instead we use a generous estimate for the size of the axial U$(1)_{\rm A}$-breaking or the lack thereof: in the latter case full chiral symmetry is present in the symmetric regime and only the $\sigma$-mode as the radial mode is massive in the chirally broken regime. Accordingly, we identify the $(\eta-a)$ part of the multiplet, or rather its couplings and masses, with that of the pions. In the former case we estimate the effect of the U$(1)_{\rm A}$-breaking with identifying the couplings and masses of the $(\eta-a)$ part of the multiplet with that of the $\sigma$-mode. For larger values of $N_f$ this is a very conservative upper bound. We shall see that the difference between these approximations is small which sustains the argument of the subleading effect of the U$(1)_{\rm A}$-breaking on the physics of many-flavour gauge-fermion systems.

We close this discussion on our approximation with the remark, that we only consider cut-off-dependent couplings and dressings in \labelcref{eq:eff4Fermi}. However, such a tensor channel also features different momentum channels. Specifically, the ($\sigma-\pi$) channel is commonly defined by the $t$-momentum channel of the (S--P) tensor structure. In the fRG approach to QCD this has been investigated in \cite{Mitter:2014wpa, Cyrol:2017ewj} and we refer the interested reader to these works for more details. 

\begin{figure}[t]
	\centering
	\includegraphics[width=.98\columnwidth]{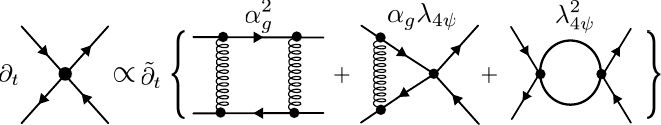}
	\caption{Cartoon of the flow equation of the four-Fermi vertex. The $\tilde\partial_t$ only hits the explicit regulator dependences of the propagators, and we have dropped the tadpole diagrams. The respective $\beta$-function of the dimensionless coupling \labelcref{eq:barlambda} is depicted in \Cref{fig:dtlambda_cartoon}. Here, $\lambda_{4\psi}$ stands for all possible four-Fermi interactions.}
	\label{fig:Generation_lambda}
\end{figure}
%

\subsubsection{RG dynamics of chiral symmetry breaking}
\label{sec:RGdynChi}

This setup allows us to provide a comprehensive discussion of necessary and sufficient conditions for $\dSSB$ in general gauge-fermion systems. The RG point of view allows for a specifically simple and precise access to the mechanisms at work. Here, we briefly report on its key ingredients with an emphasis on the properties important in the present context, specifically for the discussion of the lower boundary of the conformal window in \Cref{sec:BoundaryCBZ}. For a more detailed discussion we refer to \cite{Braun:2011pp, Dupuis:2020fhh} and references therein. 

In	\Cref{fig:Generation_lambda} we depict the structure of the flow of the scalar-pseudoscalar four-Fermi coupling. It is generated by the fermion-gauge box diagrams in the deep perturbative regime (far UV) and its strength there is proportional to $\alpha_g^2$. In the absence of any further scales such as explicit fermion masses, $\alpha_g^2$ is only multiplied by $k^2$ which follows from dimensional counting, and a dimensionless prefactor that depends on the shape of the chosen regulator function. Once generated, it triggers further diagrams being proportional to $\alpha_g  \lambda_{4\psi}$ (mixed diagrams) and $ \lambda_{4\psi}^2/k^2$ (fish diagram). Here, $\lambda_{4\psi}$ stands for all four-Fermi couplings as the flows are not diagonal and all four-Fermi couplings feed into the flow of the others. 

In the perturbative regime the four-Fermi couplings are generated by the fermion-gauge boxes, which are of the order $\alpha_g^2$. Feeding these couplings back into the other diagrams reveals that they are of higher order in $\alpha_s$ and hence are subleading in the perturbative UV-regime: the mixed diagrams are of the order $\alpha_g^3$ and the fish diagrams carry $\alpha_g^4$. It turns out that this ordering is very efficient, and the flow of the four-Fermi couplings is dominated by the fermion-gauge box for strong couplings until $\SSB$ is triggered. This is one of the properties of the four-Fermi flow that leads to a small systematic error estimate and hence to semi-quantitative precision of the current investigation, see below. We emphasise that \Cref{fig:Generation_lambda} has a very complicated substructure as the four-Fermi vertex carries different tensor structures \labelcref{eq:eff4Fermi} that all couple into each other, indicated by $\lambda_{4\psi}$. Moreover, we have left out the tadpole diagrams with a six-fermion vertex: they are of the order $\alpha_g^8$ and strongly suppressed in the perturbative regime but also lack the resonant structure that would make them relevant in the regime with chiral symmetry breaking. 

\begin{figure}[t]
	\centering
	\includegraphics[width=.9\columnwidth]{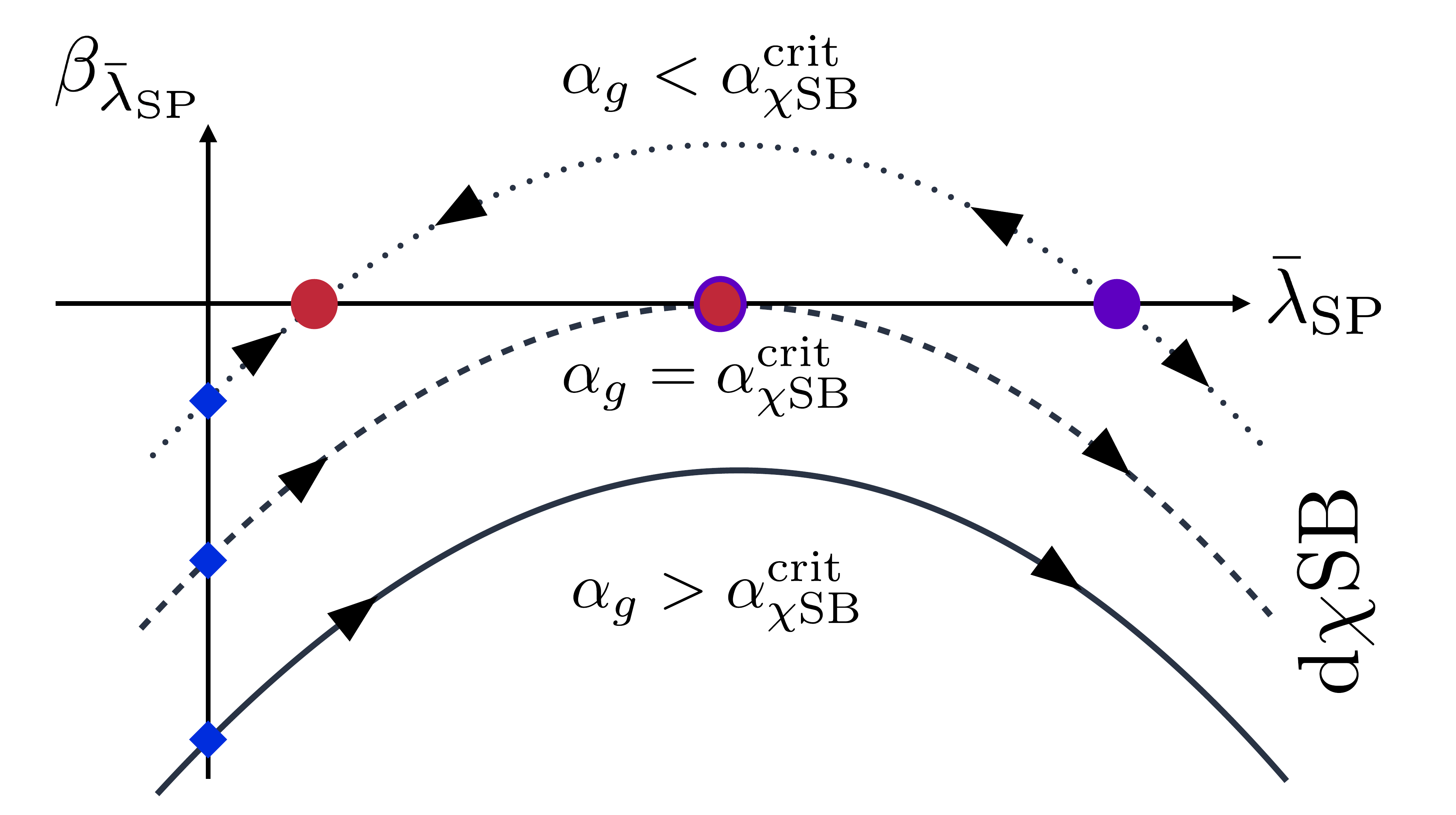}
	\caption{$\beta$-function of the dimensionless four-Fermi coupling $\bar \lambda_\textrm{\tiny{SP}}= \lambda_\textrm{\tiny{SP}} \,k^2$, see \labelcref{eq:barlambda}. The arrows indicate the direction of the flow from UV to IR. The red dot signifies the IR attractive fixed point and the purple dot signifies the UV attractive fixed point. At the critical gauge coupling $\acrit$, the two fixed points merge. For larger gauge couplings, $\dSSB$ is inevitable, see the discussion around \labelcref{eq:alphacrit}.}
	\label{fig:dtlambda_cartoon}
\end{figure}

In summary, the chiral dynamics is dominated by the diagrams shown in \Cref{fig:Generation_lambda} and we can discuss the emergence of chiral symmetry breaking in a very concise way. To that end we concentrate on the $\beta$-function of the dimensionless scalar-pseudoscalar coupling for the sake of simplicity,  
\begin{align}
	\bar \lambda_\textrm{\tiny{SP}} =  \lambda_\textrm{\tiny{SP}} k^2\,,\qquad \beta_{\bar  \lambda_\textrm{\tiny{SP}}} \equiv \partial_t \bar \lambda_\textrm{\tiny{SP}} = 2 \bar \lambda_\textrm{\tiny{SP}} + \textrm{diagrams}\,,
	\label{eq:barlambda}
\end{align}
where the diagrams are those in \Cref{fig:Generation_lambda}, divided by $k^2$. For the ensuing discussion we drop the other couplings for the sake of simplicity, $\bar \lambda_{4\psi}\to \bar \lambda_\textrm{\tiny{SP}}$. The full analysis is more complicated but similar, see e.g.~\cite{Braun:2011pp}. All the diagrammatic contributions have negative prefactors and the resulting $\beta$-function is depicted in \Cref{fig:dtlambda_cartoon}. 

For $\alpha_g=0$, the $\beta$-function in \Cref{fig:dtlambda_cartoon} collapses to that of a Nambu--Jona-Lasinio (NJL) type model. Then, it has a Gau\ss ian IR attractive fixed point at $\bar \lambda_\textrm{\tiny{SP}}=0$, and an UV attractive fixed point at $\bar \lambda_\textrm{\tiny{SP}}^*$. For initial values of the coupling at $k=\Lambda_{\rm UV}$ below the UV fixed points, $\bar\lambda_{\textrm{\tiny{SP}},\Lambda_{\rm UV}}<\bar \lambda_\textrm{\tiny{SP}}^*$, the coupling is driven to zero. Note that this accommodates all finite values of the dimensionful four-Fermi coupling $ \lambda_\textrm{\tiny{SP}}$. Accordingly, no chiral symmetry breaking takes place. In turn, for initial values above the UV fixed point, $\bar\lambda_{\textrm{\tiny{SP}},\Lambda_{\rm UV}}>\bar \lambda_\textrm{\tiny{SP}}^*$, the coupling is driven to infinity which signals the emergence of $\dSSB$. More details on this regime and full control with the emergent composite fRG approach will be provided in \Cref{sec:emergentcomposites}. 

In contrast to NJL-type four-Fermi models, the four-Fermi coupling in gauge-fermion systems is generated by the gauge-fermion dynamics as we have discussed above. Consequently, the flows of all four-Fermi couplings start with $\bar  \lambda_\textrm{\tiny{SP}} \approx 0$ at the blue squares on the $y$-axis in  \Cref{fig:dtlambda_cartoon}. Moreover, the fermion-gauge box diagrams in \Cref{fig:Generation_lambda} do not carry a dependence on the four-Fermi couplings and are proportional to $\alpha_s^2$. 
Thus, they simply shift down the NJL-type $\beta$-function. The mixed diagrams in \Cref{fig:Generation_lambda} exhibit a single one-gluon exchange and one four-Fermi coupling. 
They are proportional to $\alpha_s \bar \lambda_\textrm{\tiny{SP}}$ and lead to anomalous dimensions of $\bar \lambda_\textrm{\tiny{SP}}$ via $2  \bar \lambda_\textrm{\tiny{SP}}\to (2 - \textrm{const.}\,  \alpha_g)\bar \lambda_\textrm{\tiny{SP}}$ and only tilt the $\beta$-functions. These tilts are irrelevant for the current qualitative discussion even though they inform the value of the fixed points and also that of the critical coupling discussed in the next Section.

\subsubsection{Critical coupling for dynamical chiral symmetry breaking}
\label{sec:NecessaryChi}

This structure leads us to the first important conclusion that $\dSSB$ in a gauge-fermion system can only occur if the gauge coupling exceeds the critical coupling $\acrit$ in the flow. This coupling is defined as the minimal coupling that leads to a negative or vanishing $\beta$-function, 
\begin{align}
	\acrit= \min\left\{ \alpha_g\, |\, \beta_{\bar  \lambda_\textrm{\tiny{SP}}} \leq0\quad \forall \bar \lambda_\textrm{\tiny{SP}} \right\}\,.
\label{eq:alphacrit}
\end{align}
Only for a completely negative $\beta$-function the coupling can crossover to the regime with chiral symmetry breaking. Accordingly,  	
\begin{align}
	\alpha_{g,k}  > 	\acrit\,, 
\label{eq:dSBB-Necessary} 
\end{align} 
for some $k$-range is a necessary condition for chiral symmetry breaking. It is not sufficient as the analysis has been done in the absence of any mass scales and dynamical chiral symmetry breaking generates fermion masses and confinement generates a gauge-field mass gap. Both dynamical masses lead to a decay of the box diagrams below the respective mass scale, effectively reducing the $\beta$-function to the NJL-type in the deep IR far below the mass scales. This makes the general analysis very complicated in the regime where both scales are sufficiently close and the confinement and chiral dynamics are intertwined. In particular this holds true for the QCD-like regime with a small number of flavours.

\subsection{Emergent composites in gauge-fermion QFTs}
\label{sec:emergentcomposites}

As the avatars of the gauge dynamics strengthen towards the IR, the dressings of UV-irrelevant operators in \labelcref{eq:eff4Fermi} grow big as well. In fact, the dressing of the (S-P) channel becomes singular which signals the breaking of the chiral symmetry and the respective modes turning critical. This also entails that higher order fermion scatterings in this channel cannot be neglected any more. The divergence of $\lambda_\textrm{\tiny{SP}} $ as well as the higher order scatterings are very efficiently and reliably accommodated within the fRG with the \textit{emergent composite} formalism introduced in \cite{Gies:2001nw}.

\begin{figure}[t]
	\centering
	\includegraphics[width=.95\columnwidth]{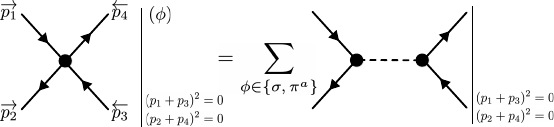}
	\caption{Reparametrisation of the four-Fermi resonant channel as the exchange of a scalar field in the particular momentum channel configuration. Figure adapted from \cite{Fukushima:2021ctq}.}
	\label{fig:Bosonisation}
\end{figure}

In this approach the four-Fermi interaction in a particular momentum channel (here the scalar-pseudoscalar channel in the $t$-momentum channel) is rewritten as the exchange of a composite bosonic degree of freedom carrying the respective Lorentz, colour and flavour structure of the corresponding fermionic bilinear, see  \Cref{fig:Bosonisation}. This field transformation is made at the level of the flow equation, and is reminiscent of a momentum-scale dependent Hubbard-Stratonovich transformation \cite{HubbardPhysRevLett.3.77,Stratonovich}. In its implementation in the flow equation it is an identity transformation. This has the important consequence, that, in contradistinction to the standard Hubbard-Stratonovich transformation, it is not troubled by potential double counting problems in \textit{any} approximation to the effective action.

In the present work we use the formulation introduced in \cite{Pawlowski:2005xe} and furthered in \cite{Fu:2019hdw, Fukushima:2021ctq, Ihssen:2024miv}. It is based on the generalised flow equation derived in \cite{Pawlowski:2005xe}, 
\begin{align}\nonumber 
		\left( \partial_t  + \int \dot\Phi \frac{\delta}{\delta \Phi} \right) \Gamma_k [\Phi ] =& \\[2ex] 
		& \hspace{-2cm}  \frac{1}{2} \textrm{Tr}\left[ \frac{1}{\Gamma^{(2)}_k[\Phi] +R_k }\,\left(  \partial_t  +2 \frac{\delta \dot\Phi}{\delta\Phi}  \right)  R_k  \right]\,,
	\label{eq:GenfRG}
\end{align}
which accounts for the scale-dependent field transformation in the function $\dot\Phi[\Phi]$. The latter is completely at our disposal and for $\dot\Phi[\Phi]=0$ the generalised flow equation \labelcref{eq:GenfRG} reduces to the Wetterich equation \labelcref{eq:Floweq}. In the present work we only bosonise the resonant $(\sigma-\pi)$ channel with the tensor structure ${\cal T}_{\left({\sigma-\pi}\right)}$ defined in \labelcref{eq:AxialSplitTSP}. This reflects the axial U$(1)_{\rm A}$ breaking and has been discussed in \Cref{sec:MatterEffAct} around \labelcref{eq:SPchannelsymmetriccouplings}. For maximal axial U$(1)_{\rm A}$ breaking this tensor structure carries the lightest composite degrees of freedom, the scalar $\sigma$ and the pseudoscalar pion $\boldsymbol{\pi}$ modes. This situation occurs in physical QCD.  For larger $N_f$ the axial U$(1)_{\rm A}$-breaking gets increasingly irrelevant and the dressing of the second tensor structure ${\cal T}_{\left({\eta-a}\right)}$, see \labelcref{eq:AxialSplitTSP}, in \labelcref{eq:SPchannelsymmetriccouplings} can be identified with that of the $\sigma-\pi$ channel. Again we refer to \Cref{sec:MatterEffAct} for the details and the ensuing systematic error estimate of our approximation. 

We proceed with the discussion of the flowing field $\dot\Phi[\Phi]$. It is chosen such that the dressing of the $(\sigma-\pi)$ channel vanishes and is absorbed entirely in the emergent mesonic sector. We collect the bosonised fields in 
\begin{align}
	\phi= (\sigma,\boldsymbol{\pi})\qquad {\rm with}\qquad  \boldsymbol{\pi}=(\pi_1,...,\pi_{N_f^2-1})\,,
\end{align}
and in the chiral invariant 
\begin{align}
	\rho= \frac{\sigma^2+\boldsymbol{\pi}^2}{2}\,.
	\label{eq:rho} 
\end{align}
The $\sigma$-mode is the radial mode in the order parameter potential and in physical QCD it is related to the $f_0(500)$ resonance with the quantum numbers $0^{++}$. Its resonance mass is about 500\,MeV and in the chiral limit in two-flavour QCD within the fRG approach its mass is about 250\,MeV, see \cite{Ihssen:2024miv}. This entails that its off-shell importance in the loops is subleading. However, as the radial mode of chiral symmetry breaking, it acquires an expectation value and our expansion of the effective action is one about the equations of motion of the theory, which is an optimal expansion point. We also consider multi-scatterings of this resonant interaction channel which are captured by an effective potential $V(\rho)$ of the chiral invariant $\rho$.

\subsubsection{Matter part of the effective action with emergent composites}
\label{sec:MatterEffActEmergentComp}

This leads us to the full approximation to  $\Gamma_\textrm{mat}$ used in the present work, 
\begin{align}\nonumber 
	 \Gamma_\textrm{mat}[A,\psi,\bar \psi, \phi]=   &\, \Gamma_{\rm D}\left[A,\bar \psi,\psi\right]+ \left.   \Gamma_{4 \psi}\left[\bar \psi,\psi\right]\right|_{ \lambda^{(\sigma-\pi)}_\textrm{\tiny{SP}} =0}\\[1ex]
	 & + \Gamma_{\psi\phi}[\psi,\bar \psi,\phi]\,,
	  \label{eq:FinalGmat}
\end{align}
with the Dirac part \labelcref{eq:DiracAction} and the four-Fermi part \labelcref{eq:eff4Fermi} and the mesonic part  
\begin{align}\nonumber 
	\Gamma_{\psi\phi}[\psi,\bar \psi,\phi] =&\,\int_x \Biggl\{ \frac{1}{2}\,Z_\phi (\partial_{\mu}\phi)^2 +V(\rho)\\[1ex] 
	&  \hspace{-1.3cm} +Z^{1/2}_\phi Z_\psi\, h_\phi(\rho)\,\bar{\psi}\left(T^0_f \sigma+\,\imag \gamma_5 \,T^a_f \pi^a\right)\psi \Biggr\}\,. 
	\label{eq:GMeson}
\end{align}
Here, $T^0_f$ is a normalised and diagonal tensor in flavour space \labelcref{eq:T0Tf} and $T^a_f$ are the generators of the SU(N$_f$) group. The Yukawa coupling $h_\phi(\rho)$ is RG-invariant as the RG-scaling of the Yukawa term is captured by the prefactor $Z^{1/2}_\phi Z_\psi$.

In the simplest version of the emergent composite approach, deep in the perturbative regime, the effective potential is Gau\ss ian and is given by a mass term 
\begin{align} 
	V(\rho) = m_\phi^2 \, \left( Z_\phi \rho\right) \,, 
	\label{eq:VGauss}
\end{align}
with the RG-invariant mass $m_\phi$, and chiral invariance dictates the $\sigma$ and pion mass functions in \labelcref{eq:VGauss} to be the same. Moreover, we take a field-independent Yukawa interaction, 
\begin{align}
\partial_\rho h_\phi(\rho) \equiv 0\,.
\label{eq:Derh=0}
\end{align}
Then, the solutions of the equations of motion (EoM) for $\phi$ are given by 
\begin{align} \nonumber 
	Z_\phi^{1/2} \sigma_\textrm{\tiny{EoM}}(p) =& -\frac{Z_\psi\, h_\phi}{ (p^2+m_\phi^2)}    \,\left[\bar\psi \,T_f^0\psi\right](p)\,, \\[1ex]
Z_\phi^{1/2} 	\pi_\textrm{\tiny{EoM}}^a(p)=&\, -\frac{Z_\psi\,h_\phi }{ (p^2+m_\phi^2) }  \,\imag\left[\bar\psi \gamma_5 T^a_f \psi\right](p)\,,
	\label{eq:EoMphi}
\end{align}
with 
\begin{align} 
\left[\bar\psi\, T_f\psi\right](p) = \int_q \bar \psi(q) \,T_f \psi(p-q)\,.
	\label{eq:Convolutionj}
	\end{align}
\Cref{eq:EoMphi} suggests that the full expectation value of the $\sigma$ at vanishing momentum, $\langle \sigma \rangle = \langle \hat \sigma(0) \rangle$ the solution of the EoM, is directly related to the chiral condensate, 
\begin{align}
	\langle \sigma \rangle \propto \int_x \left \langle  \bar\psi(x) \psi(x)\right\rangle\,,
\label{eq:sigma0ChiralCondensate}
\end{align}
the standard order parameter of $\dSSB$. Indeed, it can be shown that the relation \labelcref{eq:sigma0ChiralCondensate} holds true beyond the Gau\ss ian approximation of the effective potential, see \cite{Fu:2019hdw}. Moreover, $\langle \sigma \rangle$ is tightly related to $f_\pi$, which is obtained from the fermion two-point function, for a recent discussion in QCD see \cite{Ihssen:2024miv}. Hence, we may simply use $\langle \sigma \rangle$ as our order parameter.

\subsubsection{Effective scale of chiral symmetry breaking}
\label{sec:kchi}

The analysis in  \Cref{sec:MatterEffActEmergentComp} suggests to identify the cut-off scale $\kSSB$, below which $\langle \sigma \rangle\neq 0$ in the chiral limit, with the scale of $\dSSB$. Then 
\begin{align}
	\langle \sigma \rangle = \left\{ 
	\begin{array}{rcl} 
		=0 & \hspace{2cm} & k>\kSSB \\[1ex] 
		\neq 0& & k<  \kSSB
	\end{array}                     
	\right.\,.
	\label{eq:chiOrder}
\end{align}
with
\begin{align} 
	\kSSB = \max \{ k\,|\, \langle \sigma \rangle_k \neq 0 \}\,.
\label{eq:kchi}
\end{align}
Together with $k_\textrm{conf}$ in \labelcref{eq:kchi} this defines good proxies for two dynamical scales in IR gauge-fermion systems. We shall use them from now on, and in particular $\kSSB$, as our reference scales. 

\begin{figure}[t!]
	\centering
	\includegraphics[width=.95\columnwidth]{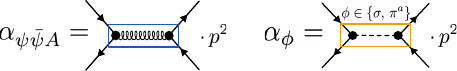}
	\caption{Diagrammatic depiction of the gauge-fermion and mesonic exchange couplings as defined in \labelcref{eq:alphapsibarpsiA,eq:alphaphi} respectively.}
	\label{fig:diagramsexchangecouplings}
\end{figure}

On the EoM \labelcref{eq:EoMphi}, the mesonic part of the effective action reduces part of the a scalar-pseudoscalar four-Fermi term
\begin{align} \nonumber 
\Gamma_{\psi\phi}[\psi,\bar\psi,\phi=\phi_\textrm{\tiny{EoM}}] = &\, \int_p Z_\psi^2 \, \Biggl\{  \\[1ex]\nonumber 
& \hspace{-2.5cm}- \lambda_\textrm{\tiny{SP}}^{\sigma}(p)
\left[\bar\psi \,T_f^0  \psi\right](p)\left[\bar\psi \,T_f^0  \psi\right](-p)\\[1ex] 
&\hspace{-2.5cm}- \lambda_\textrm{\tiny{SP}}^{\pi}(p)\left[\bar\psi \gamma_5 T_f^a   \psi\right](p) \,\left[\bar\psi \gamma_5 T_f^a   \psi\right](-p)\Biggr\}\,, 
\label{eq:SP-4Fermi}
\end{align}
with the momentum-dependent and RG-invariant dressing 
\begin{align} 
	\lambda^{(\sigma-\pi)}_\textrm{\tiny{SP}}(p) = \frac12  \frac{h_\phi^2}{p^2 + m^2_\phi} \,, 
	\label{eq:lambdasigmap}
\end{align}
with $\phi\in \{\sigma,\pi^a\}$. At $p=0$, the cut-off analogue of \labelcref{eq:lambdasigmap} can be written as the mesonic exchange coupling
\begin{align}
	\alpha_\phi  &= \frac{1}{4 \pi}\lambda^{(\sigma-\pi)}_\textrm{\tiny{SP}}(p=0)= \frac{1}{8 \pi }\frac{h_\phi^2}{ (1+ \bar m^2_{\phi})}\,.
	\label{eq:alphaphi}
\end{align}
This is depicted in \Cref{fig:diagramsexchangecouplings} along the fermion-gauge exchange couplings.  In summary, the fRG approach with emergent composites accommodates a momentum-dependent four-Fermi coupling in a specific momentum channel, and hence goes beyond the cut-off-dependent approximation \labelcref{eq:eff4Fermi}.

Now we pick up the discussion of the anomalous breaking of the axial U$(1)_{\rm A}$ symmetry discussed in \Cref{sec:MatterEffAct} around \labelcref{eq:SPchannelbroken}. If this symmetry is effectively restored, the coupling of the $({\eta-a})$ channel agrees with that of the $(\sigma-\pi)$ one, 
\begin{align}
	\lambda_\textrm{\tiny{SP}}^{({\eta-a})}(p)=	\lambda^{(\sigma-\pi)}_\textrm{\tiny{SP}}(p) =  \frac12  \frac{h_\phi^2}{p^2 + m^2_\phi} \,.
\label{eq:identificationlambdaeta}
\end{align}
In the symmetric regime this certainly is a quantitative approximation for larger $N_f$ and we will use it throughout. In the broken regime we can either 
assume that the U$(1)_{\rm A}$ symmetry is effectively restored, which renders all the mesons in the $(\eta-a)$ channel Goldstone modes: $m^2_\phi\to m^2_\pi=0$ in \labelcref{eq:identificationlambdaeta}. Alternatively, we can assume maximal anomalous U$(1)_{\rm A}$ breaking, that is explicit breaking (without Goldstones). Then, for large $N_f$ the mass of the $\sigma$-mode serves as a conservative upper value for the mass in the $\eta-a$ channel: $m^2_\phi\to m^2_\sigma$. The results presented in the following sections are based on the former identification with $m^2_\phi\to m_\pi^2$, as we are most interested in the large $N_f$ regime. We have used the identification $m^2_\phi = m^2_\sigma$ for the systematic error estimate. 

Furthermore, in the bosonised part of the effective action \labelcref{eq:GMeson}, we have included an effective potential term for the composite fields which we consider to have a polynomial form
\begin{align}
	V(\rho)= \sum_{n\geq N_\textrm{min}}^{N_{\rm max}}  \frac{\lambda_{\phi,n}}{n!}\,\left[Z_\phi \left( \rho-\rho_0\right)\right]^n \,.
	\label{eq:Veff} 
\end{align}
The powers $\rho^n$ include multi-scattering terms $(\bar \psi{\cal T} \psi)^{2n}$ of fermionic interactions, which is readily seen in an iteration of the solution to the EoM with a general potential about the solution \labelcref{eq:EoMphi} for the Gau\ss ian potential \labelcref{eq:VGauss}. In \labelcref{eq:Veff}, $\lambda_{\phi,n}$ are the expansion coefficients and $\rho_0= \langle\sigma\rangle^2/2$ is the flowing global minimum of the potential. The combinations $[Z_\phi (\rho - \rho_0)]^n$ are RG-invariant and so are the $\lambda_{\phi,n}$.  In the broken phase ($\rho_0\neq 0$) we have $N_\textrm{min}=2$, while in the symmetric phase ($\rho_0= 0$) we have $N_\textrm{min}=1$ being the first term  in the expansion the Gau\ss ian one, \labelcref{eq:VGauss}. Moreover, we employ in both limits $N_{\rm max}=5$, as this already provides well converged results even in the many flavour limit, see \Cref{app:comparisonBoundaryCBZ}. 

We close this discussion with a further remark concerning our treatment of the axial symmetry. Full U$(1)_{\rm A}$-symmetry would entail that the potential $V(\rho)$ depends on the $(\eta-a)$ part of the invariant, that is $\rho\to \rho + \rho_{(\eta-a)}$. This induces further symmetry-restoring fluctuations in the symmetric regime and may lead to an earlier onset of the conformal regime if the system admits the phenomenon of pre-condensation. This possibility is evaluated in \Cref{sec:precondensation} and the discussion there suggests that the respective systematic error is small, but these fluctuations will play a rôle at finite temperature. 

For the derivation of the flows, it is convenient to define the dimensionless form of the effective potential, 
\begin{align}
	u(\bar \rho)= \frac{V( \rho)}{k^4} \qquad {\rm with} \qquad \bar \rho= \frac{Z_\phi \rho}{k^2}\,.
	\label{eq:u} 
\end{align}
The RG-invariant curvature masses of the mesonic modes are given by the second $\phi$-derivatives (curvature) of the effective potential 
\begin{subequations}
	\label{eq:mphi}
\begin{align} 
	 m^2_\sigma &=k^2\, \left[ u'(\bar \rho_0)+2 \bar\rho_0 \, u''(\bar\rho_0)\right]\,, \\[1ex] 
	 m^2_\pi &=k^2\, u'(\bar\rho_0)\,, 
\end{align}
\end{subequations}
which reduces to $m^2_\sigma = m^2_\pi =m_\phi^2$ for the Gau\ss ian potential \labelcref{eq:VGauss}. In \labelcref{eq:mphi} we have used the expansion point 
\begin{align} 
	\phi_0 =\left(\langle \sigma\rangle \,,\,0\right)\,, \qquad \textrm{with}\qquad \langle \sigma\rangle = \sqrt{2 \rho_0}\,. 
	\label{eq:pho0rho0}
\end{align} 
Moreover, in the broken phase with $\rho_0\neq 0$ we readily conclude $m^2_\pi =0$ and  $m^2_\sigma = 2 \bar\rho \, u''(\bar\rho)$. Note also that the difference \labelcref{eq:mphi} for the masses extends to the scattering couplings $\lambda_{\sigma,n}$ and $\lambda_{\pi,n}$. While the latter are simply given by the respective $\lambda_{\phi,n}$, that is $\lambda_{\pi,n}=\lambda_{\phi,n}$, the former also get contributions from $\lambda_{\phi,n+m}$ with $m\leq n$, multiplied with the respective powers of $\rho_0$. The differences originate in strong chiral symmetry breaking as is apparent already in \labelcref{eq:mphi}.  

This analysis readily extends to the Yukawa coupling with $h_{\pi,n}(\rho_0)= h_{\phi,n}(\rho_0)=\partial_\rho^n h_{\phi}(\rho=\rho_0)$ and $h_{\sigma,n}(\rho_0)= h_{\phi,n}(\rho_0)+h_{\phi,n+m}(\rho_0) \rho_0^m-$terms. In the present work we drop the higher order terms of the Yukawa coupling and use the approximation 
\begin{align} 
	h_\sigma(\rho_0) =h_\phi= h_\pi(\rho_0) \,, \quad \textrm{with} \quad h_{\phi,n>0}(\rho_0) \approx 0\,.
	\label{eq:hphihsigmahpi}
\end{align}
Consequently we compute the uniform $h_\phi$ from the flow of the pionic Yukawa coupling $h_\pi(\rho_0)$: In the symmetric phase 
with $\rho_0=0$ both couplings agree. In turn, in the broken phase the pions are massless and lead to larger 
contributions. The quantitative reliability of this approximation has been checked in physical QCD.  Moreover, for $N_f$ flavours we have $N_f^2-1$ pions and one $\sigma$ and in the most interesting many flavour regime the pions dominate successively more. Further details can be found in \Cref{app:flowbosonisedsector}. 

We close this analysis with a discussion of the scalar part of the Yukawa term in the effective action.  It contains a $\langle \sigma\rangle$-dependent fermionic mass term with the RG-invariant mass 
\begin{align}
	m_\psi= \frac{h_\phi \, \langle \sigma\rangle}{\sqrt{2\,N_f}}\,. 
	\label{eq:mpsi}
\end{align}
which manifests the explicit appearance of a constituent fermion mass with $\dSSB$ proportional to the order parameter, see \cite{Fu:2019hdw} for a detailed discussion.  One of the aims of this work is to connect with the quantum scale-invariant limit of gauge-fermion theories. Therefore, we will focus our study on classically scale-invariant theories, where no explicit mass scale is present at the classical level (e.g., no current fermion mass as in physical QCD) and \labelcref{eq:mpsi} is purely given by the chiral condensate contribution.

\subsection{Wrap up of the approximation}
\label{sec:couplings}

We close this Section with a wrap up of our approximation for the benefit of the reader. We only consider the RG-invariant couplings which already absorb the wave function renormalisations $Z_A,Z_c,Z_\psi, Z_\phi$ of both, the fundamental and of the composite fields. 

This leaves us with the following set of renormalisation group invariant coupling parameters, 
\begin{align}
&	\lambda_{i}\,,\,m_\textrm{gap}, &&  \lambda_\pm\,,\, \lambda_\textrm{\tiny{VA}},&&& h_\phi  \,,\,V(\rho)\,,
	\label{eq:FlowingParameters}
\end{align}
with $i=A^3,A^4,c\bar c A,\psi\bar\psi A$, and $V\rho)$ stands for the set of mesonic self-scattering couplings $\lambda_{\phi,n}$ with $n\leq N_\textrm{max}$ with $N_\textrm{max}=5.$  In \labelcref{eq:FlowingParameters} we have paired the coupling parameters in the fermion-gauge sector, the fermion sector, and the fermion-meson sector. We emphasise that we only have one UV-relevant parameter, while all others are either vanishing at the initial cut-off scale $\Lambda_{\rm UV}\to \infty$, or are determined by gauge symmetry. At $k=\Lambda_{\rm UV}$ we have 
\begin{subequations}
\label{eq:Initialcondition}
\begin{align}
	\lambda_{i,\Lambda_{\rm UV}}=g_{\Lambda_{\rm UV}}\,, \qquad m_{\textrm{gap},\Lambda_{\rm UV}} = m_{\textrm{scaling},\, \Lambda_{\rm UV}}\,,  
\label{eq:Initialcondition1}
\end{align}
with the perturbative UV value $g_{\Lambda_{\rm UV}}$ of the gauge coupling. The UV value of the gauge field mass gap $m_{\textrm{gap},\, \Lambda_{\rm UV}}$ is uniquely given by $m_{\textrm{scaling},\, \Lambda_{\textrm{UV}} }$, satisfying the confinement demand of IR scaling, see \labelcref{eq:mscaling} and the discussion there. This entails, that we have one UV-relevant parameter: $g_{\Lambda_{\rm UV}}$. Note however, that its value only determines the value of $\Lambda_{\rm UV}$ in terms of the dynamical scales as chiral gauge-fermion QFTs have no inherent mass scale. 

The other couplings are UV-irrelevant and are vanishing for sufficiently large cut-off scale. Hence we use 
\begin{align}
	\lambda_{\pm,\Lambda_{\rm UV}},\lambda_{\textrm{\tiny{VA}},\Lambda_{\rm UV}}=0\,,\quad h_{\phi,\Lambda_{\rm UV}}=0 =V_{\Lambda_{\rm UV}}(\rho )\,. 
	\label{eq:Initialcondition2}
\end{align}
\end{subequations}
The couplings and propagators can be combined into exchange couplings which are more relevant for the offshell dynamics in the loops. These exchange couplings for different gauge and (S-P)-mediated correlation functions are given by the $\alpha_i$ with $i=A^3,A^4,c\bar c A, \psi\bar\psi, \phi$ and are defined in \labelcref{eq:alphasGlue,eq:alphapsibarpsiA,eq:alphaphi}. This concludes the discussion of the approximation of the effective action used in the present work.

\section{Confining and chiral dynamics in two flavour QCD}
\label{sec:InterplayLowNf}

In this Section, we benchmark the approximation to the effective action detailed in \Cref{sec:truncation} with $N_c = 3$ and $N_f = 2$ in the chiral limit, in close similarity to physical two flavour QCD. For this benchmark case we only use the $(\sigma-\pi)$-part of the scalar-pseudoscalar interaction, and the same approximation is used in \Cref{app:comparisonQCD}. This approximation includes the strong effect of the anomalous breaking of the U$(1)_{\rm A}$ in two-flavour QCD, which leads to the decoupling of the $(\eta-a)$-modes from the off-shell dynamics in the flow diagrams. The respective results are then compared with quantitative functional QCD ones and lattice data. Although this analysis is crucial for assessing the reliability and estimating the systematic error of the first-principles fRG approach to general gauge-fermion systems developed in this work, it may be skipped on a first reading. Readers can proceed directly to \Cref{sec:phasesYM+chiral,sec:BoundaryCBZ}, where we present and discuss results for general gauge-fermion systems, including the walking regime and the conformal window at a large number of flavours. 

For $N_f=2$ and 2+1 flavours there exists a plethora of quantitative results for correlation functions in Landau gauge QCD, both from functional approaches and from lattice simulations. Specifically, we will compare the results  related to confinement in \Cref{sec:Scaleconfinement} and to $\dSSB$ in \Cref{sec:ScalechiralSB}.

\subsection{Confining dynamics}
\label{sec:Scaleconfinement}

In \Cref{sec:confinement,sec:econf} we have discussed in detail confinement and its manifestation in Landau gauge QCD. Specifically, it is encoded in the mass gap $m_\textrm{gap}$ of the gluon propagator, which is directly related to the peak position $p_\textrm{peak}$ of the dressing of the propagator, see \Cref{sec:ConfWrapup}. Moreover, we have to tune the IR scaling of the gauge field dressing to the Kugo-Ojima confinement condition \labelcref{eq:kappaAk}, which requires the resolution of the respective quadratic fine-tuning condition. This and the discussion of its stability is discussed in \Cref{app:TuningConfinement+Stability}. 

The dressings of the gauge field propagator for two and 2+1 flavours with $N_c=3$ computed in the present approximation compare very well to quantitative results from functional two and 2+1 flavour QCD and from lattice simulations, see \Cref{fig:ZAcomparisonQCD}. We expect that this quantitative agreement is also present for $N_f>3$, as has been discussed in \Cref{sec:UVSimplification} and \Cref{app:flowmGap}.  Accordingly, the confining scale in the Landau gauge, $p_\textrm{peak}$, as defined in \Cref{sec:ConfWrapup} is well accounted for. 

\begin{figure}[t!]
	\centering
	\includegraphics[width=\columnwidth]{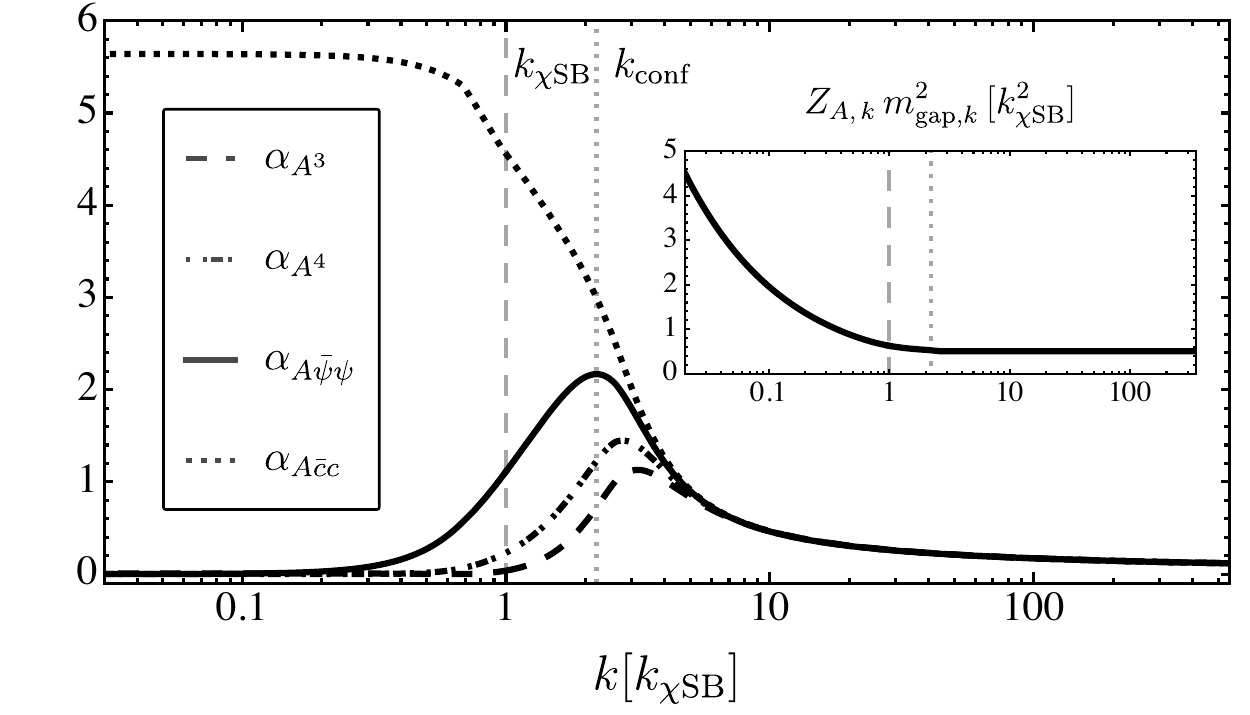}
	\caption{ Three-gauge (dashed), four-gauge (dashed dotted), ghost-gauge (dotted) and fermion-gauge (plain) exchange couplings in an SU(3) gauge theory with $N_f=2$ fundamental flavours. The momentum or cut-off scale is measured in terms of the chiral symmetry breaking scale $\kSSB$, defined in \labelcref{eq:kchi} and depicted by the vertical grey dashed line. The confinement scale \labelcref{eq:kconf} is depicted by a vertical grey dotted line. In the inlay plot, we show the gauge field two-point function at vanishing external momenta as a function of the cut-off scale.}
	\label{fig: alpha_s Nf=2 scaling}
\end{figure}

Note that in the present approximation the dressings are defined as functions of the cut-off scale and we identify $p=k$, see \labelcref{eq:Gluondressingp=k}. This identification can also be used for the exchange couplings $\alpha_i$ with $i=A^3, A^4, c\bar c A ,\psi\bar\psi A$ defined in \labelcref{eq:alphasGlue,eq:alphapsibarpsiA}, which are shown in \Cref{fig: alpha_s Nf=2 scaling}. We display them over four orders of magnitude in momentum, from the asymptotic UV regime with perturbative universality to the strongly coupled IR. At high scales, all avatars $\alpha_i$ agree due to universality and 
\begin{align} 
	\frac{1}{1+\bar m^2_{\textrm{gap},\Lambda_{\rm UV}}} \stackrel{\Lambda_{\rm UV}\to\infty}{\longrightarrow} 1\,.
\label{eq:gappingto0}
\end{align}
As the gauge dynamics strengthen towards the IR, the deviations between the avatars of the gauge coupling start growing due to the different decoupling factors and the genuine differences between the respective couplings $\lambda_i$. The exchange coupling strength increases until the confinement scale $\kconf$ is reached. For practical and illustrative purposes, we approximate the previous definition \labelcref{eq:kconf} and identify this scale with the peak of the fermion-gauge exchange coupling $\alpha_{\psi\bar\psi A}$. This exchange coupling decays at lower scales compared to the pure glue ones, as the loop corrections are not equally suppressed by the same powers as in the pure gauge case. The introduced scale serves as an approximate proxy for physical observables, such as the confinement-deconfinement critical temperature, and may also be identified with the peak of other pure-gauge exchange couplings. Notably, $\alpha_{ c\bar c A }$ freezes towards the IR in the scaling solution.

At $\kSSB$, \labelcref{eq:kchi}, ${\rm d}\chi{\rm SB}$ occurs leading to the emergence of a chiral condensate and constituent masses for the fermions. This triggers the quick decoupling of the fermion loops and will be discussed in the following \Cref{sec:ScalechiralSB}. 

\begin{figure}[t!]
	\centering
	\includegraphics[width=\columnwidth]{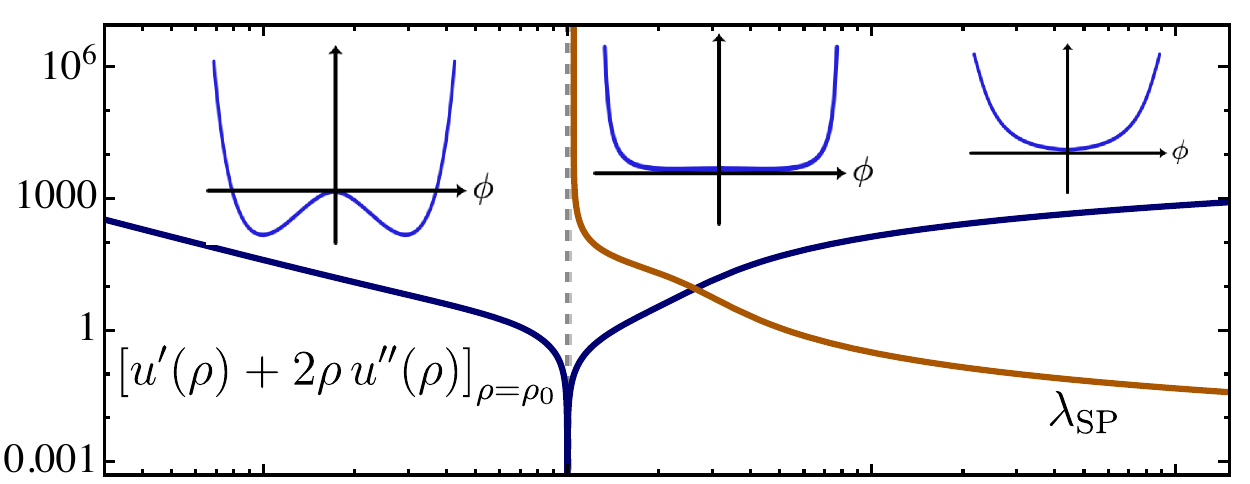}\vspace{-0.1cm}
	\includegraphics[width=\columnwidth]{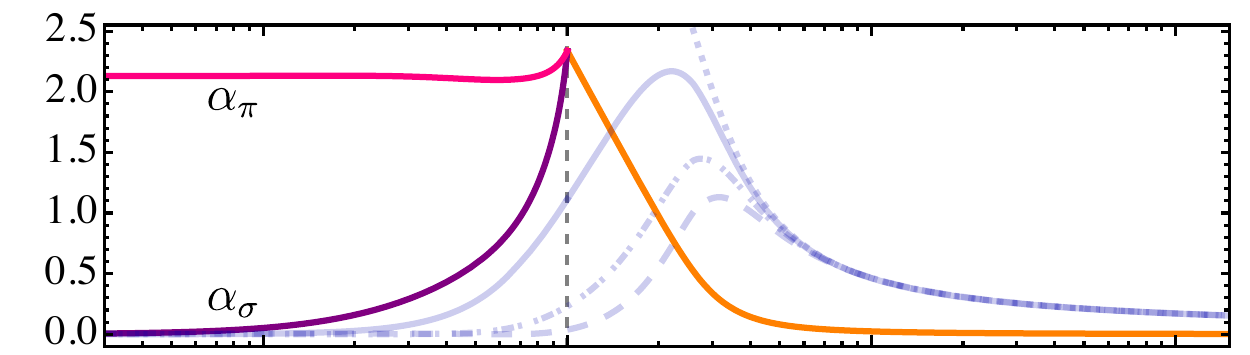}
	\includegraphics[width=\columnwidth]{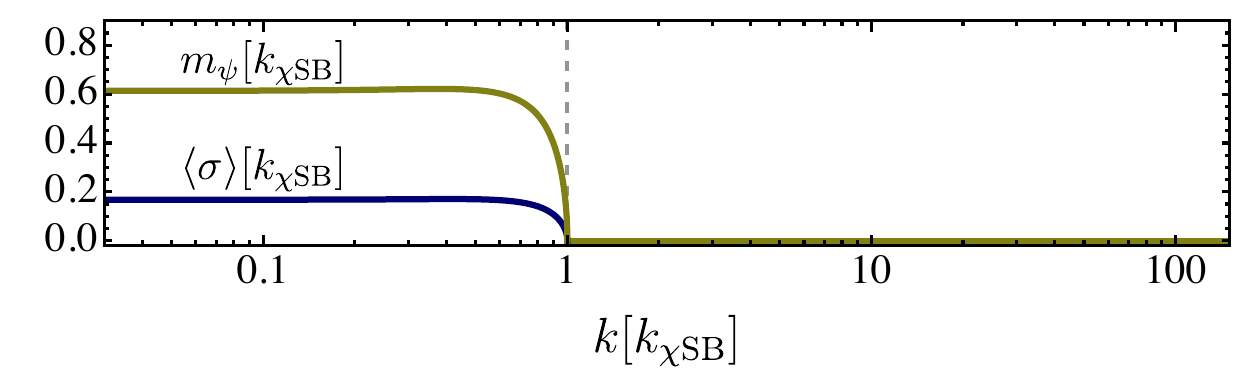}
	\caption{ In the top panel, we show the curvature of the dimensionless renormalised effective potential \labelcref{eq:u} at the minimum (blue) and the scalar-pseudoscalar four-Fermi coupling (brown) defined in \labelcref{eq:eff4Fermi}, for a SU(3) gauge theory with $N_f=2$ fundamental flavours. In the middle panel we display the two exchange couplings $\alpha_\sigma,\alpha_\pi$ defined in \labelcref{eq:alphaphi}. 	In the bottom panel, we show the development of a non-trivial minimum of the effective chiral potential ($\langle\sigma\rangle$, in blue) and the constituent euclidean fermion mass (green). Marked by vertical dashed lines, the $\kSSB$ scale is depicted.}
	\label{fig: chiralSB explained}
\end{figure}
%

\subsection{Dynamical chiral symmetry breaking }
\label{sec:ScalechiralSB}

In this Section we focus on the functional implementation of $\dSSB$ and the related observables such as the chiral condensate and the constituent fermion masses. The standard chiral order parameter is the chiral condensate which is proportional to $\langle \sigma \rangle$, see \labelcref{eq:sigma0ChiralCondensate}. Its onset defines the chiral symmetry breaking scale $\kSSB$ in \labelcref{eq:kchi}, and it is also signalled by a vanishing of the curvature of the potential, $\bar m_\sigma(\rho_0)$ at $\rho_0=0$.

In the top panel of \Cref{fig: chiralSB explained} we show the curvature of the chiral potential around the flowing global minimum $\rho_0$ as a function of the cut-off scale. In the deep UV, the curvature of the potential around the trivial minimum is very large, implying highly suppressed fermionic self-interactions. As the gauge dynamics strengthen, the chiral potential flattens until at $\kSSB$ (vertical dashed line) the potential develops a non-trivial minimum, $\langle \sigma \rangle$. In the bottom panel, we show the evolution of this parameter (blue line) which additionally is the chiral phase transition order parameter. In the same plot, we also show the fermion constituent mass ($m_\psi$, in green, see \labelcref{eq:mpsi}). It is worth noting in this context that the sharp onset of $\langle \sigma\rangle$ with the cut-off scale does not signal a second order-type behaviour, in terms of a physical momentum scale it is smooth, see e.g~the recent discussion of the fermion mass function in \cite{Ihssen:2024miv}. Still, it mirrors the second order phase chiral transition at finite temperature.

Without the emergent composites formulation, $\dSSB$ is signalled by a divergent four-Fermi coupling given that $ \lambda_\textrm{\tiny{SP}} \propto h_\phi^2/u'(\rho)$ as in \labelcref{eq:lambdasigmap}. While this approach has been made quantitative in QCD studies \cite{Fu:2022uow,Fu:2024ysj}, it is significantly more challenging when studying properties in the chirally broken sector. In particular, the access to the broken chiral regime in theories in the chiral limit is restricted to the four-Fermi formulation due to a singularity in the flow. Moreover, this strategy was traditionally employed to trace whether ${\rm d}\chi{\rm SB}$ occurs or not  \cite{Gies:2003dp,Gies:2005as} in the many-flavour limit of gauge-fermion QFTs. In the top panel of  \Cref{fig: chiralSB explained}, we show the four-Fermi coupling for the (S-P) channel  (brown line), which is the relevant one to trace the development of a non-trivial scalar condensate. It becomes apparent how the bosonised picture is more convenient as it straightforwardly allows to flow into the broken phase of the potential. We note that each independent computation leads to a slightly different $\kSSB$. 

\begin{figure}[t!]
	\centering
	\includegraphics[width=1.\columnwidth]{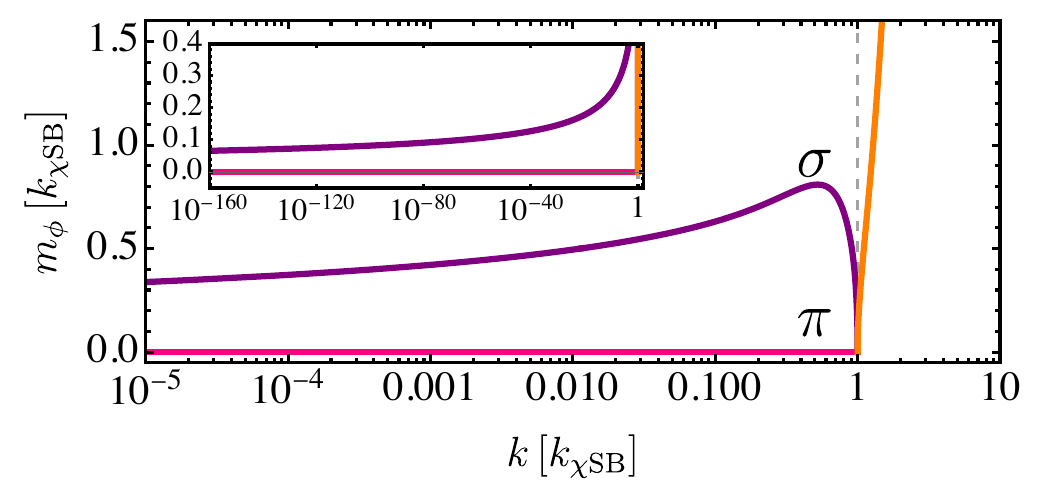}
	\caption{ For a SU(3) gauge theory with $N_f=2$ fundamental flavours in the chiral limit, the composite masses ($m_\sigma$ and $m_\pi$) in the symmetric (orange) and broken (purple and magenta respectively) phases of the chiral potential. In the inlay, we zoom deep into the deep IR scales. 
	}
	\label{fig:mphi_Nf2}
\end{figure}

The curvature of the mesonic potential displayed in \Cref{fig: chiralSB explained} is equivalent to the mass of the scalar degrees of freedom, see \labelcref{eq:mphi}. In the UV, the curvature is very large leading to very large masses and a decoupling of these modes in the dynamics. In \Cref{fig:mphi_Nf2} we show the euclidean masses of the bosonised channels at scales approximately and below $\kSSB$. For $k>\kSSB$, the symmetry is not broken and hence the Goldstone bosons and radial modes are not distinguishable. In the broken phase, the Goldstone bosons are exactly massless and the $\sigma$-mode behaves analogous to the Higgs boson. Its mass in the chiral limit grows until the fermionic loops decouple as $k\sim m_\psi$. For lower scales, the only remaining dynamical degrees of freedom are the massless pions. While the mass of the $\sigma$-mode is subject to discussion and is an open case of research \cite{Ihssen:2024miv,Santowsky:2020pwd,Eichmann:2020oqt,Santowsky:2021bhy,Santowsky:2021ugd,Santowsky:2021lyc}. In the present truncation the bosonic contributions to the flow of the curvature of the potential lead to a mild change which over many orders of magnitude finally freezes around $m_\sigma \approx 0.07\, \kSSB$. The precise determination of this quantity is delicate and subject to the truncation employed on the effective chiral potential \labelcref{eq:Veff}. Similar computations employing a more advance  non-polynomial type of potential have been performed in \cite{Ihssen:2024miv}  and display a freeze-out if the mass at $m_\sigma \approx  0.53 [\kSSB]$. 

On the other hand, the results here obtained are well understood in the non-chiral limit and physical QCD, see eg. \cite{Ihssen:2024miv}. The consideration of explicit fermion masses breaks chiral symmetry at the classical level. This leads to pions as Goldstone bosons of the global chiral symmetry and hence massive. Consequently, their contribution decouples at a finite cut-off scale $k\sim m_\pi$ in the broken chiral phase, leading to a heavier $m_\sigma$.  For example, in the case of $m_\pi\sim 0.1 \,\kSSB$ this will lead to $m_\sigma\sim 0.6\, \kSSB$. 

\begin{figure*}[t]
	\includegraphics[width=1\textwidth]{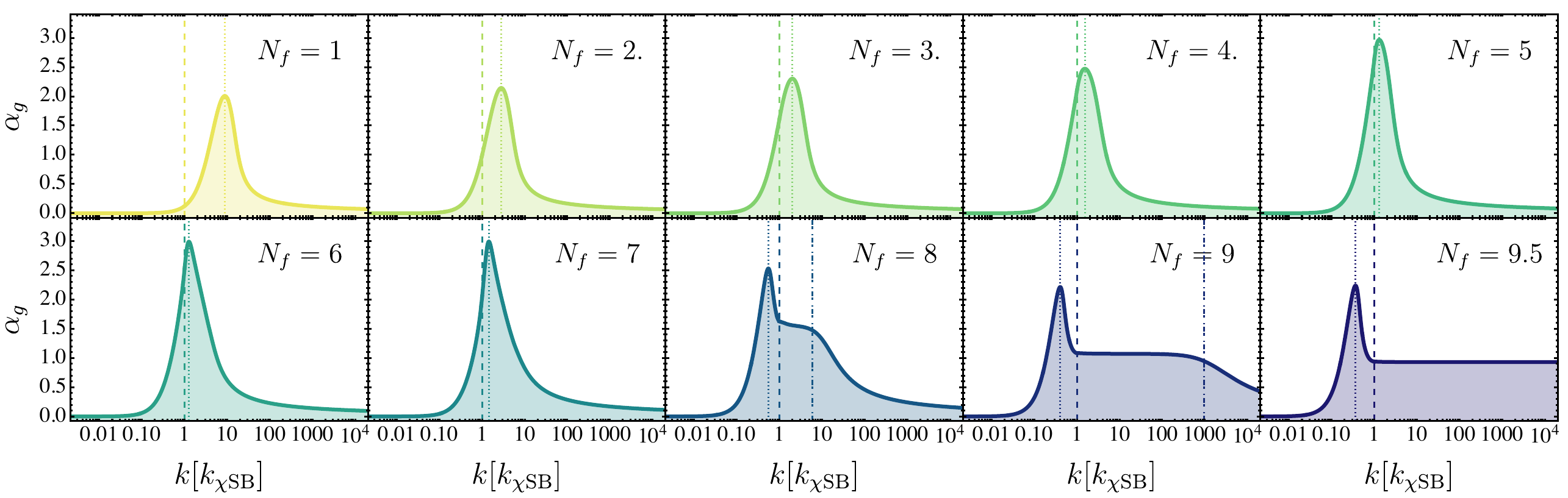}
	\caption{Gauge exchange couplings for a SU(3) gauge theory with different number of $N_f$ fundamental flavours.  From right to left and top to bottom, we increase the number of flavours. The couplings are shown  in units of the respective $\kSSB$ scales. In each panel, the respective confinement (dotted) and chiral symmetry breaking (dashed) scales are marked by vertical lines. For theories in the close-conformal regime $N_f\gtrsim8$, the approximate onset of the walking regime $k_{\rm walk}$ is marked by a vertical dot-dashed line. In this Figure we have employed the single-avatar truncation introduced in \Cref{sec:phasesYM+chiral} and detailed in \Cref{app:ManyFlavour}. }
	\label{fig: alpha_g semipert diff Nf}
\end{figure*}
%

\section{Towards the conformal window}
\label{sec:phasesYM+chiral}

In this Section we evaluate the interplay between the chiral and confining dynamics of gauge-fermion systems for different number of flavours including the QCD-like, just discussed in \Cref{sec:InterplayLowNf}, and the vicinity to the conformal window. The results in the previous Section confirmed the semi-quantitative nature of the present approximation of the effective action and the flow even in the strongly correlated QCD-like regime. This IR regime is successively getting smaller and less correlated for larger $N_f$ and the reliability of the current approximation is even improved. 

In the many flavour limit, we observe the emergence of conformality, signalled by a rapidly increasing walking regime: the doubly gapped IR regime with $\dSSB$ and confinement is moved far towards the IR. This goes hand in hand with a widening gap of the characteristic scales $\kSSB$ for chiral symmetry breaking and confinement $\kconf$ with $\kSSB/\kconf>1$. This is but one of the dynamical signatures of the 't Hooft anomaly matching argument \cite{tHooft:1979rat,Ciambriello:2024xzd} for $N_f\geq 3$: no confinement without chiral symmetry breaking, see \cite{HPW} for more details. In turn, the momentum cut-off regime, which carries the approach to the gapped regime, is getting increasingly more prolonged stretching over several orders of magnitude. Accordingly, its accurate description including also higher orders of perturbation theory is mandatory for a quantitative description close and inside the conformal regime. In summary, while the distinction of the different avatars of the strong coupling gets successively less important, the accuracy of the cut-off running of the strong coupling in the regime without chiral and confining gaps gets more important. 

On the basis of this structure we improve our approximation for the fermion-gauge  coupling $\alpha_{\psi\bar \psi A}$ to capture higher orders perturbation theory: this is relevant if running over several orders of magnitudes in the perturbative and semi-perturbative regimes and connecting to the conformal limit. We accompany this improvement with a simplification of the approximation that reduces the fine-tuning problem that increases with flowing over successively more orders of magnitude. Moreover, we identify all other avatars of the strong coupling with $\alpha_{\psi\bar \psi A}$, 
\begin{align}
	\alpha_i = \alpha_g\,,\qquad \textrm{with} \qquad \alpha_g =\alpha_{\psi\bar \psi A}\,,
	\label{eq:alphai=alpha}
\end{align}
and $i= A^3,A^4,c\bar c A$. This approximation is detailed in \Cref{app:largeNfapproximation}, and the unique gauge coupling flow is given in \labelcref{eq:gsImproved}. Effectively, the improved approximation incorporates the full perturbative multi-loop resummations, while still accommodating the non-perturbative IR dynamics discussed in \Cref{sec:InterplayLowNf}.  We have checked explicitly that the identifications of all couplings has a subleading impact on the large flavour results, see the discussion in \Cref{app:ManyFlavour} around \Cref{fig: alpha_s Nf=123456 scaling} for $N_f=1,..,5$. In summary, the approximation described here governs both regimes well, the perturbative and semi-perturbative running for $k\gtrsim k_\textrm{conf}, \kSSB$, and the strongly correlated IR regime with $k\lesssim k_\textrm{conf}, \kSSB$. 

The interplay between confinement and $\dSSB$ can be divided into three sub-regimes in $N_f$ for all $N_c$. In the following, we will restrict our analysis to $N_c = 3$ with $N_f$ fundamental flavours, where the regimes are the following:
\begin{itemize} 
\item[(i)]\underline{\textit{QCD-like regime:}} small flavour numbers $N_f\lesssim 4$, where confinement, signalled by the gauge field mass gap, emerges at a larger momentum scale than the chiral symmetry breaking one, given by the onset of the chiral condensate. This regime is discussed in \Cref{sec:Nf<4} and at the end of it as well as in \Cref{app:ManyFlavour}, we discuss why the single avatar approximation lacks quantitative precision in this regime given the very different dynamics are accounted by a single gauge avatar. 

\item[(ii)] \underline{\textit{Locking regime:}} intermediate flavour numbers $5\lesssim N_f\lesssim 8$, discussed in \Cref{sec:4<Nf<8}. In this regime, confinement and chiral symmetry breaking are \textit{locked} and are triggered at similar momentum scales for all flavours in this regime. This is a novel phenomenon, and to our best knowledge it has not been observed before. Since the confining and chiral dynamics are fully intertwined, we expect that the current approximation is only semi-quantitative.  

\item[(iii)] \underline{\textit{Walking regime:}} large flavour numbers $8\lesssim N_f \lesssim N_f^\textrm{crit}$ close to the conformal window signalled by the absence of confinement and $\dSSB$. At the interface with the regime {(ii)} the locking is overcame by the dynamics leading to a rapid decay of the ratio before settling in the scaling in the walking regime. The walking is signalled by the close to vanishing gauge flow and $\dSSB$ occurs at successively higher scales, and confinement follows suit at an increasing relative distance. This regime is dissected in detail in \Cref{sec:EmergentWalking} and we expect quantitative validity of the approximation given its perturbative character and the separation of the dynamical scales. 
\end{itemize}
\begin{figure}[t!]
	\centering
	\includegraphics[width=1\columnwidth]{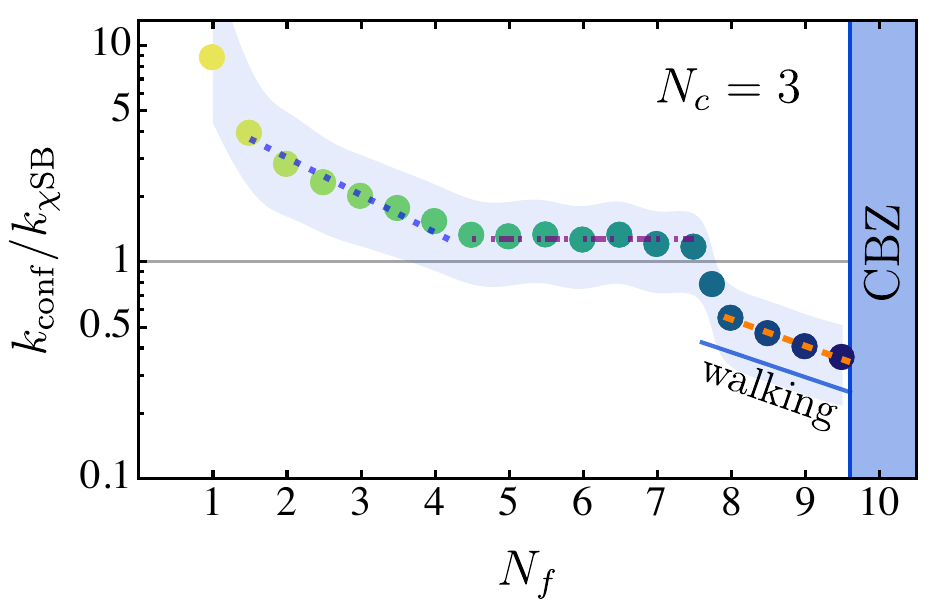}
	\caption{Ratio of the confinement scale $k_{\rm conf}$ and the chiral symmetry breaking scale $\kSSB$ for an SU(3) gauge theory coupled to different numbers of fundamental flavours. 
		The lines indicate the scaling of the ratio in the QCD-like regime {(i)} with \labelcref{eq:DecayQCD} (dotted blue), that in the regime with the locking of confinement and $\dSSB$ {(ii)} with \labelcref{eq:DecayIntertwine} (dot-dashed purple), and the walking regime {(iii)} with  \labelcref{eq:DecayWalking} (dashed orange). 
		The shaded blue bands indicate relative 40\% variation with a slight enhancement in the few flavour limit as discussed in \Cref{app:errorestimateSingleavatar}. The vertical blue line marks the onset of the conformal CBZ window discussed in \Cref{sec:BoundaryCBZ}. 
	}
	\label{fig:kconfkSB_diffNf}
\end{figure}

In the following, we evaluate the onset of the different dynamical sectors with the scales $k_\textrm{conf}$ and $\kSSB$. We emphasise that these scales should be understood as approximate proxies for the physical ones, even though they are directly related to the physical confinement and chiral symmetry breaking scales. 

In \Cref{fig: alpha_g semipert diff Nf} we display the fermion-gauge couplings for an SU($N_c=3$) theory coupled $N_f$ fundamental flavours with $N_f=1,...,9$. For more flavours the gauge coupling enters the conformal regime analysed in \Cref{sec:BoundaryCBZ}. The confinement and chiral symmetry breaking scales are depicted as vertical lines, where $k_{\rm conf}$ is a dotted line and $\kSSB$ is a dashed one. We can clearly distinguish the three regimes (i-iii) which are discussed in the following Subsections:

\subsection{QCD-like regime for $N_f \lesssim 4$} 
\label{sec:Nf<4} 

For few $N_f \lesssim 4$, the confining mass gap proportional to $k_\textrm{conf}$ is sizeably larger than $\kSSB$, see the upper panel of \Cref{fig: alpha_g semipert diff Nf}. In this regime the theory is clearly QCD-like and  we see an approximately linear decrease with $N_f$, 
\begin{align} 
	\frac{k_\textrm{conf}}{\kSSB} \approx  \, ({\rm const.} -\gamma^{\textrm{(i)}} N_f) \qquad \textrm{with}\qquad \gamma^{\textrm{(i)}} \approx 0.6\,.
	\label{eq:DecayQCD}
\end{align}
This ratio is depicted in \Cref{fig:kconfkSB_diffNf} which also illustrates the clear separation of the three regimes (i-iii). 

The large value of the gluon mass gap effectively switches off the glue dynamics and $\alpha_g$ decays towards the IR as $k\lesssim k_\textrm{conf}$. Conceptually, this picture allows for an scenario without chiral symmetry breaking if  $\alpha_g$ peaks too early with a relatively small peak value. This intricacy is responsible for the fact that few-flavour theories are very sensitive to the value of the coupling and its momentum dependence. In short, physical QCD and even more the $N_f=1$ case, are \textit{living on the (chiral) edge}, see \cite{Dupuis:2020fhh,HPW}. 

\begin{figure}[t]
	\centering
	\includegraphics[width=1\columnwidth]{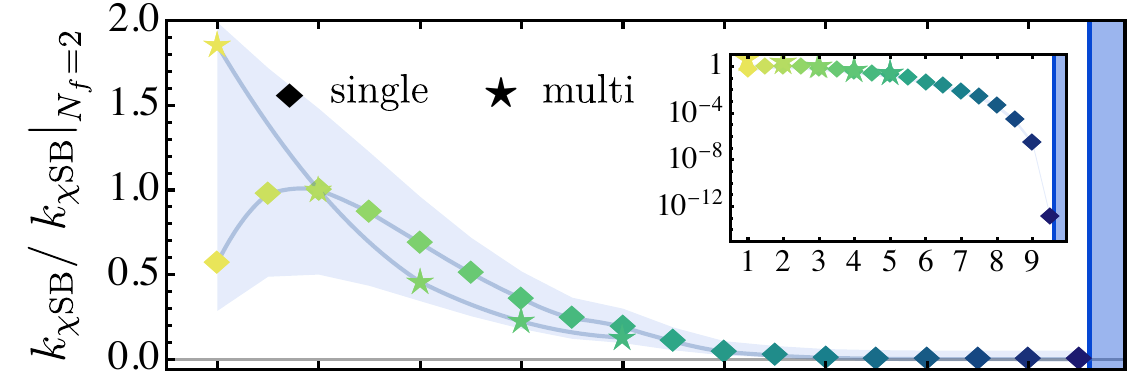}
	\includegraphics[width=1\columnwidth]{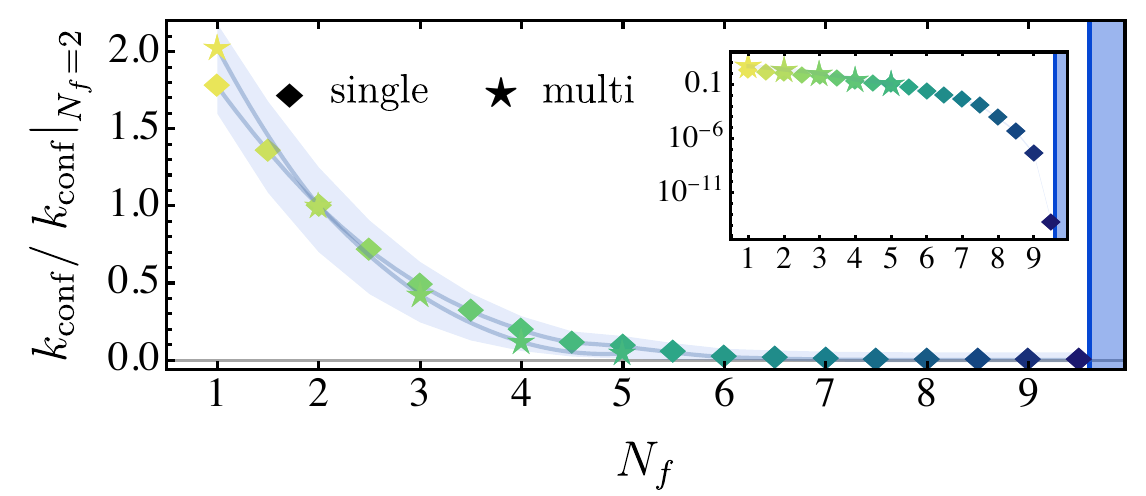}
	\caption{ $N_f$ dependence of the $\kSSB$ (top) and $\kconf$ scales (bottom) in absolute UV scale (where all couplings are initiated with $\alpha_g\approx0.08$) for a SU(3) gauge theory with $N_f$ fundamental  flavours. We show the results in the single (diamonds) and multi-avatar (stars) approximations detailed in \Cref{app:largeNfapproximation} and \Cref{sec:InterplayLowNf}, respectively. The shaded areas depict the estimated errors discussed in \Cref{app:errorestimateSingleavatar}. In the inlay plots, the same quantities in a logarithmic scale are depicted. In \Cref{fig:mpsi_sigma_largeNf} we display the equivalent dependence of $m_\psi$ and $\langle\sigma\rangle$ on $N_f$.}
	\label{fig:kSSBkconf_largeNf}
\end{figure}

Furthermore, in terms of an absolute UV scale, the onsets of confinement and chiral symmetry breaking are delayed with an increasing number of flavours. This can be seen in \Cref{fig:kSSBkconf_largeNf} where we have depicted the relevant scales in units of an absolute UV scale for different $N_f$. We show the results in the single (diamonds) and multi-avatar (stars) approximations detailed in \Cref{app:largeNfapproximation} and \Cref{sec:InterplayLowNf}, respectively. For $N_f=1$, the single-avatar approximation displays a dip in $\kSSB$ in comparison to the multi-avatar truncation. This is an artefact of the latter truncation which displays a too low $\kSSB$ in the $N_f\lesssim 2$ limit. This is a straightforward consequence of only accounting for a single avatar (the fermion-gauge coupling) which is demanded to describe both, qualitatively different confining and chiral dynamics. As shown in \Cref{fig: alpha_s Nf=2 scaling}, the pure gauge avatars are the responsible for confinement and peak much lower than the gauge-fermion one, responsible for $\dSSB$. Identifying all to the same, requires a longer exposure to the gauge dynamics in order to trigger $\dSSB$ which causes a delay in the appearance of $\kSSB$.  Moreover, this defect propagates into $\kSSB$-dependent quantities such as the constituent fermion mass or the chiral condensate, which both show a smaller magnitude in units of $\kconf$ (see \Cref{fig:mpsi_sigma_largeNf}). Consequently, it is well understood why the single-avatar truncation in the few flavour limit only provides a qualitative description. On the other hand, we point out the remarkable agreement in $\kconf$ between both truncations. In shaded blue in \Cref{fig:kSSBkconf_largeNf,fig:mpsi_sigma_largeNf}, we depict the error estimate on these quantities accounting for the knowledge from both approximations each successful in a different regime. For a detailed discussion we refer to \Cref{app:largeNfapproximation} and \Cref{app:Nfscaling}.

\begin{figure}[t]
	\centering
	\includegraphics[width=1\columnwidth]{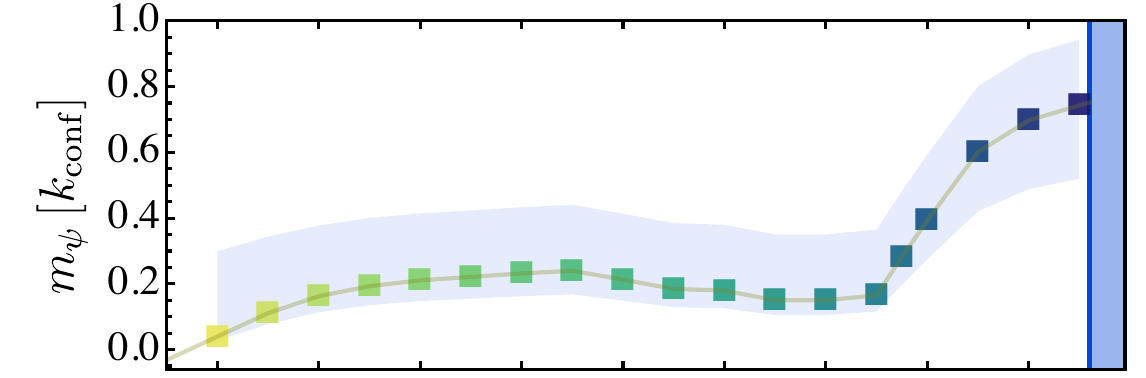}
	\includegraphics[width=1\columnwidth]{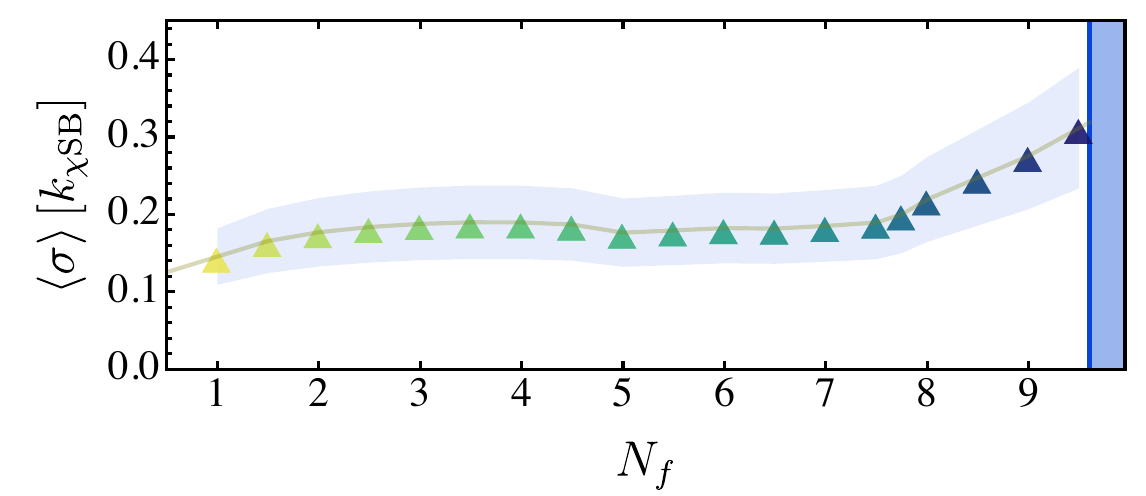}
	\caption{Constituent fermion masses (squares in the top panel) and chiral condensate $\langle \sigma\rangle$ (triangles in the bottom panel) for $N_f$ fundamental flavours charged under a SU(3) gauge theory. The fermion masses are shown in units of the respective confinement scales and the chiral condensate in those of the chiral symmetry breaking scales. The blue bands show the estimated error discussed in \Cref{app:errorestimateSingleavatar}. In \Cref{fig:mqmsigmasigma_diffNf_appendix} we show the same quantities in different units along with the mass of the $\sigma$-mode.}
	\label{fig:mqmsigmasigma_diffNf}
\end{figure}
%

\subsection{Locking regime for  $5\lesssim N_f \lesssim 7$} 
\label{sec:4<Nf<8} 

In this regime we encounter a novel phenomenon, the \textit{locking} of confinement and chiral symmetry breaking, see \Cref{fig:kconfkSB_diffNf}: the ratio of the confinement and $\dSSB$ scales is approximately unity for all $N_f$ in this regime, 
\begin{align} 
	\frac{k_\textrm{conf}}{ \kSSB} \approx 1 \,.
	\label{eq:DecayIntertwine}
\end{align}
This is in stark contrast to the decrease of the ratio in the QCD-like regime and the scaling in the walking regime, discussed in the next \Cref{sec:EmergentWalking}, see \labelcref{eq:DecayWalking}. Its detection in the present work is closely linked to the differential scale resolution 
inherent to the fRG. In this strongly correlated regime, fermion self-interactions are more enhanced than in the QCD-like regime, causing typically subdominant channels to become relevant in the chiral dynamics. In fact, in this regime, we find evidence for potential critical channels at $k < \kSSB$, particularly the second-leading (V-A)$_{\rm adj}$ channel. A finer resolution of the dynamics is provided by \Cref{fig:mqmsigmasigma_diffNf}, where we show the constituent fermion mass in units of the confinement scale, $m_\psi[k_\textrm{conf}]$ (squares), and the chiral condensate in units of the chiral symmetry breaking scale, $\langle \sigma\rangle[\kSSB]$ (triangles). Both quantities are approximately constant for $3\lesssim N_f\lesssim 7$, also providing further evidence for our definitions of $k_\textrm{conf}$ and $\kSSB$ as good proxies for the confinement and $\dSSB$ scales. We defer the reader to \Cref{app:Nfscaling} for further details. 

To our knowledge, this phenomenon has not been observed before and this exciting locking of the confining dynamics and that of $\dSSB$ deserves a detailed analysis as it certainly leads to new insights in both of them. In this context we remark that a similar phenomenon has been observed in finite temperature QCD  \cite{Braun:2012zq,Braun:2011fw}, where it was shown that the order parameters of both phase transitions, the Polyakov loop and the chiral condensate, are dynamically locked as are the phase transitions. While being fascinating it goes beyond the scope of the present analysis and we deferred to future work, where we will also use a fully momentum-dependent approximation which facilitates this analysis.

\subsection{Walking regime for $8 \lesssim N_f < N_f^\textrm{crit}$} 
\label{sec:EmergentWalking}

As discussed before, at the interface between the locking and walking regime {(iii)}, the chiral dynamics overpowers the confining one. Hence, the locking is released and the ratio changes rapidly within less than one additional flavour until it settled in the walking regime, see \Cref{fig:kconfkSB_diffNf}. Indeed, this behaviour illustrates the locking itself.  

We see in \Cref{fig: alpha_g semipert diff Nf}, showing the momentum-scale dependence of the fermion-gauge coupling $\alpha_g$, that a growing momentum window with close-conformal scaling emerges in the regime $8 \lesssim N_f < N^{\rm crit}_f$. The scaling window starts at a momentum scale $k_{\rm walk}$ where the coupling growth slows down. This scale is approximately determined by the minimum of the gauge coupling flow in the regime $k>\kSSB$ and its marked by a dot-dashed vertical line in \Cref{fig: alpha_g semipert diff Nf}. This near-root in the gauge flow is approached before the onset of $\dSSB$ and indicates the appearance of \textit{walking} spanning over a range of scales. The walking window extends rapidly towards the IR as the number of flavours approaches the critical and $k_\textrm{conf}/\Lambda_{\rm UV}$ and $\kSSB/\Lambda_{\rm UV}$ in terms of the UV cut-off scale $\Lambda_{\rm UV}$, rapidly decay. The critical flavour number where conformality appears is $N_f^\textrm{crit}\approx 9.6$  for $N_c=3$  and its determination will be the topic of \Cref{sec:BoundaryCBZ}. 

The emergence of walking triggers a particular scaling of fundamental parameters shown in \Cref{fig:kconfkSB_diffNf,fig:mqmsigmasigma_diffNf,fig:mqmsigmasigma_diffNf_appendix}. This is directly caused by a progressive separation of the dynamical scales $\kconf/\kSSB\ll1$ given by the walking regime emerging at weaker values of $\alpha_g$ as $N_f$ increases. Moreover, after fermions decouple at $k \lesssim m_\psi \sim \kSSB$, the gauge dynamics are approximately those of a pure gauge theory. Hence, $\alpha_g$ grows until the Yang-Mills-like gluon mass gap appears. Then, in the regime $k\lesssim \kSSB$, the difference between close-conformal theories is given by the boundary condition of the gauge coupling at $\kSSB$. For increasing $N_f$ at a fixed $N_c$, we observe an exponential decay (dashed orange line in \Cref{fig:kconfkSB_diffNf})  
\begin{align} 
	\frac{k_\textrm{conf}}{ \kSSB  } \propto e^{-\gamma^{\textrm{(iii)}}  N_f} \qquad \textrm{with}\qquad \gamma^{\textrm{(iii)}}  \approx  0.28\,
	\label{eq:DecayWalking}
\end{align}
for the ratio of the two dynamical scales. 

The decay of the confinement scale relative to that of $\dSSB$ suggests that the spectrum of resonances in the close-conformal limit features glueballs that are increasingly lighter than the mesonic and baryonic states, except for the massless Goldstone modes, in contradistinction to QCD-like theories. Being short of a complete analysis that would necessitate solving bound state equations in the current setup, we use the current proxies for the dynamical mass scales: to begin with, the mass scale of chiral symmetry breaking is the chiral condensate, which is proportional to $\langle \sigma\rangle$, for a detailed discussion in the fRG approach see e.g.~\cite{Fu:2019hdw, Braun:2020ada}. The constituent fermion mass is proportional to the product of the chiral condensate and the Yukawa coupling, see \labelcref{eq:mpsi}. If we use it as a proxy for the characteristic hadron mass scale, its ratio with the confinement scale provides an insight in the development of the mass ratios of glueballs and, say, baryons. Surprisingly, this ratio freezes in close to the conformal window, even though the ratio of the chiral condensate and the confinement scale rises even stronger than the ratio of $\kSSB$ and the latter. This hints at a quite non-trivial dynamics in the vicinity of the conformal window, which deserves further investigation. 

\begin{figure}[t]
	\centering
	\includegraphics[width=.975\columnwidth]{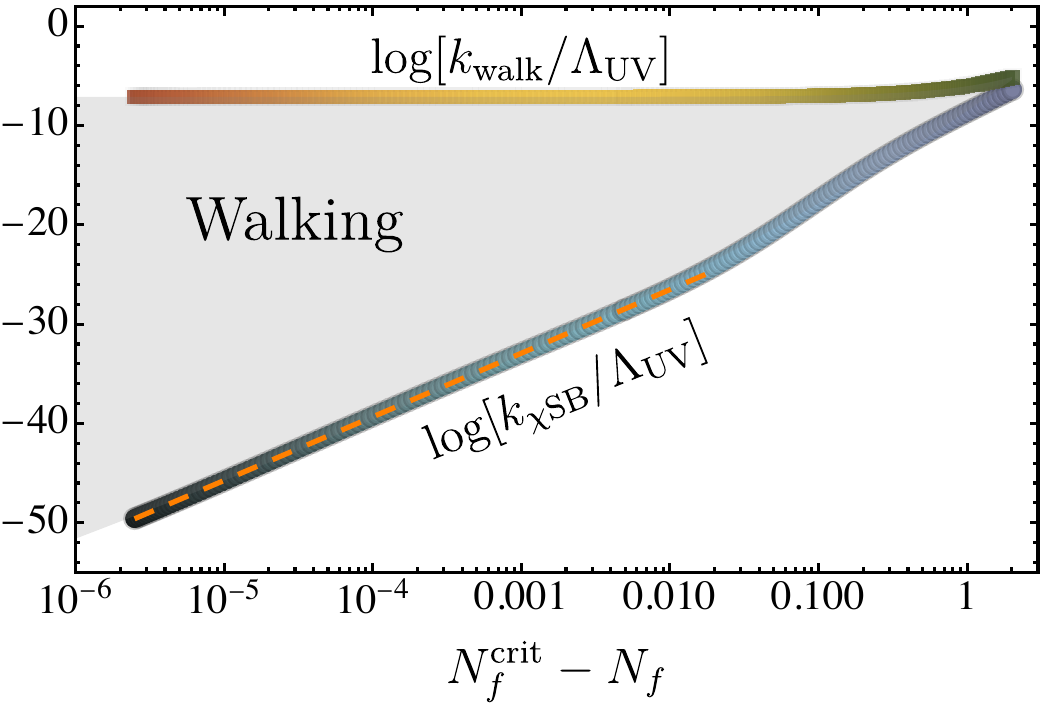}
	\caption{Scaling of the onset scales of $\dSSB$ ($\kSSB$, bottom circles) and walking ($k_{\rm walk}$, top squares) in units of an absolute UV scale $\Lambda_{\rm UV}$ in the near-conformal regime. The grey-shaded regime marks the length of the walking regime. The dashed line depicts the fit in \labelcref{eq:Miransky2} describing near-conformal scaling.}
	\label{fig:Miransky}
\end{figure}

Finally, we investigate the scaling of the walking interval with $N_f$.  In \Cref{fig:Miransky} we show $\kSSB$ in units of the UV scale (lower blue circles) zooming into the conformal regime. Additionally, we display the onset scale of the walking regime ($k_{\rm walk}$, top squares) in absolute units. The shaded regime displays the orders of magnitude over which $\alpha_g$ walks. For $\Delta N^{\rm crit}_f=N^{\rm crit}_f-N_f\lesssim 10^{-2}$, that is close to the conformal window, we find
\begin{align}
	{\rm ln}[\kSSB/\Lambda_{\rm UV}]&\sim {\rm const}+2.78\,\, {\rm ln}\,\Delta N^{\rm crit}_f\,,
	\label{eq:Miransky2}
\end{align}
see the orange dashed line in \Cref{fig:Miransky}. This is in very good agreement with Miransky scaling \cite{Braun:2009ns,Braun:2010qs,Braun:2005uj,Kotov:2021hri,Miransky1997,Miransky:1998dh} and equivalent to a Berezinskii-Kosterlitz-Thouless transition in two dimensions \cite{Kosterlitz:1974sm}. Notably, in the approach to the conformal window, the size of the walking regime expands towards infinity characterising the nature of the conformal phase transition.

Determining the size of the walking regime is pivotal for beyond the SM model building. It requires an approach sensitive to the appearance of fixed points and capable of describing the non-perturbative emergence of $\dSSB$,  as the employed in the present computation.  \Cref{fig:walking regimes all Nc} displays the size of the walking regime, $\Delta k_{\rm walk}=k_{\rm walk}/\kSSB$, as a function of the number of fundamental flavours and for different colours. We emphasize the novelty of this first-principles result and its importance for existing models \cite{Sannino:2016sfx,Goertz:2023nii,Evans:2005pu,DelDebbio:2010hx,DelDebbio:2010hu,Foadi:2007ue,Gudnason:2006ug}. 

A detailed determination of the critical exponents associated to the conformal transition and the size of the walking regime for different gauge groups and fermionic representations will be provided in an independent work.

\subsection{Pre-condensation at large $N_f$} 
\label{sec:precondensation}

The fRG approach with emergent composites allows to investigate the interesting question of the occurrence of regimes with pre-condensation. This phenomena typically happen for second order phase transitions with relatively small condensates. A simple example are Weiss domains in ferromagnets where the condensate points in different spacial areas with a characteristic radius $\xi_\textrm{\tiny pc}$, towards a specific directions in the symmetry group, in our case SU($N_f$). However, averaging over all of these gives a vanishing global condensate. In the fRG approach to gauge-fermion systems this is signalled by a condensate emerging at scales below $\kSSB^+=\kSSB$, which vanishes below a smaller cut-off scale $\kSSB^-$. This translates into,
\begin{align}
	\frac{1}{\cal V_{\xi_{\textrm{\tiny pc}}}}\int\limits_{\| \vec x\|\lesssim \xi_{\textrm{\tiny pc}}} \langle \sigma(x)\rangle    \approx  \int\limits^{\kSSB^+}_{\kSSB^-} \frac{dk}{k} \partial_t \sigma_\textrm{EoM}  \neq 0\,.
	\label{eq:Precondensate}
\end{align}
with the characteristic radius $\xi_\textrm{\tiny pc}$ of the domains ${\cal V_{\xi_\textrm{\tiny pc}}}$, 
\begin{align} 
	\xi_\textrm{\tiny pc}\propto \frac{1}{\kSSB^-}\,,\qquad {\cal V_{\xi_\textrm{\tiny pc}}}= \frac{4 \pi}{3} \xi_\textrm{\tiny pc}^3\,.
	\label{eq:xi+Vxi}
\end{align}
For a related discussion in two-colour QCD see \cite{Khan:2015puu}. Such a pre-condensation phenomenon in the presence of a second order phase transition requires of strong IR dynamics dominated by massless modes, leading to an IR melting of the condensate. Generically this happens at finite temperature $T$, where the IR dynamics in a four-dimensional theory is enhanced significantly, as it is three-dimensional below the temperature scale $2 \pi T$. This is the case for the diquark condensate in two-colour QCD as described in \cite{Khan:2015puu}, but the mechanism is universal. 

For gauge-fermion systems in the chiral limit, pre-condensation could even be present in the vacuum. In QCD-like theories with a small number of flavours the pion corrections are not strong enough to trigger this phenomenon. The pion decay constant $f_\pi\sim \langle \sigma\rangle$, drops from  $\sim93$\,MeV in the physical case where $m_\pi \approx 140$\,MeV to approximately $\sim88$\,MeV in the chiral limit with $m_\pi=0$. This is depicted in \Cref{fig: chiralSB explained,fig:mphi_Nf2} and mirrors the fully quantitative analysis with the fRG approach to QCD in \cite{Ihssen:2024miv}. On the other hand, pre-condensation could be expected  for large flavour numbers as with a growing $N_f$, the infrared dynamics of the pion is enhanced with the increasing number of Goldstone bosons. Despite of this, we do not find evidence for its occurrence in the present vacuum computation as shown in the bottom panel of \Cref{fig:mqmsigmasigma_diffNf}. 

In this context we note that in the present approximation we have reduced the symmetry-restoring effects of the light mesonic degrees of freedom by our approximation of the effective potential: as discussed in \Cref{sec:kchi} below \labelcref{eq:Veff}, the effective potential only depends on $\rho=(\sigma^2 + \boldsymbol{\pi}^2)/2$, hence missing the ($\eta-a$)-fluctuations. This is an approximation that maximally breaks the axial U$(1)_{\rm A}$-symmetry in the effective potential. We expect that the axial U$(1)_{\rm A}$-symmetry is effectively restored for large $N_f$, basically doubling the chiral symmetry restoring fluctuations. However, the current analysis suggests that this only has a minimal impact on the pre-condensation mechanism in the vacuum, but it is important at finite temperature. This will be explored elsewhere.

Nonetheless, the present approach with emergent composites allows us for the first time to resolve between two competing scenarios in the large-$N_f$ limit where either
\begin{itemize}
	\item[(1)] pre-condensation occurs before conformality or,
	\item[(2)] the conformal bound is reached first and pre-condensation is absent.
\end{itemize} 
We emphasise that in scenario (1) the emergence of a chiral condensate at $\kSSB$ at intermediate momentum scales is no signature of $\dSSB$ at $k=0$ as the global condensates vanishes, $\langle \sigma\rangle =0$. However, as the condensate is non-vanishing at length scales $x\lesssim \xi_\textrm{\tiny pc}$, global conformal invariance is violated. The picture realised in this scenario would display an effectively scale invariant regime for length scale $x> \xi_\textrm{\tiny pc}$ while broken at smaller. Furthermore, this concludes that the appearance of $\dSSB$ either triggering pre-condensation or a global condensate leads to the breaking of quantum scale invariance and hence provides a good criterion for the diagnosis of the loss of conformality. This also sustains non-trivially previous functional analyses of the conformal window \cite{Gies:2005as} where scenario (2) was assumed.

\section{Dissecting the conformal window}
\label{sec:BoundaryCBZ}

In this Section we address the conformal limit of gauge-fermion QFTs introduced in \Cref{sec:CBZ}. In particular, we provide a quantitative estimate for the lower boundary of the conformal window $N_f^\textrm{crit}(N_c)$, improving qualitatively and quantitatively on previous estimates from functional approaches. This improvement is based on the emergent composites formulation which allows us to systematically incorporate resonant massless and close-massless composites. Importantly, this also includes the multi-scattering effects in these channels that are known to be important in the vicinity of phase transitions, or more generally, in scaling regimes. In \Cref{sec:CC} we begin with a discussion of the quantitative criterion for conformality, completing the evaluation initiated in \Cref{sec:NecessaryChi}. This analysis is used in \Cref{sec:BZResults} for delimiting the conformal phase boundary of gauge-fermion QFTs.  In order to keep the present analysis accessible to the reader, we have deferred  many of the technical details to \Cref{app:comparisonBoundaryCBZ}. This concerns the chiefly important discussion of the systematic error estimates in the scaling regime, including the non-perturbative effects relevant in the quantitative determination of the boundary.

\subsection{Criterion for loss of conformality}
\label{sec:CC}

In QFT, conformal symmetry requires quantum scale invariance \cite{Wetterich:2019qzx}. This is characterised by an RG fixed point of the theory. Here, a given system exhibits no fixed scale such as a finite mass. In turn, the presence of any fixed scale implies the loss of the fixed point and hence conformality.
 
In \Cref{sec:NecessaryChi}, we have discussed a necessary condition for the dynamical emergence of chiral symmetry breaking: the gauge coupling must exceed the critical coupling strength $\acrit$ in \labelcref{eq:alphacrit}, satisfying \labelcref{eq:dSBB-Necessary}. In this Section, we revisit this argument and assess when this condition also serves as a sufficient criterion. 

For a sufficient criterion it is crucial to analyse all scenarios that entail the disappearance of the CBZ fixed point for small enough $N_f$. We distinguish two scenarios which also encompass those introduced in \Cref{sec:CBZ}: In the first scenario, the CBZ fixed point disappears at values larger than $\acrit$, either towards infinite or via a merger in the strong limit. In the second scenario, the CBZ fixed point disappears at values smaller than $\acrit$, most likely via a merger. Each of these two scenarios leads to a qualitatively different dynamical behaviour of theories with  $N_f\lesssim N^{\rm crit}_f$.  

In the following, we review both scenarios and discuss the sufficient condition for loss of conformality and the impact on the dynamics of the non-conformal theories.

\subsubsection{Walking regimes} 
\label{sec:Walking+Conformality}

In the case where the CBZ fixed point disappears at values larger than $\acrit$, a walking regime necessarily appears at $N_f\lesssim N^{\rm crit}_f$, signalled by close-zero $\beta$-functions for the gauge coupling and the dimensionless four-Fermi coupling, to wit, 
\begin{align}
	&\beta_{\alpha_g}(\alpha_\textrm{walk})\to 0^-\quad {\rm and}
	&\beta_{\bar  \lambda_\textrm{\tiny{SP}}}(\alpha_\textrm{walk} ) \to 0^-\,.
	\label{eq:lambdacrit}
\end{align}
Note that in this scenario the walking regime of the gauge coupling is achieved first and the four-Fermi coupling is dragged towards its walking regime in a finite flow time. The smallness of the $\beta$-functions in the walking regime is inversely proportional to its lengths, and $\dSSB$ occurs at a finite flow time as the $\beta$-function of the dimensionless four-Fermi coupling is negative for all $\bar \lambda_\textrm{\tiny{SP}}$, see \Cref{fig:dtlambda_cartoon} in \Cref{sec:RGdynChi} and the related discussion. The occurrence of $\dSSB$ also terminates the walking of the gauge coupling and confinement follows suit. 
 
As $N^{\rm crit}_f$ is infinitesimally approached from below, the walking coupling $\alpha_\textrm{walk}$ tends to the critical one from above, $\alpha_\textrm{walk}\to \acrit$. This opens the limit where an infinite walking regime occurs as both couplings approach the critical one, $\alpha_\textrm{walk}\to \alpha^*_g\to \acrit$. 
This leads to the disappearance of the $\dSSB$ scale, $\kSSB$, towards zero and an infinite size walking regime as the conformal window is approached, as depicted in \Cref{fig:Miransky}. The described mechanism explains the smooth and continuos disappearance of dynamics and the appearance of quantum scale invariance.

In conclusion, in this scenario the introduced condition \labelcref{eq:dSBB-Necessary} is both, necessary and sufficient. Hence, the boundary between dynamics and conformality is given by the theories satisfying
\begin{align} 
	\acrit=\alpha_g^*: \quad \beta_{\bar \lambda_\textrm{\tiny{SP}}}=0= \beta_{\alpha_g}\,. 
	\label{eq:boundarycondition}
\end{align}
A plausible but unlikely sub-scenario within the present class may be given when the $\dSSB$ scale disappears at finite values as the conformal window is reached. This occurs if the CBZ fixed point profile displays a discontinuity in $N_f$. Then for $N_f<N_f^{\rm crit}$, $\alpha_{\rm walk}$ does not cross $\acrit$ but remains at larger values. Then, the discontinuity at $N_f\ge N_f^{\rm crit}$ continuos with a fixed point at $\alpha_g^*<\acrit$, realising conformality.  Here, the conformal-non-conformal phase transition is of first order. 

\subsubsection{Fixed-point mergers} 
\label{sec:Merging+Conformality}

The second possibility involves the CBZ fixed-point solution disappearing at values smaller than $\acrit$, most likely via a fixed-point merger. This mechanism, reviewed in \Cref{sec:CBZ}, is indeed present in the perturbative five-loop $\overline{\text{MS}}$ $\beta$-function for $\alpha_g$ \cite{Herzog:2017ohr,Baikov:2014qja,Baikov:2016tgj}. Up to this order, and without further improvement of the perturbative series, the CBZ fixed point disappears at weak coupling values, much smaller than $\acrit$. In this scenario, the lower boundary of the conformal window is determined by the merger rather than the appearance of a dynamical scale. 

Furthermore, theories below the conformal window do not exhibit a quasi-root in the flow of the gauge coupling, indicating the absence of walking for $N_f < N_f^{{\rm crit}}$. The emergent walking regime discussed in \Cref{sec:EmergentWalking} now transitions to a regime where a similar scenario to the $N_f = 6$ case extends up to $N_f^{{\rm crit}}$. The transition between the conformal and dynamical phases in this cases matches that of the latter sub-scenario in \Cref{sec:Walking+Conformality}, is of first-order type.

While this scenario cannot be entirely ruled out, we consider it unlikely. Notably, the fixed-point merger is absent after a (Padé or Borel) resummation \cite{DiPietro:2020jne} and does not appear in lower-loop orders of the coupling. The fRG inherently generates such a resummation via the $[\Gamma^{(2)}_k + R_k]^{-1}$ structure, which supports the resummation results. Moreover, lattice computations do not observe this scenario, instead finding indications of walking regimes. In summary, this suggests that fixed-point mergers are artefacts of the polynomial expansion of the $\beta$-function in powers of $\alpha_g$, where it is well-known that such expansions may not only introduce fixed-point mergers but also additional fixed points.

\begin{figure*}[t!]
	\centering
	\includegraphics[width=.9\columnwidth]{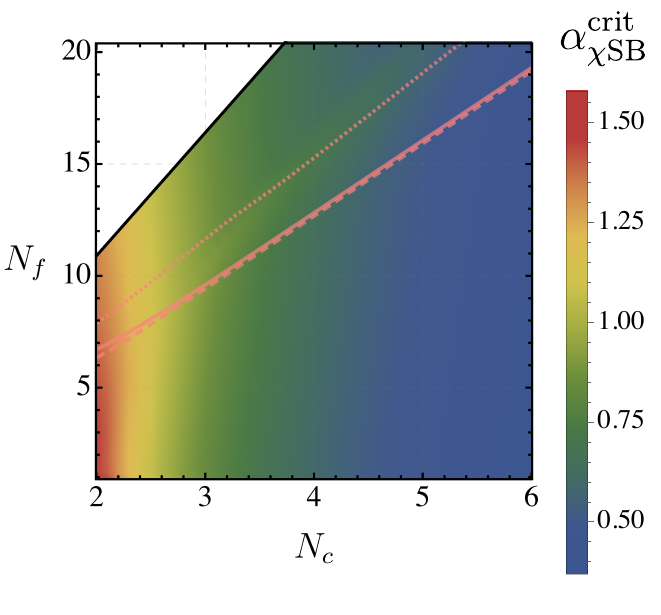}\hspace{.5cm}
	\raisebox{0.12\height}{\includegraphics[width=1.05\columnwidth]{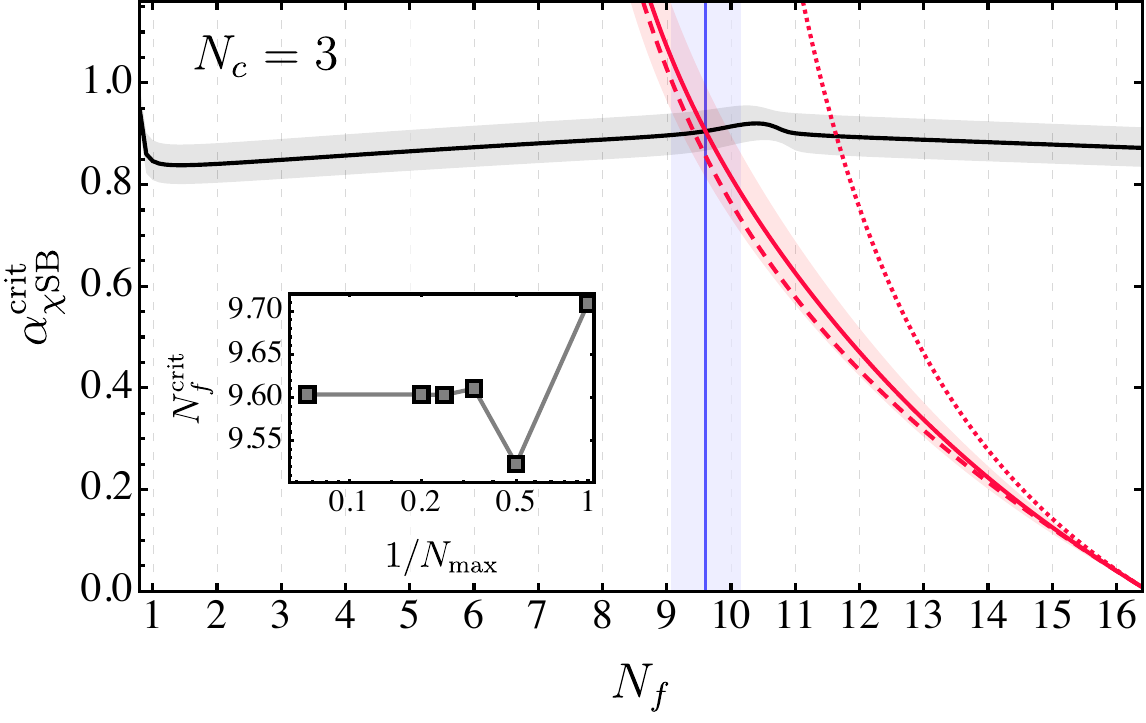}}
	\caption{On the left-side panel, the critical exchange gauge coupling $\alpha^{\rm crit}_{\chi{\rm SB}}$ for SU($N_c$) gauge theories with $N_f$ flavours in the fundamental representation of the gauge group is displayed. The upper black line marks the loss of asymptotic freedom and the pink ones the lower boundary of the CBZ window considering two- (dotted), three- (dashed) and four-loop (plain) $\overline{\rm MS}$ results. On the right-side panel, the same quantity is shown projected on $N_c=3$. The grey-shaded area depicts the error on the $\alpha^{\rm crit}_{\SSB}$ determination and the red-shaded one, the error considered on the fixed-point value. The vertical blue line marks \labelcref{eq:Nfcritbest} and the respective shaded area the estimated error on its determination. See \Cref{app:comparisonBoundaryCBZ,fig:alphas_crit Comparisons} for details on the systematic approach. In the inlay plot, we display $N_f^{\rm crit}$ as a function of the highest-order fermionic operators accounted for in \labelcref{eq:NmaxfourFermi}, see \Cref{fig:alphas_crit Comparisons Nmax} for a more detailed analysis. }
	\label{fig:alphas_crit}
\end{figure*}
%

\subsection{Boundary of the conformal window}
\label{sec:BZResults}

In \Cref{sec:CC}, we have discussed the sufficient criteria for determining the conformal regime in various scenarios. In this Section, we focus on the likely scenario discussed in \Cref{sec:Walking+Conformality}, where the CBZ fixed point disappears at larger values than the critical coupling $\acrit$. The boundary between the conformal and dynamical regimes is then given by the equality in \labelcref{eq:boundarycondition}. For a quantitative determination of $N_f^{{\rm crit}}(N_c)$, where this relation is satisfied, we require two pieces of information: the precise dependence of $\acrit$ and the CBZ fixed point $\alpha_g^*$ on the number of flavours.

In the present approach with emergent composites, the absence of $\dSSB$ (signalled by $\beta_{\bar{\lambda}_{\tiny{\rm SP}}} = 0$) appears as a globally positive curvature of the effective chiral potential \labelcref{eq:Veff} at $k = 0$. More precisely, $\left.\partial_{\bar{\rho}} V \right|_{\bar{\rho}_0 = 0} > 0$, or equivalently, the first expansion coefficient $\lambda_{\phi,1,\,k=0} > 0$. The gauge dynamics strength that nullifies these quantities at $k = 0$ precisely determines $\acrit$.

In particular, for the first time we consider higher order mesonic scattering effects. These effects are captured in the effective potential including high-order interactions in the mesonic fields up to order $N_{\text{max}}$. These terms accommodate key non-perturbative effects encoded in fermionic self-interactions up to power
\begin{align}
	\left( \bar \psi\, {\cal T}_{\left( {\rm S-P} \right)}\, \psi \right)^{2 N_{{\rm max}}}\,,
	\label{eq:NmaxfourFermi}
\end{align}
with the resonant and dominant scalar-pseudoscalar tensor structure ${\cal T}_{\left( {\rm S-P} \right)}$ defined in \labelcref{eq:4FermiTensors}. We emphasise that non-perturbative contributions are accounted for in particular for $N_{\text{max}} > 1$. These effects are crucial for a quantitative determination of $N_f^{{\rm crit}}$, see the inlay in the right panel of \Cref{fig:alphas_crit}, and are one of the novel advances of the present work. In the following analysis we employ $N_{{\rm max}}=5$ which shows well converged results as discussed in detail in \Cref{app:convergenceCBZ} and shown in the inlay of the right plot of \Cref{fig:alphas_crit}.

The second ingredient for the determination of $N_f^{\rm crit}$ with \labelcref{eq:boundarycondition} is the CBZ fixed-point values as a function of $N_f$. For a quantitative determination, we include higher loop orders of the $\alpha_g$ beta function derived in perturbative computations. Accordingly, we infuse this running into the hybrid non-perturbative approach employed in \Cref{sec:phasesYM+chiral} and \Cref{app:largeNfapproximation} that includes both, the multi-loop perturbative running for small couplings and the emergence of $\dSSB$ and confinement for larger values. This heuristically amounts to substituting the anomalous dimensions and $\beta$-function prefactors in the diagrams by those of higher-order perturbation theory. This procedure can be applied systematically and  here we do a first step in this direction using directly perturbative $\beta$-functions in the $\overline{\rm MS}$ scheme, available up to several loop orders \cite{Chetyrkin:2017bjc,Ritbergen1997,Herzog:2017ohr,Vermaseren:1997fq,Ritbergen1997,Ruijl:2017eht}. This approximation ignores the relative RG scales, which manifests itself, amongst other effects, in different $\Lambda_\textrm{QCD}$ in the $\overline{\rm MS}$  scheme and that used (implicitly) in the fRG, the MOM${}^2$ scheme, see \cite{Gao:2021wun}. A detailed discussion of the respective errors, the systematic improvement of this scheme, and the incorporation of results from other perturbative approaches, such as the mMOM scheme \cite{Gracey:2013sca,vonSmekal:2009ae}, super-YM-inspired beta functions \cite{Novikov:1983uc,Ryttov:2007cx,Pica:2010mt,Kim:2020yvr}, and others, as well as the application of improvement techniques like Pad\'e and Borel resummations \cite{Pica:2010mt,DiPietro:2020jne}, will be addressed in a separate work.

With these preparations, we are set up to predict the boundary of the conformal window. In the left panel of \Cref{fig:alphas_crit}, we depict $\acrit$ for arbitrary SU($N_c$) gauge theories with $N_f$ fundamental fermions. Its profile decreases with $N_c$ and is approximately constant in $N_f$. We mark the line of theories satisfying the critical condition \labelcref{eq:boundarycondition} employing two- (dotted), three-(dashed) and four-loop (plain) $\overline{\rm MS}$ results in the determination of $\alpha^{*}_{g}$. On the right panel, we show the projection on to the $N_c=3$ plane (black line) and display the fixed-point values for each of the perturbative beta functions studied (red lines). The critical coupling derived in the best approximation of the bosonised system matches the fixed-point value of the four-loop $\overline{\rm MS}$ scheme at 
\begin{align}
	N^{\rm crit}_f(N_c=3)= 9.60^{+0.55}_{-0.53} \,,
	\label{eq:Nfcritbest}
\end{align} 
providing a non-perturbative and quantitative prediction for the boundary of conformality. A detailed analysis of error estimate determination as well as a comparison to different truncations and previous computations can be found detailed in \Cref{app:comparisonBoundaryCBZ}. In the following, we briefly discuss the two main sources of error. 

First, the present approach uses input from high-order perturbative computations for the beta-function coefficients and fixed-point values which are scheme dependent and subject to an error. For the employed beta functions we estimate an error of $10\%$ in the magnitude of the fixed-point value (red shaded area in the right plot in \Cref{fig:alphas_crit}). We find this to be a good estimate given the relative difference of the four- and  three-loop ones is less than $3\%$. 

The second source of error in \labelcref{eq:Nfcritbest} comes from the determination of $\acrit$ in the fRG. A detailed analysis as well as numerous checks of the truncation have been performed in \Cref{app:comparisonBoundaryCBZ}, see  \Cref{fig:alphas_crit Comparisons,fig:alphas_crit Comparisons Nmax}, concluding that the results presented are very stable and all the aspects of the truncation are well converged. However, we consider a 5\% error (grey band) for unaccounted tensor structures in the gauge-fermion vertex, see discussion in \Cref{app:comparisonQCD}.

\begin{figure}[t!]
	\centering

	\includegraphics[width=.95\columnwidth]{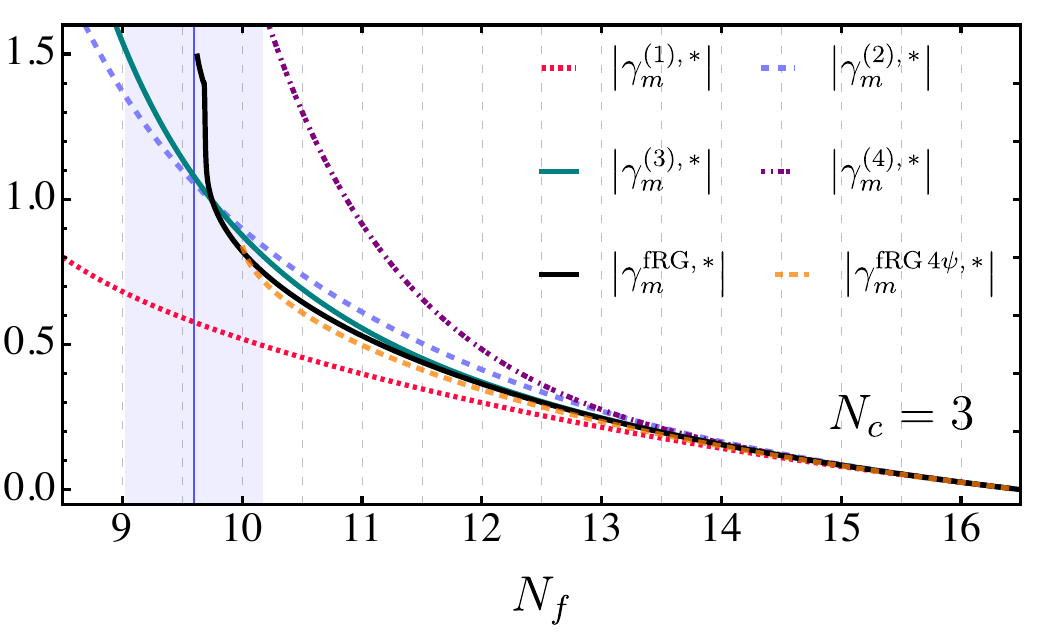}
	\caption{Comparison between the absolute values of the non-perturbative fRG fermion mass anomalous dimension ($\gamma^{\rm fRG}_m$ in \labelcref{eq:gammamfull}, black line), the obtained in the pure four-Fermi language  ($\gamma^{{\rm fRG\, 4}\psi}_m$, orange dashed) and the up to one- ($\gamma_m^{(1)}$, red dotted), two- ($\gamma_m^{(2)}$, light blue dashed), three- ($\gamma_m^{(3)}$,  green plain) and four-loop ($\gamma_m^{(4)}$, dark purple dot-dashed) $\overline{\rm MS}$ expressions \cite{Vermaseren:1997fq} evaluated on the four-loop $\overline{\rm MS}$ fixed-point values. As a vertical blue line, we depict the critical number of flavours \labelcref{eq:Nfcritbest} with the estimated error (shaded band). }
	\label{fig:gammam}
\end{figure}

We close this Section with a discussion on a commonly used approach for diagnosing the appearance of $\dSSB$. It is based on a perturbative computation of the anomalous dimension $\gamma_m$ of the fermion mass, see \labelcref{eq:gammampert} in \labelcref{sec:CBZ}. It is suggestive that $\dSSB$ occurs for a fixed-point value  $|\gamma^*_m |\sim 1$. The non-perturbative fRG approach used allows us to assess the accuracy and reliability of such a determination of the boundary of the conformal regime. The details of the computation of $\gamma^{\rm fRG}_m$ are presented in  \Cref{app:gammam}. We are led to \labelcref{eq:gammamfull} for the anomalous dimension $\gamma^{\rm fRG}$ which effectively incorporates higher orders of perturbation theory as well as the dynamics of chiral symmetry breaking absent in the standard purely perturbative setup. We quote the result on the four-loop $\overline{\rm MS}$ fixed-point values for $N_c=3$ and $N_f^{\rm crit}$ in \Cref{app:gammam}, 
\begin{align}
	\left|\gamma^{\rm fRG,\,*}_m\right|(N_f^{\rm crit}) = 1.49\,. 
\label{eq:gammamcrit}
\end{align}
In \Cref{fig:gammam}, the anomalous dimensions are evaluated at the four-loop $\overline{\rm MS}$ fixed point for different $N_f$. The functional results lie  closely to the three-loop anomalous dimensions for $N_f \gtrsim 10$, before non-perturbative effects become significant. Indeed, in the non-perturbative fRG approach we observe a increase growth in the as the critical flavour number is approach with a mild non-trivial flattening in $\left|\gamma^{\rm fRG,\,*}_m\right|$. This effect is driven by high-order mesonic interactions leading to $\bar{m}^*_\phi \lesssim 10$ and manifests itself in the thresholds encoded in \labelcref{eq:gammamfull}. In summary, this highlights the importance of a non-perturbative treatment of gauge-fermion systems even within the conformal window. Moreover, it provides non-trivial evidence for the potential existence of strongly interacting conformal field theories at the lower boundary of the CBZ window, which may have interesting phenomenological consequences. 

This result has to be contrasted with the perturbative one-, two-, three-, and four-loop $\overline{\rm MS}$ results \cite{Vermaseren:1997fq}, 
\begin{align}
	\left|\gamma^{(i),\,*}_m\right|(N_f^{\rm crit}) = \{0.58,\, 1.05,\, 1.08,\, 2.64\} \,,
\label{eq:gammamcritpert}
\end{align}
respectively. Interestingly, there are large differences between \labelcref{eq:gammamcritpert} and \labelcref{eq:gammamcrit}, if comparing approximations that accommodate the same order of perturbation theory deep in the conformal regime. We conclude from \Cref{fig:gammam} that this difference stems from the rapid rise of $\gamma^{\rm fRG}_m$ close to the lower boundary of the conformal regime. This rise originates in the strong chiral dynamics close to the boundary which cannot be captured within perturbation theory due to its non-perturbative nature. In our opinion this emphasises the lack of quantitative precision of the perturbative estimate. However, the present functional determination also shows that the perturbative estimate employing the two- and three-loop expressions provides a successful rough estimate.

\section{Summary and conclusions}\label{sec:conclusions}

\begin{figure}
	\centering
	\includegraphics[width=.9\columnwidth]{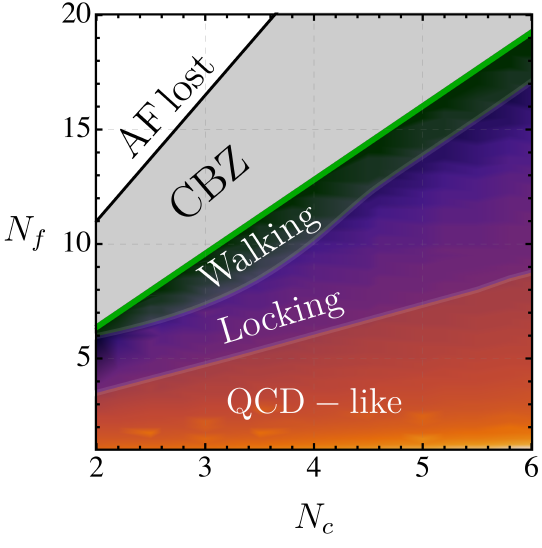}
	\caption{Regimes of SU($N_c$) gauge theories with $N_f$ fundamental flavours. Above the top black line at $N_f \approx 5.5 N_c$, asymptotic freedom is lost. The CBZ conformal window is depicted by a shaded grey region. Below the green line at $N_f \approx 3.2 N_c$, confining and chirally breaking dynamics appear. In the close-conformal region a regime with near conformal scaling is present where $\kconf/\kSSB\lesssim1$ and for fewer flavours, chiral and confining dynamics dominate in the intertwined and QCD-like regimes.} 
	\label{fig:SummaryBZ}
\end{figure}

We have studied the emergence and interplay between colour confining and chiral symmetry breaking dynamics in the landscape of gauge-fermion theories. We use the functional renormalisation group (fRG) for resolving the quantum fluctuations in a non-perturbative fashion. For the first time, we address the confinement and chiral symmetry breaking dynamics in the many flavour limit within a first principle continuum approach to gauge-fermion theories. Specifically, our results connect the QCD-like limit, well in agreement with data from lattice and quantitative fRG computations in QCD, to the conformal CBZ regime. At the root of this analysis is a new simplified and semi-analytical treatment of confinement in the functional RG approach with quantitative power. This has allowed us to study chiral and confining dynamics self-consistently and to depart towards the 
uncharted many-flavour--many-colour regime. With a modest truncation we investigate different gauge phases and the impact on fundamental parameters such as the fermion constituent mass, the chiral condensate and the relevant scales. The respective details can be found in \Cref{sec:econf,sec:InterplayLowNf} and \Cref{app:flowmGap,app:TuningConfinement+Stability}.

Below we emphasise the main physics results and refer for the detailed discussion of the results to \Cref{sec:phasesYM+chiral,sec:BoundaryCBZ}. 

In particular, we have studied the interplay of gauge and chiral dynamics in the plane of gauge-fermion theories for the first time. We explored the landscape of theories from the QCD to the conformal limit, see \Cref{fig:SummaryBZ}. We analysed the change in the scales relevant for confinement and $\chi{\rm SB}$ dynamics as well as the value of fundamental parameters such as the chiral condensate, the constituent fermion mass and the mass of the $\sigma$-mode. For this we have developed an improved a semi-perturbative approximation which combines the virtues from both perturbative and functional approaches. This framework naturally allows access to theories displaying non-trivial features such as walking. We provide for the first time estimates to a intricate quantities such as the size of the walking regime and study the close-conformal scaling, see also discussion in \Cref{sec:phasesYM+chiral} and \Cref{app:ManyFlavour}.

A further important result is the quantitative estimate of the lower boundary of the conformal window. The functional approach to emergent composites allows to precisely diagnose the presence of ${\rm d}\chi{\rm SB}$. Adapting $\overline{\rm MS}$ scheme results, we expect $N_f^{\rm crit}\simeq 9.6$ for $N_c=3$,  see  \Cref{eq:Nfcritbest}. In this context, we also assess the standard approach to approximately diagnose $\chi{\rm SB}$ via $\gamma_m$ and present evidence for potentially relevant non-perturbative effects within the conformal region. For details see  \Cref{sec:BoundaryCBZ} and \Cref{app:systematicsCBZboundary}. 

In summary, we have provided a top-bottom framework to study extensions of the SM with novel strong sectors. One potential application is the dynamical generation of the electroweak scale or of dark matter. Such inherently non-perturbative extensions can now be easily investigated and self-consistently tested on functional implementations of the complete SM \cite{Pastor-Gutierrez:2022nki,Goertz:2023pvn}. We hope to report on this and related subjects in the near future. 

\section*{Acknowledgements}
We thank Jens Braun, Wei-jie Fu, Fei Gao, Holger Gies, Chuang Huang, Joannis Papavassiliou, Franz R.~Sattler, Zi-ning Wang, Jonas P.~Wessely, Masatoshi Yamada and Shi Yin for discussions. This work is funded by the Deutsche Forschungsgemeinschaft (DFG, German Research Foundation) under Germany’s Excellence Strategy EXC 2181/1 - 390900948 (the Heidelberg STRUCTURES Excellence Cluster) and the Collaborative Research Centre SFB 1225 (ISOQUANT). It is also supported by EMMI.

\newpage
\appendix
\section*{Appendices}
%

\section{Details of the full effective action}\label{app:Effective action}

In this Appendix we provide further details on the truncation of the effective action discussed in \Cref{sec:econf,sec:emergentcomposites}. The pure gauge sector part of \labelcref{eq:G-glue-matter} reads
\begin{widetext} 
	\begin{align}\nonumber
		\Gamma_{\textrm{glue},k}[A,c,\bar c]  =&\,
		\frac12 \int_p \, A^a_\mu(p) \,  \left[  Z_{A,k}\,\left( p^2+m^2_{\textrm{gap},k} \right)\Pi^\perp_{\mu\nu}(p) 
		+\frac{ 1}{\xi}Z^\parallel_{A,k}\,\left( p^2 + m^2_{\textrm{\tiny{mSTI}},k}\right)  \frac{p_\mu p_\nu}{p^2} \right] \, A^a_\nu(-p)  \\[2ex]\nonumber
		&\hspace{-2cm}  + \frac{1}{3!} \int_{p_1,p_2} Z_{A,k}^{3/2} \lambda_{A^3,k}  \left[ {\cal T}_{A^3}^{(1)}(p_1,p_2)\right]^{a_1 a_2 a_3}_{\mu_1\mu_2\mu_3}
		\prod_{i=1}^3 A^{a_i}_{\mu_i}(p_i)  
		+  \frac{1}{4!}\int_{p_1,p_2,p_3} \hspace{-.3cm}Z_{A,k}^{2} \lambda_{A^4,k}   \left[ {\cal T}_{A^4}^{(1)}(p_1,p_2,p_3)\right]^{a_1 a_2 a_3 a_4}_{\mu_1\mu_2\mu_3\mu_4}
		\prod_{i=1}^4 A^{a_i}_{\mu_i}(p_i)    \\[2ex]
		&\hspace{-2cm}	+ 	\int_p   Z_{c,k} \, \bar c^{\,a}(p) p^2 \delta^{ab}c^b(-p) + \int_{p_1,p_2} Z_{c,k}  Z_{A,k}^{1/2}\lambda_{c \bar c A,k}  \left[ {\cal T}_{c\bar c A}^{(1)}(p_1,p_2)\right]^{a_1 a_2 a_3}_{\mu}
		\bar c^{a_2} (p_2) c^{a_1}(p_1) A^{a_3}_\mu(-p_1-p_2)   \,.
		\label{eq:Gglue}
	\end{align}
\end{widetext} 
Here we have pulled out the wave functions of the fields, rendering the respective vertex dressings RG-invariant. In contradistinction to \labelcref{eq:GglueA2}, we have kept the cut-off dependencies of dressings and couplings explicit.  ${\cal T}^{(i)}_{{\cal O}}$ stand for the tensor structures of the different operators containing the colour ($a_i$) and Lorentz ($\mu_i$) indices and momentum structure. The first line includes the transversal and longitudinal components of the gauge two-point function with the gauge-fixing part. The second line contains the gauge-three and four-point functions with the classical tensor structures. The last line contains the terms corresponding to the ghost part of the action. For further details we refer to \cite{Ihssen:2024miv} where the same truncation for the glue sector is employed.

The matter part of the full effective action includes the Dirac term in \labelcref{eq:DiracAction}, the four-fermion interactions in \labelcref{eq:eff4Fermi} and the high-order scalar-pseudoscalar fermionic interactions in the form of a mesonic action in  \labelcref{eq:GMeson}.  The four-Fermi action in \labelcref{eq:eff4Fermi} accounts for a Fierz complete flavour and colour basis \cite{Jaeckel:2003uz,Mitter:2014wpa, Ihssen:2024miv} with the vector and pseudovector tensor structure 
\begin{subequations}
\label{eq:4FermiTensors}
\begin{align} \nonumber 
	{\cal T}_{\left({\rm V-A}\right)}&= \left(\bar \psi \gamma_\mu T^0_f \psi\right)^2 +\left(\bar \psi \gamma_\mu \gamma_5  T^0_f \psi\right)^2 \,,\\[1ex] \nonumber 
	{\cal T}_{\left({\rm V+A}\right)}&= \left(\bar \psi \gamma_\mu  T^0_f \psi\right)^2 -\left(\bar \psi \gamma_\mu \gamma_5  T^0_f \psi\right)^2  \,,\\[1ex] 
	{\cal T}_{\left({\rm V-A}\right)^{\rm adj}}&= \left(\bar \psi \gamma_\mu  T^0_f T^a_f \psi \right)^2 +\left(\bar \psi \gamma_\mu \gamma_5 T^0_f T^a_f  \psi \right)^2 \,, 	
\end{align}
with
\begin{align}\label{eq:T0Tf}
	T^0_f&= \frac{\mathbbm{1}_{i,\,j}}{\sqrt{2 N_{f}}}\quad {\rm with }\quad \tr \left[(T^0_f)^2\right]=1/2\,,
\end{align}
\end{subequations} 
and $T^a_f$ being the generators of the SU(N$_f$) flavour group.	

In the presence of anomalous U$(1)_{\rm A}$-breaking, the dominant scalar pseudo-scalar (S-P) channel has to be written as a sum of tensor structures of the ($\sigma-\pi$) and ($\eta-a$) modes, 
\begin{subequations} 
	\label{eq:AxialSplitTSP} 
\begin{align}
	{\cal T}_{\left({\rm S-P}\right)}&= {\cal T}_{\left({\sigma-\pi}\right)}+{\cal T}_{\left({\eta-a}\right)}\,,
	\label{eq:SplitTspApp} 
\end{align}
with 
\begin{align}\nonumber 
	{\cal T}_{\left({\sigma-\pi}\right)}&= \left(\bar \psi \,T^0_f \psi\right)^2 -\left(\bar \psi \gamma_5 T^a_f \psi\right)^2 \\[1ex]
	{\cal T}_{\left({\eta-a}\right)}&=  \left(\bar \psi \,T^a_f \psi\right)^2 -\left(\bar \psi \gamma_5 T^0_f \psi\right)^2  \,.
	\label{eq:SplitTsigmaeta} 
\end{align}
\end{subequations}
%

\section{Regulators}
\label{app:regs}	

The regulator $R_k$ is a block diagonal matrix in field space with the matrix elements 
\begin{align}
	R^{AA} =R_A\,, \quad R^{c\bar c} =R_c\,, \quad R^{\psi\bar \psi} =R_\psi\,,
\end{align}
where we suppressed the momentum dependence as well as the Lorentz, Dirac, flavour and gauge group indices in the notation. All other entries are vanishing or follow by crossing symmetry, leading to 
\begin{align}
	R_k= \begin{pmatrix}
		R_A & 0 & 0& 0 & 0\\[1ex]
		0 & 0 & -R_c & 0 & 0\\[1ex]
		0 & R_c  &  0  & 0 & 0\\[1ex]
		0 & 0 & 0  & 0 & -R_\psi \\[1ex]
		0 & 0 & 0 & R_\psi & 0 
	\end{pmatrix}\,.
	\label{eq:QCD-RegulatorMatrix}
\end{align}
One of the goals of the present computation is to provide an analytic access to the computations. Therefore, we employ the flat or Litim regulators introduced in \cite{Litim:2001up} which allows for an analytic treatment of the flow equations, 
\begin{align} \nonumber 
	R_{A,c}(p^2) =& \,Z_{\varphi}\, p^2 \,  r_{A,c} (x)\,,\\[1ex] 
	R_{\psi}(p^2) =& \,\imag\,  Z_{\psi}\,  \gamma_\mu p_\mu \,  r_{\psi} (x)\,,
	\label{eq:Regs}
	\end{align}
with $x=p^2/k^2$ and 
\begin{align} \nonumber 
	r_{A,c} (x) =&  \,(1/x-1)\,\theta\left(1-x\right)\,,\\[1ex]
	r_{\psi} (x) =& \, (1/\sqrt{x}-1)\,\theta\left(1-x\right)\,.
	\label{eq:regxdeff}
\end{align}
for the gauge field and ghost $A,c$, and the fermion $\psi$ respectively.

\section{Flow equations}\label{app:flowequations}

In this Section we provide the derivation of the flow equations employed in this work. The respective flows have been derived from the flow of different two-, three- and four-point functions. All flows have been derived at the symmetric point configuration and vanishing momenta.

\subsection{Anomalous dimensions}
\label{app:AnomalousDimensions}

The anomalous dimensions of any component $\varphi=\Phi_i$ of the superfield \labelcref{eq:Superfield}, 
\begin{align}\label{eq:anomalousdimensiondefinition}
\eta_\varphi = -\frac{\partial_t Z_{\varphi,\,k}}{Z_{\varphi,\,k}}
\end{align}
have been derived from the respective two-point functions by applying the suitable projection,
\begin{align}\label{eq:anomalousdimensionderivation}
\partial_t Z_{\varphi,\,k}= \left. \partial_{p^2}\partial_t \Gamma^{(\varphi \varphi)}_k(p^2)\right|_{p=0}\,.
\end{align}
In this work we obtain a system of anomalous dimensions including $\eta_A,\,\eta_c,\, \eta_\psi,\, \eta_\pi$ and $\eta_\sigma$. Its resolution involves a resummation, and the resummed anomalous dimensions are used in the flows.

\subsection{Flow of the gauge avatars}
\label{app:flowgaugecouplings}

In \labelcref{eq:alphapsibarpsiA,eq:alphasGlue} we have defined the exchange couplings for each of the gauge avatars of the effective action. The flow for the fermion-gauge coupling reads
\begin{align}
	\partial_t \lambda_{A \bar \psi \psi} &= \left( \frac{1}{2} \eta_A + \eta_\psi \right)\lambda_{A \bar \psi \psi} \notag\\[2ex]
	&\hspace{.5cm}+ \frac{1}{Z^{1/2}_A Z_{\psi} }\left. \frac{\tr   \left[ {\cal P}^{(A \bar \psi \psi)}\,\partial_t \Gamma_k^{(A \bar \psi \psi)}\right]}{\tr \left[\left({\cal P}^{(A \bar \psi \psi)}\right)^2\right]}\right|_{p=0}\,,
	\label{eq:flowgApsipsibar}
\end{align}
Here, ${\cal P}^{(A \bar \psi \psi)} = \imag \gamma_{\mu} \delta_{a_1\,a_2}T_{i j}$ is the projection operator carrying the classical tensor structure and the trace sums over gauge group, flavour, Lorentz and Dirac indices. The flows of the other avatars has been obtained analogously to \labelcref{eq:flowgApsipsibar} from the projection with the classical tensor structures of the flows of the respective $n$-point functions. Their derivation as well as their explicit forms in the same truncation employed in this work can be found in \cite{Ihssen:2024miv,Fu:2019hdw,Braun:2014ata}. We close this discussion with the remark that the tensor bases are not orthogonal, and the projection procedure above also includes the dressings of other tensor structures.

\subsection{Flow of the gauge two-point function}
\label{app:flowmGap}

In covariant gauges confinement is carried by the mass gap in the gauge propagator as explained in \Cref{sec:confinement,sec:econf}. One of the novel achievements in this work is the inclusion of the confining mass gap in a rather simple and semi-analytical manner, enabled by an appropriate parametrisation and approximation of the gauge boson propagator, see \labelcref{eq:GAk2,eq:ZAk2} in \Cref{sec:econf}. As explained there, the approximation \labelcref{eq:GAk2} is well justified for the propagators in all loop diagrams which are evaluated at vanishing external momenta, $p_i=0$, or more generally for momenta $p_i^2\lesssim k^2$. For the following discussion we restrict ourselves to  $p_i=0$, the case also taken in the explicit computations in the present work. Then, all propagators only depend on the loop momentum $q$, which only takes momentum values $q^2\lesssim k^2$, due to the insertion $\partial_t R_k(q)$. This allows for a Taylor expansion about any expansion point in this regime. The best expansion point is at $q^2 = k^2$, as the integrands of all diagrams peak close to this momentum. However, such an expansion point requires a numerical treatment of the loop integral. In turn, for an expansion at $q = 0$ we obtain analytic flows. This facilitates the access to the physics dynamics of the large set of coupled flows equations considered here. Moreover, it leads to a significant speed-up of the computation and is the backbone of the present versatile framework to investigate a large number of theories. We remark that fully momentum-dependent computations will be considered elsewhere, based on \cite{Ihssen:2024miv}.

\subsubsection{Projection operators for $m^2_\textrm{gap}$ and $Z_A$}
\label{eq:ProjectmgapZA}

It is left to specify the projection procedure on the pair $Z_{A},m_{\textrm{gap}}$ which optimises the accuracy and provides maximal stability of the scheme. As discussed in \Cref{sec:econf}, the momentum dependent propagator $G_A(p)$ with   
\begin{align} 
G_A(p) = \frac{1}{ Z_{A} } \frac{1}{ p^2 \left[1+  r_A(p^2/k^2) \right]+ m^2_{\textrm{gap}} }\,,
\end{align}
is dominated by its momentum-independent part for $p^2\lesssim k^2$, 
\begin{align} 
	G_A(p)-G_A(0)  = G_A(p) 
	 \frac{1 - x\left[1+ r_A(x)\right]}{1+\bar m^2_{\textrm{gap}} }\,,
	 \label{eq:GAdiff}
\end{align}
with $x=p^2/k^2$ and the dimensionless mass gap $\bar m^2_\textrm{gap}= m^2_{\textrm{gap}}/k^2$, see  \labelcref{eq:DimlessMassGap}.  Note that the structure of  \labelcref{eq:GAdiff} is also present when using the full momentum-dependent propagator. The difference in  \labelcref{eq:GAdiff} is proportional to the propagator itself and the second factor is a ratio of a universal numerator that does not depend on the parameters $Z_{A}, \bar m^2_{\textrm{gap}} $ and the full mass gap at vanishing momentum, $1+ \bar m^2_{\textrm{gap}}$. Due to the denominator, this term  rapidly decays for all momenta in the confining regime $k\lesssim k_\textrm{conf}$ with $\bar m^2_\textrm{gap} \to \infty$. In turn, for $k\gg k_\textrm{conf}$, the denominator tends towards unity and only the universal numerator is left. It is regulator-dependent, which allows us to minimise the difference \labelcref{eq:GAdiff}, hence optimising the approximation of the propagator used in the loops. For the flat regulator \labelcref{eq:regxdeff} used in the present work we get 
\begin{align}
\left[	1 - x + x\,r_A(x)\right]_{x\leq 1}  \equiv 0\,. 
\label{eq:0Diff}
\end{align}
In conclusion, the flat regulator is tailor-made for the current approximation as it minimises the error of the approximation. Moreover, if using the flow at the gauge two-point function at vanishing momentum for  deriving $Z_{A,k} m^2_{\textrm{gap},k}$, \labelcref{eq:0Diff} also holds true for the difference between the full momentum-dependent propagator and the approximated one at $p=0$. Consequently we are led to 
\begin{align}
	\partial_t \left( Z_{A}\, m^2_{{\rm gap}} \right) = 
	\left. \frac{\tr\left[ {\cal P}^{(A A)} \partial_t\Gamma^{(AA)}_k\right]}{\tr\left[{\left({\cal P}^{(A A)}\right)}^2\right]  }\right|_{p=0}\,, 
	\label{eq:flowmGap}
\end{align}
and the final analytic equation reads 
\begin{align}
		\partial_t \left( Z_{A} m^2_{{\rm gap},k}\right) = Z_A k^2\, \overline{\textrm{Flow}}_{AA}(0)\,,
	\label{eq:flowmAQuatraticDiv}
\end{align}
where we have separated a dimensionful part $Z_A k^2$  of the right-hand side of 	\labelcref{eq:flowmGap}, where $Z_A$ carries the RG-running of the gauge two-point function. The remaining part $\overline{\textrm{Flow}}_{AA}(0)$ is RG-invariant, and is evaluated at vanishing momentum. After performing the traces and loop integrations we are led to 
\begin{align}
\overline{\textrm{Flow}}_{AA}(0)	&=\,\frac{1}{4\,(4\pi)^2}\Bigg[-\frac{8 N_c \, \lambda_{A\bar c c }^2} {6}\left(1-\frac{\eta_c}{8} \right)\notag\\[1ex] 
	&\hspace{-1.6cm} +\frac{8 N_c \, \lambda_{A^3 }^2} {\left(1+\bar m_{\rm gap}^2\right)^3}\left(1-\frac{\eta_A}{8} \right)-\frac{18 N_c \, \lambda_{A^4 }^2} {2 \left(1+\bar m_{\rm gap}^2\right)^2}\left(1-\frac{\eta_A}{6} \right) \notag\\[1ex] 
	&\hspace{-1.6cm}  +\frac{4 N_f \, \lambda_{A \bar \psi \psi}^2\left(1+ 3 \bar m_\psi^2\right)} { \left(1+\bar m_\psi^2\right)^3}\left(1-\frac{\eta_\psi}{5} \right) \Bigg]\,.
	\label{eq:flowmA}
\end{align}
It is left to fix the projection on the flow of the wave function $Z_{A}$. Here we shall use its sub-leading nature to our advantage: for large cut-off scales with $k\gg k_\textrm{conf}$, the approximation leaves us with the choice to store the perturbative marginal running of the wave function either in $Z_{A}$ or solely in $m^2_{{\rm gap}}$, see \labelcref{eq:ZAk2}. In view of momentum dependences it is more natural to store it in $ Z_{A}$. Moreover, this choice leads to better results if reading out full momentum dependences from the flows as done in \cite{Ihssen:2024miv}. This leads us to 
\begin{align}
 \eta_A = -\frac{\partial_t Z_{A,k}}{Z_{A,k}}= -\left.\frac{\partial_{p^2}\,\tr\left[ {\cal P}^{(A A)} \partial_t\Gamma^{(AA)}_k\right]}{Z_{A,k}\tr\left[{\left({\cal P}^{(A A)}\right)}^2\right] }\right|_{p=0}\,,
\label{eq:etaAkUV}
\end{align}
for $k\gtrsim k_\textrm{conf}$. In turn, for $k\lesssim k_\textrm{conf}$, the propagator is dominated by the mass gap with $\bar m_{\textrm{gap}} \to \infty$ for $k\to 0$. In this regime the choice of $Z_{A}$ is dictated by computational convenience and its occurrence in the regulator: For example, a choice of $Z_{A}$ with $\eta_A > 2$ deforms the low momentum regime of the regularised theory with $p^2\lesssim k^2$ and $k\to 0$. In this regime the regulator diverges at $p=0$ with $k^{2-\eta_A}\to\infty$. We note in passing that $\eta_A < -2$ is also a problem as it indicates a dispersion of the gauge field with $1/p^{2-\eta_A}$ which is not a tempered distribution as it diverges with more than $1/p^4$. 

While $\eta_A>2$ is not problematic for approximations with full momentum dependences, it destabilises Taylor expansions about vanishing or small momenta. For this reason we choose a constant $Z_{A}$ with 
\begin{align}
	\eta_A = 0 && {\rm for }&&& k\lesssim k_\textrm{conf}\,.
		\label{eq:etaAkIR}
\end{align}
Choosing different freezing scales does not have an impact on the correlation functions, as all information on $Z_A$ at these scales is carried by the cut-off dependence of the mass gap. This closes the discussion of the projection procedures for $m^2_{\textrm{gap}}$ and $ Z_{A}$.

\subsubsection{Confining $m_\textrm{gap}^2$}
\label{eq:FineTuningUV}

The Kugo-Ojima criterion for a confining solution is satisfied for a unique choice of the initial mass gap parameter $m^2_{\textrm{gap},\Lambda_\textrm{UV}}$ or rather its dimensionless version $\bar m^2_{\textrm{gap},\Lambda_\textrm{UV}}$ at the initial cut-off scale $\Lambda_\textrm{UV}$, see \Cref{app:TuningConfinement+Stability} for more details.  Here we discuss the more technical part of simplifying the related tasks. An obvious one concerns the fine-tuning problem, for more details see \cite{Cyrol:2016tym}, and a further one concerns the gauge consistency over several orders of magnitudes, for respective discussions see \cite{Pawlowski:2022oyq}. We start this analysis with the discussion of the flow of  $\bar m^2_{\textrm{gap}}$, which is readily deduced from \labelcref{eq:flowmAQuatraticDiv}, 
\begin{align}
	\partial_t \bar m^2_{\textrm{gap}} = \left( -2 +\eta_A\right) \bar m^2_{\textrm{gap}}+ \overline{\textrm{Flow}}_{AA}(0)\,,
			\label{eq:flowbarm}
\end{align}
with \labelcref{eq:flowmA}. We conclude that the mass gap has a quadratic running with the cut-off scale for $\overline{\textrm{Flow}}_{AA}(0)\neq 0$, leading to a quadratic fine-tuning problem: the Kugo-Ojima criterion for the confining solution is achieved for the unique initial condition $m^2_{\textrm{gap},\Lambda_\textrm{UV}}=m^2_{\textrm{scaling},\Lambda_\textrm{UV}}$, which is proportional to $\Lambda_\textrm{UV}^2$. 

This difficult task can be marginalised with the following observations: as discussed in \Cref{sec:UVSimplification}, the transverse mass agrees with the longitudinal mass for cut-off scales larger than the scale $k_\textrm{conf}$, at which the confining dynamics kicks in, see \labelcref{eq:mgap=mSTI}. In the present work we have approximated this scale by the peak scale of the gauge field propagator, thus providing a lower bound. Moreover, we may choose a combination or regulators $R_k$ in \labelcref{eq:QCD-RegulatorMatrix} such that the longitudinal mass vanishes for all cut-off scales. 
This is an STI-optimised choice and the existence of such regulator combinations can readily be inferred from \labelcref{eq:flowmA}: the different terms in \labelcref{eq:flowmAQuatraticDiv} come with different signs. The  three-gauge vertex and fermion-gauge terms come with positive signs while the ghost-gauge and in particular the four-gauge term come with a negative one. For the regulators used here, see \Cref{app:regs}, the latter term dominates the pure gauge part of the flow, which comes with a negative sign. This is seen in the quantitative YM computation in \cite{Cyrol:2016tym}, where the regulator shape functions $r_{A,c}$ and the cut-off scales of gluons and ghosts can be used to achieve the exact cancellation of $m^2_{{\rm mSTI}, k}$. 

In gauge-fermion systems, the positive terms will start dominating for sufficiently many flavours, in the present approximation and combination of regulators this happens for $N_f> N_c$. However, due to the different signs of the diagrams we can use combinations $R_k$ in \labelcref{eq:QCD-RegulatorMatrix} of regulators and cut-off scales such, that $\overline{\textrm{Flow}}_{AA}(0)$ vanishes in the perturbative regime. This leaves us with 
\begin{align}
\overline{\textrm{Flow}}_{AA}(0)(R_k)\stackrel{!} {=}0\,, \qquad \textrm{for}\qquad k\gtrsim k_\textrm{conf} \,,  
\label{eq:Flowmgap0}
\end{align}
and the flow $\partial_t (Z_{A} m^2_{{\rm gap},k})=0$ vanishes in this regime and we arrive at 
\begin{align}
	m_{\textrm{gap},k_\textrm{conf}} = m_{\textrm{gap},\Lambda_\textrm{UV}}= m_{\textrm{scaling},\Lambda_\textrm{UV}}\,,
\label{eq:mgap=mscal}
\end{align}
as depicted in \Cref{fig:MassDifferenceManyFlavour}. With \labelcref{eq:mgap=mscal} the fine-tuning problem is resolved as the quadratic running in the UV-regime over many orders of magnitude is absent. For smaller cut-off scales, \labelcref{eq:mgap=mSTI} fails and \labelcref{eq:flowmA} is genuinely non-zero for all regulator choices. 

\begin{figure}
	\centering
	\includegraphics[width=\columnwidth]{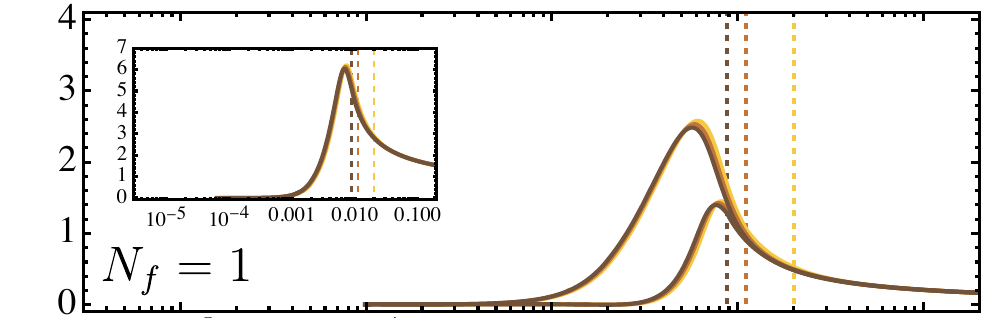}
	\includegraphics[width=\columnwidth]{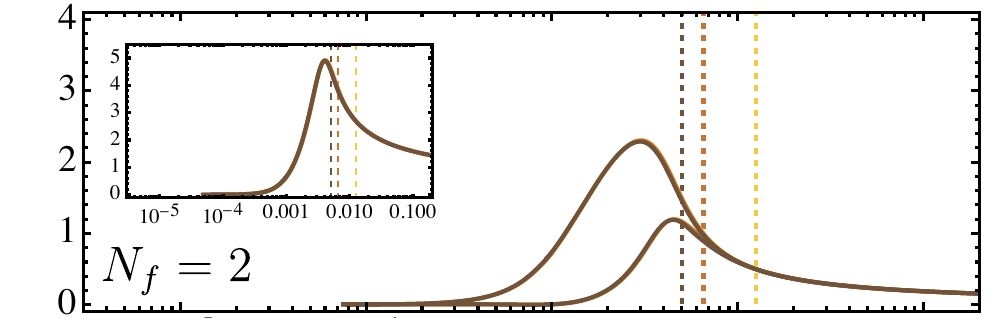}
	\includegraphics[width=\columnwidth]{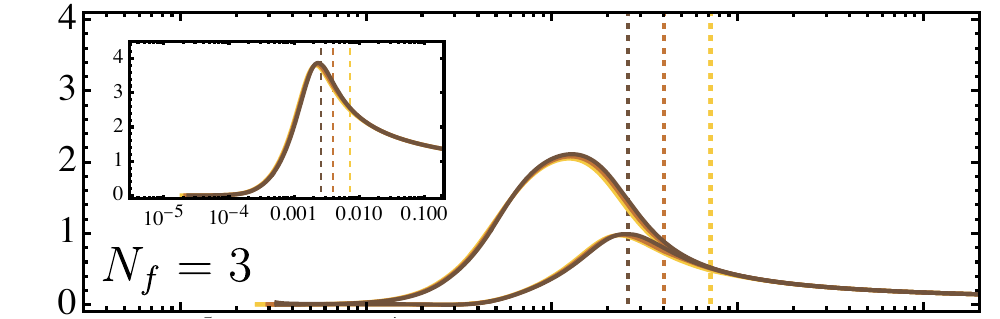}
	\includegraphics[width=\columnwidth]{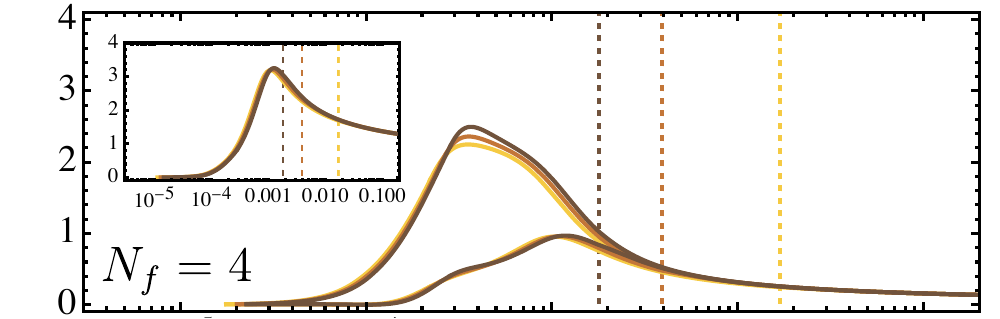}
	\includegraphics[width=\columnwidth]{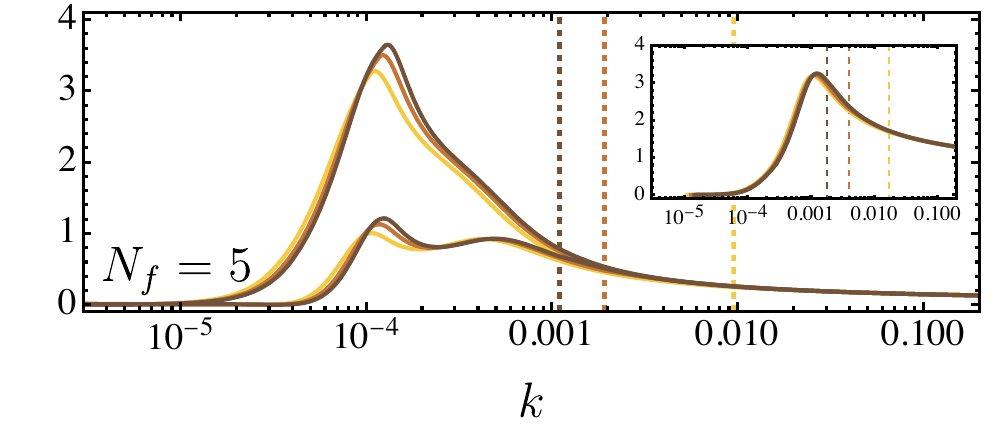}
	\caption{ Gauge-fermion (highest peaking curves) and three-gauge (lowest peaking curves) exchange couplings as defined in \labelcref{eq:alphasGlue,eq:alphapsibarpsiA} in an SU(3) gauge theory with $N_f=1$ (top panel) to 5 (bottom panel) fundamental flavours with different onset scales (vertical dashed lines) of the power-law running in the flow of the gluon mass gap. All flows are started at the same UV perturbative scale with same initial conditions.  In the inlay plots, we display the gauge field wave function renormalisation and the onset scales (vertical dashed lines) defined in \labelcref{eq:RegimeSwitch}.}
	\label{fig:Nf2 alphag diff QD onset}
\end{figure}

While the adjustment of \labelcref{eq:Flowmgap0} with an appropriate choice of $R_k$ or different cut-off scales of $A,c,\psi$ poses no technical challenges, we resort to a more hands-on procedure: for $k\gtrsim k_\textrm{conf}$ the mass gap may be put to zero in the flows for any choice of $R_k$ as its impact is subleading. In turn, for $k\lesssim k_\textrm{conf}$ it has to be taken into account as $\bar m^2_{\textrm{gap},k}\to \infty$ for $k\to 0$. This leads us to 
\begin{subequations}
	\label{eq:EffmGap}
\begin{align}
	m^2_{\textrm{gap},k} = \theta(\gamma_\textrm{conf}\, k_\textrm{conf} - k) \,m^2_\textrm{scaling}\,, 
\end{align}
with 
\begin{align}	
		 1 \leq  \gamma_\textrm{conf}  \lesssim 2\,, 
	\label{eq:RegimeSwitch}
\end{align} 	 
\end{subequations}
where $m^2_{\textrm{scaling}}(\gamma_\textrm{conf}\, k_\textrm{conf} )$ is the $\gamma_\textrm{conf}$-dependent unique choice of the mass gap at $\gamma_\textrm{conf} \,k_\textrm{conf}$, leading to scaling in the IR. In 	\labelcref{eq:kconf}, $k_\textrm{conf}$ has been defined by the peak position rather than the cut-off scale, where the confinement dynamics does kicks in. Clearly $\gamma_\textrm{conf}$ is larger than unity, but the confinement dynamics starts close to this scale. We consider a maximal value of $2 k_\textrm{conf}$ as a conservative estimate for the upper value and this estimate is informed by the analysis in \cite{Cyrol:2016tym}. 

Moreover, as $N_f$ is increased for a fixed $N_c$, the fermionic contributions in \labelcref{eq:flowmA} lead to a significant linear growth of  $m_{\textrm{\tiny mSTI}}$ linearly with $N_f$ in the perturbative regime. However, as argued in \Cref{sec:UVSimplification}, $m_{\textrm{\tiny mSTI}}$ vanishes identically  at $k=0$ and its removal within the flow requires full an approximation with momentum dependences of the correlation functions of the gauge field. Consequently, the importance of these momentum dependences grows with the successive increase of  $m_{\textrm{\tiny mSTI}}$. Furthermore, the Schwinger mechanism with its dynamical generation of the gauge field mass gap only leads to a mild linear growth of $m_\textrm{gap}$, approximately an order of magnitude smaller than the growth of $m_{\textrm{\tiny mSTI}}$. Consequently, within the present simplified approximation, we account for these subtle dependences by setting $N_f = N_c$ for $N_f > N_c$ in the fermionic contribution to the quadratic part of the flow in \labelcref{eq:flowmA}. We note that the full flavour dependence is still encoded in the dressings and anomalous dimensions.

We emphasise that the procedure detailed above is self-adjusting for the approximation artefacts:    $m^2_{\textrm{gap},k}=0$ for $k\geq \gamma_\textrm{conf}\, k_\textrm{conf}$ and $ m^2_\textrm{scaling}(\gamma_\textrm{conf}\, k_\textrm{conf})$ 
is chosen such that scaling is obtained. 
Moreover, we check this insensitivity of this approximation by varying $\gamma_\textrm{conf}$ and the results of this check are depicted in \Cref{fig:Nf2 alphag diff QD onset}. 
There we show the gauge-fermion (highest peaking curves) and three-gauge (lower peaking curves) exchange couplings for a SU(3) gauge theory with $N_f=1$ to 5 flavours. The inlays contain the respective dressing functions of the gauge field. 
All flows are initiated with the same boundary conditions at a $\Lambda_{\rm UV}$ scale and only the onset scale (vertical dashed line) of the power-law term in the flow of the mass is varied. An early onset (yellow lines) triggers a separation between the avatars and forces a deviation from the mSTI at higher scales. This can be noted for example in the $N_f=5$ case.  
For low number of flavours $N_f\lesssim3$, we observe that different onsets do not lead to significant effects in the couplings.

\subsection{Flow of the four-Fermi couplings}
\label{app:flows4Fermi}

Part of the high-order correlations of the full effective action are carried by the four-Fermi interactions which are accounted for in the Fierz complete basis in \labelcref{eq:eff4Fermi}. The flow of the respective dimensionless RG-invariant couplings $ \bar \lambda_j= \lambda_j\, k^{2}$ can be obtained by adequately projections of the flow of the fermionic four-point function 
\begin{align}\label{eq:flow4F}
\partial_t \bar \lambda_j = (2+\eta_\psi) \bar \lambda_j + \tr\left[ {\cal P }^{( \bar \psi \psi \bar \psi \psi )}_j \partial_t \Gamma_k^{( \bar \psi \psi \bar \psi \psi )}\right]\,,
\end{align}
where  $j= \,_\textrm{\tiny{SP}}^{\left({\sigma-\pi}\right)},\,\,_\textrm{\tiny{SP}}^{\left({\eta-a}\right)},\,\pm\,\, {\rm and}\, {\rm VA}$. In particular, a basis of orthonormal projectors can be constructed such that 
\begin{align}
{\cal P }^{( \bar \psi \psi \bar \psi \psi )}_j \,\Gamma_k^{( \bar \psi \psi \bar \psi  \psi)}= - Z^2_\psi \lambda_j  \,, 
\end{align}
This set of projection operators can be systematically constructed and we refer the interested reader to e.g.~ \cite{TensorBases,Ihssen:2024miv,Gehring:2015vja} for the procedure. Furthermore, we note as an important benchmark check that the results for these analytic flows are in exact agreement with those in~\cite{Gies:2005as,Fu:2024ysj,Fu:2022uow}, where similar sets of operators has been considered.

\subsection{Flows of the bosonised sector}
\label{app:flowbosonisedsector}

In this Appendix, we discuss the derivation of the flow equations for the Yukawa coupling $h_\phi$ between fermions and mesons, as well as the chiral effective potential. Combined with the meson kinetic term, these flows generate higher-order fermionic self-interactions in the scalar-pseudoscalar channel; see \Cref{sec:emergentcomposites}.

If using the emerging composite formulation for the ($\sigma-\pi$) channel, higher-order fermionic interactions in this channel are represented as bosonic self-interactions. They are accounted for in the mesonic effective potential \labelcref{eq:Veff}.  We have considered a polynomial expansion of the potential about the flowing minimum similar to that in \cite{Fu:2019hdw}, see \labelcref{eq:Veff}. We have studied its convergence and have taken into account a sufficiently high order to guarantee it: $N_\textrm{max}=5$. 

The effective potential is simply the effective action, evaluated for constant fields and divided by the space-time volume, 
\begin{align} 
	V(\rho) =\frac{1}{{\cal V}_4} \Gamma[0,\phi_c]\,,
\end{align}
where $\phi_c$ indicates the constant mesonic field, and $0$ indicates that the other fields are set to zero. Hence its flow is that of the effective action, divided by ${\cal V}_4$. Below we provide the flow equation of the dimensionless renormalised version \labelcref{eq:u}, 
\begin{align}\nonumber 
	\overline{\text{Flow}}_V =& \, \frac{\partial_t|_{\rho}	V(\rho)}{k^4}=\left.\partial_{t}\right|_{\rho}u(\bar\rho)+ 4 \, u (\bar\rho)	\\[2ex]
	=& \, \frac{1}{(4\pi)^2} \bigg[  \frac{3\,  N_c (1- \eta_A/6 )}{2 (1+ \bar m_{\rm gap}^2)}- N_c (1- \eta_c/6 )\notag\\[1ex]
	&+\frac{(N_f^2-1)(1-\eta_\pi/6)}{2(1+\bar m_\pi^2)}+\frac{(1-\eta_\sigma/6)}{2(1+\bar m_\sigma^2)}\notag\\[1ex]
	&- \frac{2\, N_f N_c (1- \eta_\psi/5 )}{(1+ \bar m_\psi^2)}\bigg]\,,  
	\label{eq:flowudiagramatic}
\end{align}
where no functional field derivatives have been performed. \Cref{eq:flowudiagramatic}  grants access to the flow of the dimensionless and renormalised expansion coefficients, $\bar \lambda_{\phi,n} =\lambda_{\phi,n} k^{2n-4}$, by performing $\bar \rho$ derivatives
\begin{align}\nonumber 
	\partial_t  \lambda_{\phi,n} =& \Big\{ \partial^n_{\bar \rho}\Bigl[ \overline{\text{Flow}}_V - 4 u (\bar \rho) \Bigr]\\[1ex] 
	&\hspace{-1cm}+ (2+ \eta_\phi)\partial^n_{\bar \rho}\left[ \bar \rho \,\partial_{\bar \rho} u (\bar \rho) \right]\Big\}_{\bar \rho=\bar{\rho}_0} +\partial_t \bar\rho_0 \,  \lambda_{\phi,n+1}\,,
	\label{eq:flowlambdas}
\end{align}
and to the flow of  renormalised and dimensionless minimum of the potential,
\begin{align}\nonumber 
	\partial_{t} \bar \rho_0 =& -\frac{1 }{\partial^2_{\bar \rho} u (\bar \rho)} \Biggl\{\partial_{\bar \rho} \Bigl[ \,\overline{\text{Flow}}_V - 4 u (\bar \rho) \Bigr]\\[1ex]
	&\hspace{1.5cm}+ (2+ \eta_\phi)\partial_{\bar \rho}\Bigl[ \bar \rho \,\partial_{\bar \rho} u (\bar \rho) \Bigr] \Biggr\} _{\bar \rho =\bar{\rho}_0}\,,
	\label{eq:flowrho0}
\end{align}
from which to the order parameter \labelcref{eq:chiOrder} is obtained. For the anomalous dimensions in \labelcref{eq:flowlambdas,eq:flowrho0} we have employed the respective to the pion modes, $\eta_\pi$ as $\eta_\phi(\rho) = \eta_\pi(\rho)$, while $Z_\sigma$ also includes $2 \rho \partial_\rho Z_\phi$. 

Next, we discuss the flow of the Yukawa coupling. We consider a single avatar of this coupling, which is an immediate consequence of dropping the $\rho$-derivatives of $h_\phi$ in the minimum, see \labelcref{eq:Derh=0,eq:hphihsigmahpi} and the discussion there. As for the anomalous dimension $\eta_\phi$ discussed above, we compute the uniform $h_\sigma=h_\phi=h_\pi$ from the flow of the fermion two-point function projected on $T^0_f$ as in \cite{Braun:2014ata,Fu:2019hdw,Ihssen:2024miv}, which provides $h_\phi(\rho)$. 

The respective flows from the the fermion-meson three-point function give $h_\phi$ for the pions and $h_\phi+ 2\rho h'_\phi$ for the $\sigma$. We note in this context that the projection on the flow of the pion-fermion vertex in \labelcref{eq:flowh} not only generates the tensor structure projected but also an additional one proportional to the fully symmetric tensor flavour tensor $d^{abc}$,
\begin{align}
	\Gamma_k^{( \pi \bar \psi  \psi )}	\propto    \gamma_5  \left(  T^a_f +d^{abc} T_f^b T_f^c \right)\,,
\end{align}
which turns relevant in the large $N_f$ limit. For $N_f=2$ the fully symmetric tensor is exactly zero and hence such contributions exactly vanishes. This renders the flow of the pion-fermion and $\sigma$-fermion the same in the symmetric regime. 

We proceed with the flow equation for $h_\phi$ derived from the projection onto the scalar part of the quark two-point function,  
\begin{align}\nonumber 
	\partial_t  h_\phi =&\, \left(\frac{1}{2}\,\eta_\pi +\eta_\psi \right)  h_\phi + \bar m^2_\pi \,\dot{\bar{A}}_\phi\\[1ex] 
	  &\hspace{2.cm}+\left.\frac{ \tr \left[ T^0_f \,\partial_t\Gamma_k^{( \bar \psi \psi)}\right]}{\langle \sigma\rangle Z_{\psi}\, \tr \left[(T^0_f)^2\right]}\right|_{p=0}\,, 
	\label{eq:flowh}
\end{align}
with 
\begin{align}
\dot{\bar{A}}_\phi=- \frac{1}{ h_\phi} \left.\tr \left[{\cal P^{(\bar \psi \psi \bar \psi \psi)}_{(\sigma-\pi)} } \partial_t\, \Gamma_{k}^{(\bar \psi \psi \bar \psi \psi)} \right]\right|_{p=0}\,. 
\label{eq:flowAbar}
\end{align}
\Cref{eq:flowAbar} carries the flow of the four-Fermi coupling $\partial_t \bar{\lambda}^{(\sigma-\pi)}_{\textrm{\tiny{SP}}}$ from \labelcref{eq:flow4F} that is stored in the meson exchange in the fRG approach with emergent composites. The choice \labelcref{eq:flowAbar} eliminates the respective flow of the scalar-pseudoscalar dressing, $\partial_t \lambda^{(\sigma-\pi)}_{\textrm{\tiny{SP}}} = 0$. Since the coupling is set to zero at the initial cutoff scale, see \labelcref{eq:Initialcondition2}, this leads to $ \lambda^{(\sigma-\pi)}_{\textrm{\tiny{SP}}} \equiv 0$ for all cutoff scales. We refer to  \cite{Braun:2011pp,Gehring:2015vja,Braun:2018bik,Braun:2019aow,Fukushima:2021ctq,Ihssen:2024miv} for the full derivation, complete analytic expressions, and further details.

\begin{figure}
	\centering
	\includegraphics[width=\columnwidth]{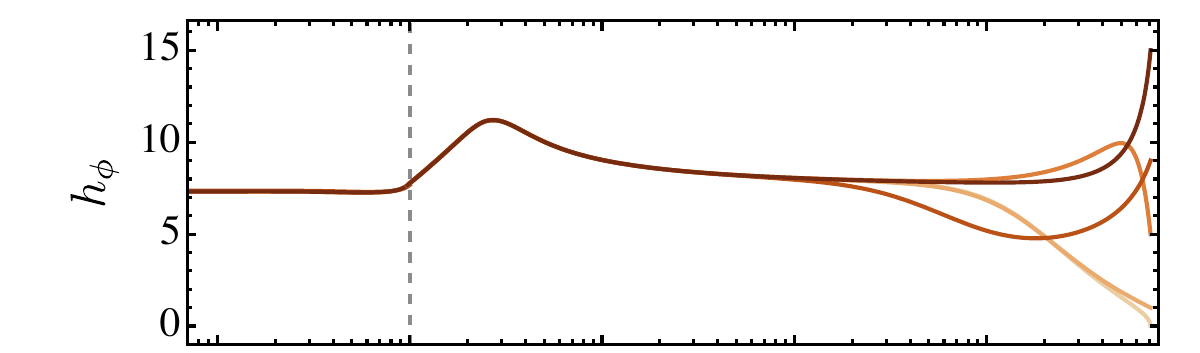}
	\includegraphics[width=\columnwidth]{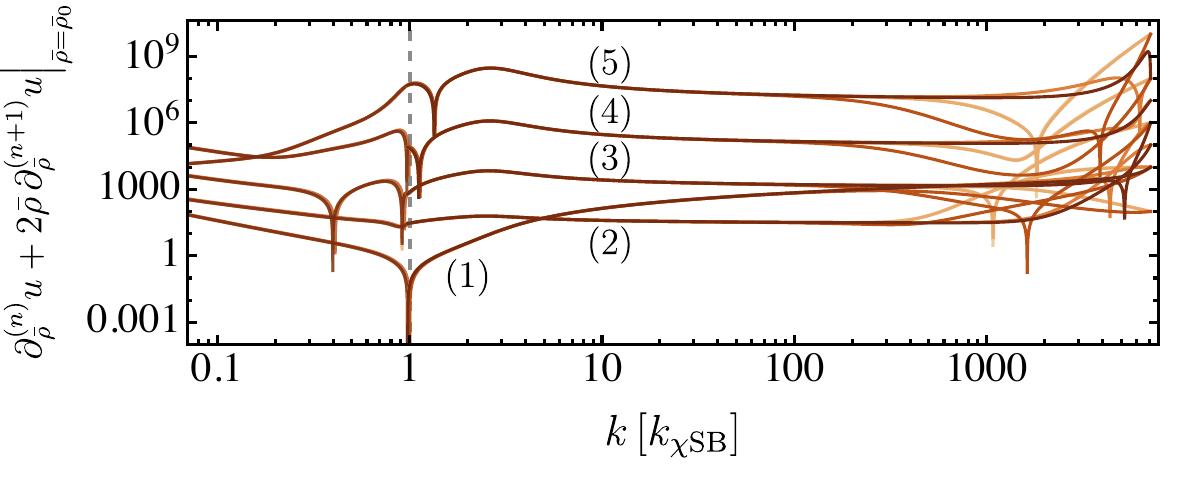}
	\caption{In the top panel, the Yukawa coupling between fermions and resonances and in the bottom panel, different orders of the effective potential for very different UV boundary conditions are displayed. All trajectories converge to the same IR values. See text for details.}
	\label{fig: independence UV}
\end{figure}

In \Cref{fig: independence UV} we show the integrated flow of the Yukawa coupling $ h_\phi$ and the expansion coefficients of $u(\bar \rho)$ for various random boundary initial conditions at the $\Lambda_{\rm UV}$ scale. Independent of their UV values, all trajectories are attracted towards the same one defined by the gauge-fermion interactions. This entails that the chiral sector is purely generated by the gauge dynamics and does not host further relevant parameters.

\section{Comparison to 2 and 2+1 flavour (chiral) QCD}
\label{app:comparisonQCD}

In this Appendix we compare the results in the chiral limit for few flavours discussed in \Cref{sec:InterplayLowNf,sec:phasesYM+chiral} to existing ones in the QCD literature at physical quark masses, obtained with high precision functional computations \cite{Ihssen:2024miv,Mitter:2014wpa,Fu:2019hdw,Braun:2014ata} and Monte Carlo lattice simulations \cite{Zafeiropoulos:2019flq,Sternbeck:2006rd}. The purpose of this Section is to evaluate the reliability of the present truncation and estimate the error.

This error estimate is based the LEGO$^{\copyright}$ principle introduced in \cite{Ihssen:2024miv}. There, it has been argued and demonstrated that the systematics of the full computation can be separated into the analysis of different sectors. In the functional approach to QCD, we identify two detachable sectors: the pure gauge and the matter.

To assess the quantitative reliability of the glue sector, in \Cref{fig:ZAcomparisonQCD} we have compared the gauge field propagator derived in the best approximation of this work with results from quantitative fRG studies \cite{Mitter:2014wpa, Cyrol:2017ewj, Ihssen:2024miv} and lattice data for $N_f=2$ and $N_f=3$ flavours \cite{Sternbeck:2006rd,Zafeiropoulos:2019flq}. While these benchmark results have been obtained at physical pion masses, it has been shown in \cite{Cyrol:2017ewj} that the pion mass-dependence of the gluon propagator is negligible. 

As discussed in \Cref{sec:confinement}, this quantity encodes the confining dynamics and we find in both theories with $N_f=2,3$ an excellent agreement of the gluon propagator for momentum scales $p\gtrsim p_\textrm{peak}$. For smaller momenta the agreement is still excellent for $N_f=3$, while it is semi-quantitative for $N_f=2$. Note however that the gluon decouples in this momentum regime due to  the mass gap and this small error does not feed in the matter sector. Note also, that the lattice data \cite{Sternbeck:2006rd} flattens at $p\gtrsim 5\, {\rm GeV}$ due to the lack of a conversion to continuum momenta. In summary, this comparison shows the impressive semi-quantitative nature of the present simplified and versatile approach to confinement put forward in \Cref{sec:econf}.  
\begin{table}[t!]
	\centering
	\begin{tabular}{ l cc} 	\toprule[.2ex] 
		{$m_\psi \left[\kSSB\right]$ \qquad\qquad\qquad\qquad\qquad\qquad}&  {\quad$N_f=2$\quad }  &  {\quad$N_f=3$\quad } \\
		\midrule \midrule 
		{QCD \cite{Ihssen:2024miv}} & 0.80 & 0.74\textcolor{black}{\bf *} \\[1ex]
		{\rm Multi-avatar: Enhanced} & 0.74 & 0.53\\ [1ex]
		{\rm Multi-avatar: Not-enhanced} & 0.63 & 0.50\\ [1ex]		
		{\rm Single-avatar } & 0.64& 0.52 \\[1ex]
		\hline
	\end{tabular}
	
	\caption{$m_\psi$ obtained in the different approximations put forward in \Cref{sec:truncation} (multi-avatar (not-)enhanced) and in \Cref{app:largeNfapproximation} (single-avatar) as well as with existing results from high precision functional computations in the chiral limit (fQCD) \cite{Ihssen:2024miv}. The \textcolor{black}{\bf *} highlights that the fQCD $N_f=2+1$ results are obtained with two chiral flavours and one heavy, justifying the large $m_\psi$ value.}
	\label{table:comparison mpsi}
\end{table}
\begin{figure*}[th]
	\centering
	\includegraphics[width=\columnwidth]{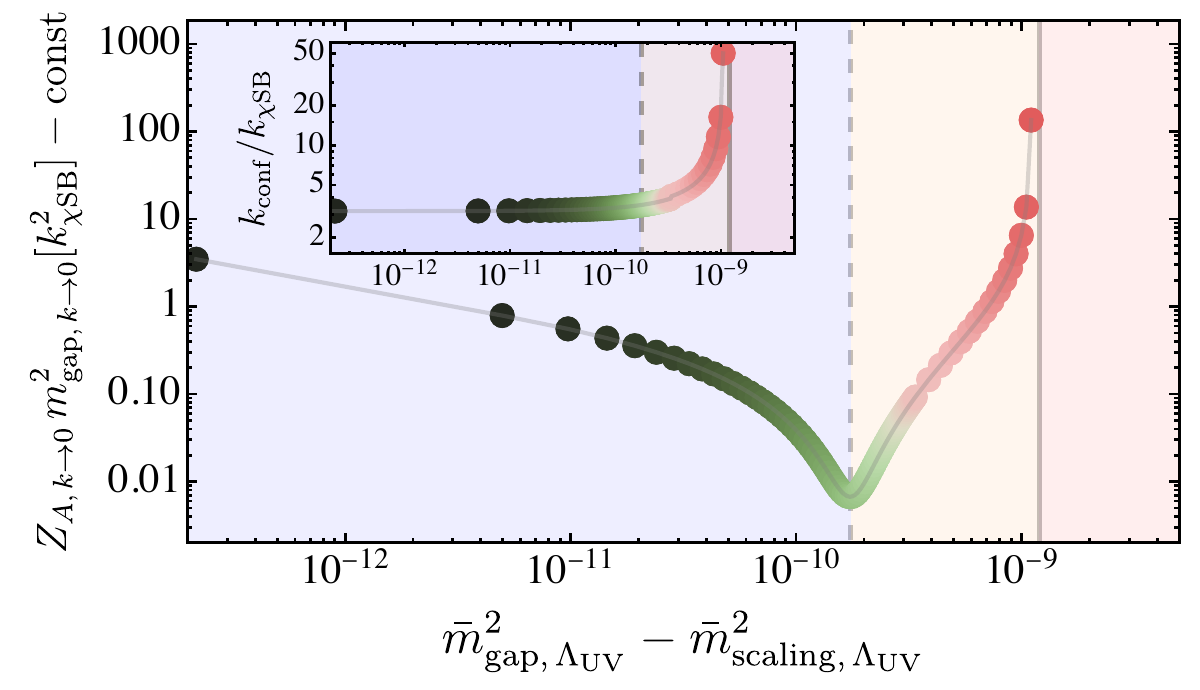}\hspace{.3cm}
	\includegraphics[width=\columnwidth]{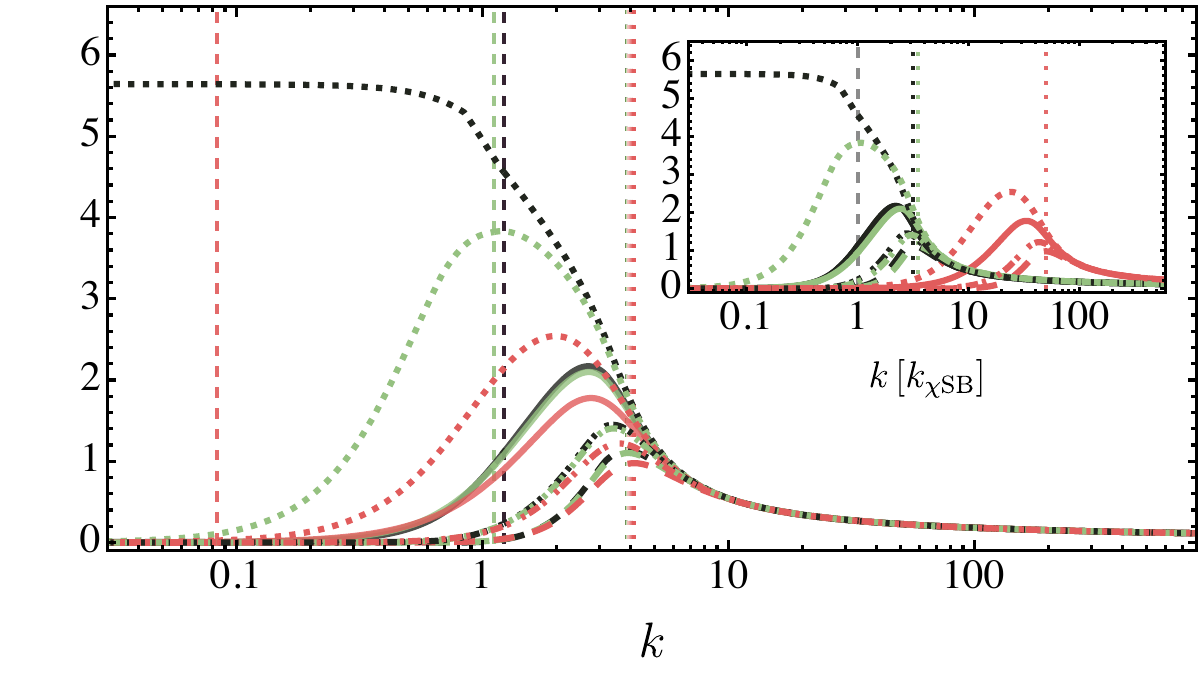}
	\caption{In the left panel, the IR-value of the gauge field two-point function \labelcref{eq:GA0INverse} at vanishing momentum is shown for different UV values of the dimensionless mass gap $\bar m^2_{{\rm gap},\,\Lambda_{\rm UV}}$ in a SU(3) gauge theory with two massless flavours in the fundamental representation. We have subtracted a constant for better visibility of the results and in the inlay we show the respective $k_{\rm conf}/\kSSB$ ratios. The vertical lines denote the {\it maximal decoupling solution} (dashed) and the {\it critical solution} (plain), explained in the main text, and the corresponding exchange couplings are depicted in the right panel in green and red, respectively. The left-most point (black) corresponds to the {\it scaling solution} plotted in \Cref{fig: alpha_s Nf=2 scaling} and in the right panel in black. In the right panel, the exchange couplings are displayed for the same initial conditions at $k=\Lambda_{\rm UV}$ and the respective $\kSSB$ and $k_{{\rm conf}}$ scales are indicated by vertical dashed and dotted lines respectively.  In the inlay plot we show the same integrated flows rescaled to their respective $\kSSB$. }
	\label{fig:YM phases  Nf2 gamma2}
\end{figure*}

In the matter sector, we can identify two sub-sectors: one including the fermion-gauge and another with pure fermionic interactions. In the first, the full fermion-gauge vertex in the effective action includes eight different tensor structures
\begin{align}\label{eq:AbarpsipsiFull}
\Gamma^{(A \bar \psi \psi)}(p_1,-p_2)=\sum_{i=1}^8 \lambda_{A \bar \psi \psi}^{(i)} \left[{\cal T}^{(i)}_{A \bar \psi \psi}(p_1,-p_2)\right]_\mu\,,
\end{align}
where we have dropped the colour and flavour structures and the dominant \cite{Mitter:2014wpa, Cyrol:2017ewj, Gao:2021wun} ones are
\begin{subequations}
\label{eq:Abarpsipsidominant tensors}
\begin{align} 
\left[{\cal T}^{(1)}_{A \bar \psi \psi}(p_1,p_2)\right]_\mu=&\, - \imag \gamma_{\mu}\,,\\[1ex]  
\left[{\cal T}^{(4)}_{A \bar \psi \psi}(p_1,p_2)\right]_\mu=&\, (\slashed{p}_1+\slashed{p}_2) \gamma_{\mu}\\[1ex]
\left[{\cal T}^{(7)}_{A \bar \psi \psi}(p_1,p_2)\right]_\mu=&\,\frac{\imag}{2} \left[\slashed{p}_1,\slashed{p}_2\right]\gamma_{\mu}\,
\end{align}
\end{subequations}
It has been shown \cite{Mitter:2014wpa, Cyrol:2017ewj, Gao:2020qsj, Gao:2021wun} the classical tensor structure ${\cal T}^{(1)}_{A \bar \psi \psi}$ is the leading one by far. However, also ${\cal T}^{(4)}_{A \bar \psi \psi}$ and ${\cal T}^{(7)}_{A \bar \psi \psi}$ are important for the quantitative determination of the fermion gauge dynamics in $N_f=2$ and  2+1 QCD, while the other five transverse tensors are irrelevant. In the present computation we only account for the classical tensor structure ${\cal T}^{(1)}_\mu$ included in \labelcref{eq:DiracAction}. For a comparison with full QCD we have implemented the effect of the missing leading tensor structures by an enhancement of the gauge-fermion coupling, as done in \cite{Fu:2019hdw, Ihssen:2024miv}. With the same parameters as in \cite{Ihssen:2024miv} we obtain the values displayed in \Cref{table:comparison mpsi} and labelled as ``Multi-avatar: Enhanced''. We compare to the values obtained in the not-enhanced and the single-avatar truncations discussed in \Cref{app:ManyFlavour}. 

The second sub-sector includes the pure-fermionic interactions. Here, we have used a Fierz-complete basis of momentum-independent four-Fermi operators. Moreover, we account for higher-order scatterings in the dominant scalar-pseudoscalar channel. For this we considered a polynomial mesonic potential \labelcref{eq:Veff} expanded to $N_\textrm{max}=5$, as in \cite{Fu:2019hdw}. The pure fermionic sector shows well converged results, and hence we consider the respective error to be subleading.

In conclusion, we find the truncation employed shows good agreement with results derived in quantitative functional  \cite{Cyrol:2017ewj, Fu:2019hdw, Ihssen:2024miv} and Monte Carlo investigations \cite{Zafeiropoulos:2019flq, Sternbeck:2006rd}, supporting the semi-quantitative nature of our results.

\section{Tuning confinement}
\label{app:TuningConfinement+Stability} 

In this Appendix we provide further technical details on the bootstrap approach to confinement discussed in \Cref{sec:econf}, and in particular on  the tuning of the confining scaling solution in the present approximation. We also evaluate the decoupling regime of gauge-fermion systems. Specifically we show, that observables and RG-invariant quantities in the fermionic sector and the intrinsic dynamical scales have no sizeable dependence on the chosen solution. In combination with the independence on the variation of the cut-off scale $\gamma_\textrm{conf} k_\textrm{conf}$ of the onset of the confinement dynamics, this analysis supports the reliability of the bootstrap approach to confinement used here.

\begin{figure}[t!]
	\centering
	\includegraphics[width=\columnwidth]{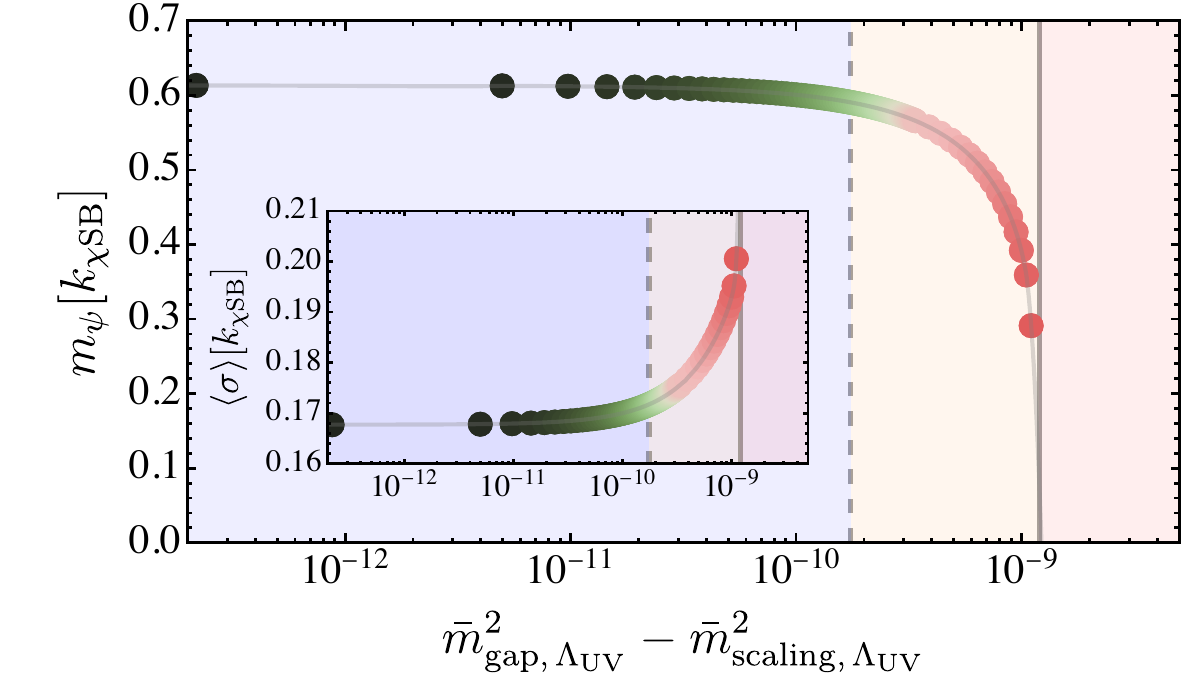}
	\caption{ Constituent fermion masses $m_\psi$ and the chiral condensate $\langle \sigma \rangle$ (inlay plot) in units of $\kSSB$ for different values of the dimensionless mass gap at the UV scale $\bar m^2_{{\rm gap},\,\Lambda_{\rm UV}}$ in a SU(3) gauge theory with two flavours in the fundamental representation. The vertical lines and shaded regions are described in the caption of \Cref{fig:YM phases  Nf2 gamma2}. }
	\label{fig: YM phases  Nf2 masses}
\end{figure}

\begin{figure}
	\centering
	\includegraphics[width=\columnwidth]{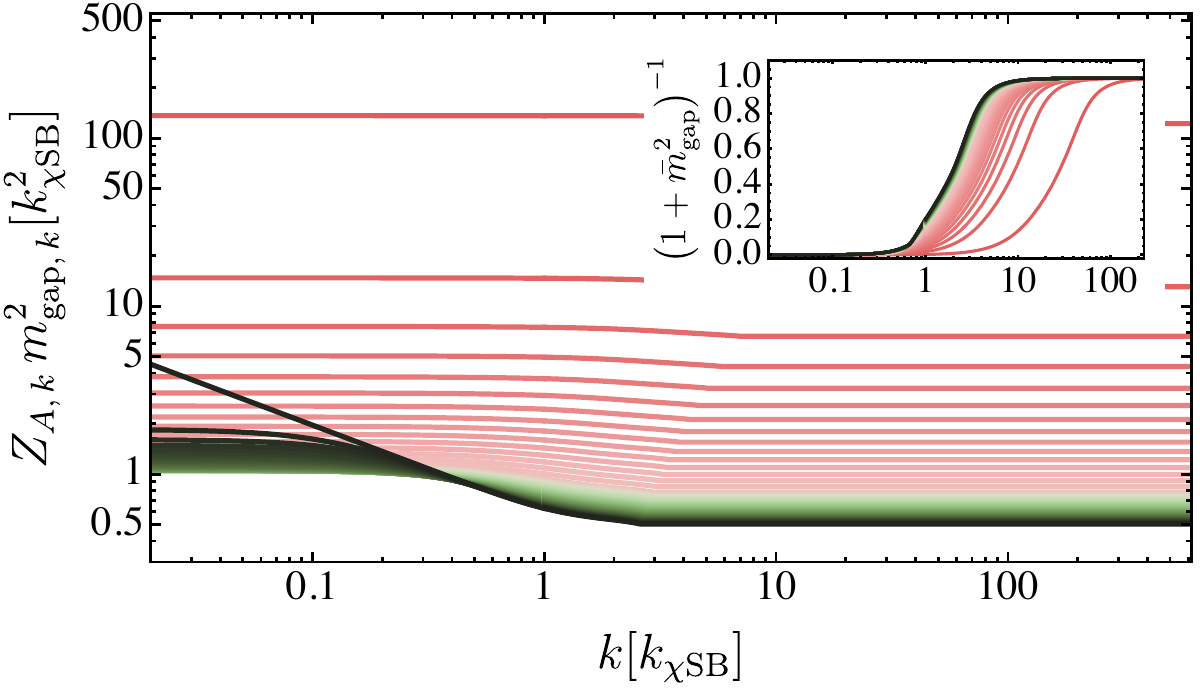}
	\caption{ Integrated flows for the gauge field renormalised mass gap for a SU(3) gauge theory with $N_f=2$ chiral flavours. The IR values are shown in \Cref{fig:YM phases  Nf2 gamma2} with the same colour coding.}
	\label{fig:mass flows  Nf2}
\end{figure}

In the left panel of \Cref{fig:YM phases  Nf2 gamma2} we display the IR value of the scalar part of the gauge field two-point function within the approximation \labelcref{eq:GglueA2}, evaluated at $p=0$ and $k\to 0$, 
\begin{align} 
\lim_{k\to 0} Z_{A,k} \, m^2_{\textrm{gap},k}\,, 
	\label{eq:GA0INverse}
\end{align}
as a function of the UV-values of $m^2_\textrm{scaling}$ in \labelcref{eq:RegimeSwitch} for an SU(3) gauge theory with two massless flavours in the fundamental representation. Similarly to \cite{Cyrol:2016tym} we can identify three different phases: 
\begin{enumerate}
	
	\item  \underline{\it Scaling solution}:  
	it is given by the left-most point (black) in the left panel of \Cref{fig:YM phases  Nf2 gamma2} and the trajectories are displayed in the right panel of the same Figure as well as in \Cref{fig: alpha_s Nf=2 scaling}. 
	The scaling solution is uniquely defined by the scaling of correlation functions as explained in \Cref{sec:confinement}. This is also reflected by the flattening of the gauge-ghost avatar towards the IR. Moreover, in the inlay of \Cref{fig: alpha_s Nf=2 scaling} and in \Cref{fig:mass flows  Nf2}, we show the integrated flow of the renormalised and dimensionful mass gap.  
	
	\item \underline{\it Decoupling regime}: when increasing the value of the gluon mass gap in the UV from the scaling one, confining decoupling or massive solutions are found. For these, the gauge field propagator has a finite limit in the IR. This region of equivalent solutions (shaded in light blue) spans until the {\it maximally decoupling} solution. This solution is approximately defined by the minimum of the IR value of the inverse propagator, marked by a vertically dashed grey line. The exchange couplings for this solution are displayed in \Cref{fig:YM phases  Nf2 gamma2,fig: YM phases  Nf2 masses,fig:mass flows  Nf2} in green. All couplings except the ghost-gluon are in quantitative agreement with the scaling solution. The latter avatar does not freeze towards the IR, as in the scaling solution, but decays. This is a consequence of a slightly too large mass gap. Note that the respective differences do not show in any physical observable computed so far in QCD, thus supporting the interpretation of an IR gauge ambiguity.  

	\item  \underline{\it Massive YM or Higgs regime:} For even larger UV masses, we enter a regime where the theory simply is massive YM (shaded orange and red in \Cref{fig:YM phases  Nf2 gamma2,fig: YM phases  Nf2 masses,fig:mass flows  Nf2}), or can be understood as an approximation of a gauge-fermion system coupled to a scalar field in the Higgs phase. Within this realm, we identify a critical UV mass (plain grey line) value beyond which ${\rm d}\chi{\rm SB}$ is lost. This is induced by the successive decrease of the gauge coupling, finally not approaching the critical coupling. We denote the boundary case as critical and display its running coupling in \Cref{fig:YM phases  Nf2 gamma2} in red. We emphasise that these theories also do not show confinement which is obvious for asymptotically large gluon masses. In this regime, the peak position of the gluon dressing function simply reflects the explicit gluon mass. 
\end{enumerate}

In \Cref{fig: YM phases  Nf2 masses}, we display the values of the constituent fermion masses $m_\psi[ \kSSB]$ in dependence on the gauge mass at the UV scale $\bar m^2_{{\rm gap},\, \Lambda_{\rm UV}}$ with respect to the scaling one. In the inlay, the value of the chiral condensate $\langle\sigma\rangle[\kSSB]$ is shown. This illustrates, that observables are stable in the confining regime covering the scaling and decoupling solutions.

\section{Many-flavour truncations}
\label{app:ManyFlavour}

In this Appendix we extend the analysis of the interplay of confinement and chiral symmetry breaking for different flavours, discussed in  \Cref{sec:phasesYM+chiral}, with further details. 
In \Cref{app:multiavatar many flavour} we discuss the many flavour results using the multi-avatar truncation introduced in \Cref{sec:truncation} and benchmarked in \Cref{sec:InterplayLowNf} and \Cref{app:comparisonQCD,app:TuningConfinement+Stability}. In \Cref{app:largeNfapproximation} we provide the details of the improved single-avatar truncation which results have been discussed in \Cref{sec:phasesYM+chiral}. Last, in \Cref{app:Nfscaling} we provide more results on the scaling of fundamental parameters in the many-flavour limit.

\subsection{Multi-avatar results}\label{app:multiavatar many flavour}
\begin{figure}[t]
	\centering
	\includegraphics[width=\columnwidth]{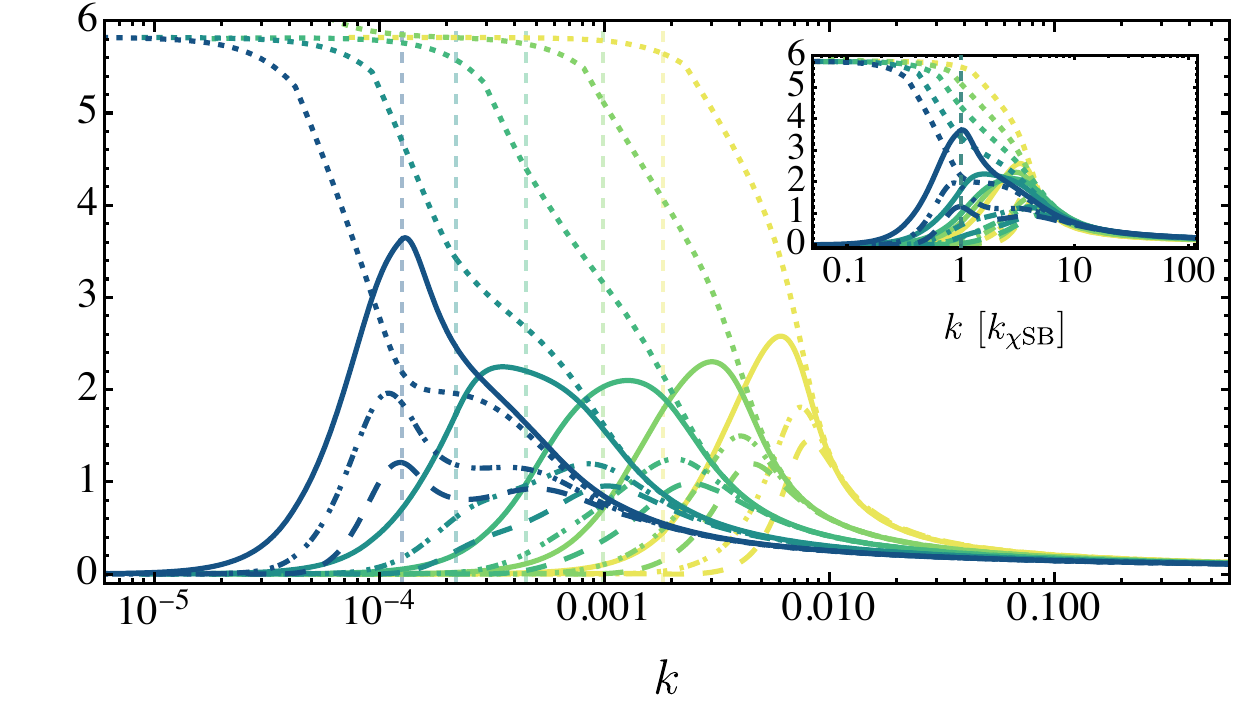}
	\caption{ Gauge exchange couplings as defined in \labelcref{eq:alphasGlue,eq:alphapsibarpsiA} for the pure gauge (dashed), fermion-gauge (plain) and ghost-gauge (dotted) processes in an SU(3) gauge theory with $N_f=1$ (yellow) to 5 (dark blue) flavours. All flows are started at the same perturbative scale and with same initial conditions.  The same curves scaled in units of the respective $\kSSB$ are shown in the inlay. }
	\label{fig: alpha_s Nf=123456 scaling}
\end{figure}

In this Appendix we discuss results within the multi-avatar approximation put forward in \Cref{sec:InterplayLowNf}. In \Cref{fig: alpha_s Nf=123456 scaling} we show the gauge-fermion (plain), gauge-ghost (dotted), three-gluon (dashed)  and four-gluon (dot dashed) exchange couplings (see \labelcref{eq:alphasGlue,eq:alphapsibarpsiA}) for different number of flavours from $N_f=1$ (yellow) to 5 (dark blue) in a $N_c=3$ theory. All flows are initiated at the same UV scale with the identical boundary conditions. It is appreciable how fermionic corrections soften the growth of the gauge dynamics towards the IR delaying the emergence of both confinement and ${\rm d}\chi{\rm SB}$. The peak of the three-gluon vertex is rather constant and rather independent of the number of flavours. We infer from this that the critical coupling for confinement is rather $N_f$-independent. Moreover, the ghost-gauge coupling in the deep IR reaches always the same constant coupling given it is only colour-group dependent. A mild fermionic dependence on this coupling is found in the for large number of flavours which enters through the fermionic effects to the gauge anomalous dimension given the non-renormalisability of the ghost-gluon vertex, see \Cref{app:flowgaugecouplings}. Additionally we observe two non-trivial features already present in the single-avatar results discussed in \Cref{sec:phasesYM+chiral}: a growth with $N_f$ of the gauge-fermion exchange peak and an earlier appearance of $\kSSB$ with respect to the confinement scale, as expected when entering the locking regime. 

As discussed in \Cref{sec:Scaleconfinement}, the pure gauge avatars are the most sensitive to the decoupling of the gauge corrections given by the mass gap. The roughly constant magnitude in the peak of $\alpha_{A^3}$ and $\alpha_{A^4}$ signals how the pure gauge sector and the onset of confinement are rather independent of the number of flavours. On the other hand, the gauge-fermion coupling peak magnitude increases with $N_f$. The larger the number of flavours the more the flow of this $n$-point function is enhanced. The diagrams responsible for this enhancement are at the same time the least sensitive to the gluon mass gap given that mostly encode fermionic ones. Consequently this appears in an expected increment of the gauge-fermion avatar which has the largest impact on the bosonised sector. Moreover, a stronger value of this avatar causes a flattening of the chiral potential and consequently an earlier $\kSSB$. Altogether this leads to the observed effect on increment in the coupling peak and both scales $k_{\rm conf}$ and $\kSSB$ coming together, signalling the appearance of the locking regime. 

%

\subsection{Improved single-avatar truncation} 
\label{app:largeNfapproximation}

A precise resolution of the perturbative higher orders of the running coupling is key in the many-flavour limit of gauge-fermion QFTs. In the conformal and near-conformal limits, the exact $N_c$-$N_f$ dependence of the gauge beta function is crucial for quantitatively determining the CBZ fixed-point value. This is relevant for describing walking regimes as well as for the precise determination of the lower boundary of the conformal window.

To that end, we improve the non-perturbative truncation introduced in \Cref{sec:fRG} and employed in \Cref{sec:InterplayLowNf} to incorporate the perturbative contributions while preserving the non-perturbative character of the fRG and its sensitivity to thresholds. Furthermore, we improve the gauge-fermion avatar, as it is most sensitive to fermionic effects and is responsible for $\dSSB$, to incorporate all perturbative and non-perturbative features. In the many-flavour limit, the distinction between different avatars becomes increasingly irrelevant as the dynamical scales where confinement and $\dSSB$ occur, shift towards the IR. Consequently, we identify the remaining avatars with the fermion-gauge one, see \labelcref{eq:alphai=alpha}, thereby reducing the complexity of the gauge system to a single avatar $\alpha_g = g^2 / 4\pi$ of the gauge coupling. The flow of this single gauge avatar reads,
\begin{widetext}
	\begin{align}\label{eq:gsImproved}
		\partial_t g &= \partial_t \lambda_{A \bar \psi \psi}^{(1)}+ \frac{\partial_t \lambda_{A \bar \psi \psi}^{(>1)}}{\left.\partial_t \lambda_{A \bar \psi \psi}^{(>1)}\right|_{\{\bar m_{\rm gap},\bar m_{\psi}\}=0}} \sum_{n=2}^{{\rm N_{max}}} \Bigg\{ {\tiny \sum_{i=1}^{i_{\rm max}}}\frac{1}{(1+i_{\rm max})}\frac{1}{ (1+\bar m^2_{\rm gap})^{3n+i}} \left(\beta^{(n)}_{A} +  \left[\frac{(2+\bar m^2_\psi)}{2\,(1+\bar m^2_\psi)^{3}}\right]^{N_f {\rm\,  in \,loop }} \beta^{(n)}_{ \psi} \right)\Bigg\}\,.
	\end{align}
\end{widetext}
The first term $\partial_t \lambda_{A \bar \psi \psi}^{(1)}$ is the one-loop component of the full fRG flow of the gauge-fermion avatar \labelcref{eq:alphapsibarpsiA} in \labelcref{eq:flowgApsipsibar}. Moreover, this coupling is the dominant in the large $N_f$ limit given as shown in \Cref{fig: alpha_s Nf=123456 scaling}. The second term in \labelcref{eq:gsImproved} is proportional to $\partial_t \lambda_{A \bar \psi \psi}^{(>1)}= \partial_t \lambda_{A \bar \psi \psi}-\partial_t \lambda_{A \bar \psi \psi}^{(1)}$ which carries all remaining higher-order effects in a non-perturbative fashion (eg. as higher powers in $g$ as well as via the four-Fermi and through the rebosonised couplings $h$ and $u(\bar \rho)$). In the denominator of this term, we have the high-order effects of the gauge-fermion avatar evaluated at vanishing thresholds, $\bar m_{\rm gap},\,\bar m_{\psi}=0$. In the UV perturbative regime, chiral symmetry is unbroken ($\bar m_\psi=0$) and the dimensionless gluon mass gap is negligible $\bar m_{\rm gap }\approx0$ (see inlay in \Cref{fig:mass flows  Nf2} as an example), hence this ratio is unity. 

Inside the curly brackets we include the perturbative contributions adjusted with the respective threshold functions for the gauge and fermionic loops of different power. Here, $\beta^{(n)}_{A}$ and $\beta^{(n)}_{\psi}$ stand for the pure gauge and the fermionic at the $n$-loop level. For these, we have employed four-loop $\overline{\rm MS}$ results \cite{Herzog:2017ohr,Vermaseren:2000nd} but other schemes will be implemented in future works. The respective perturbative contributions have been properly gapped by adjusted powers of the number of propagators which are the functions encoding the $\bar m^2_{\rm gap}$ and $\bar m_\psi$  dependencies. We have added this normalised summation in $i$ to account correctly for the progressive decoupling of different orders. We have checked there is no significant dependence of the results discussed on the powers of the threshold functions. 

It is left to implement the confining dynamics reflected by the emergence of the gluon mass gap. In the present  single-avatar approximation we employ the same truncation as in the multi-avatar setup, and we refer for the details description to \Cref{sec:confinement} and \Cref{app:flowmGap}. In particular, we use the same effective flow for $m^2_{\textrm{gap},k}$ defined in \labelcref{eq:EffmGap}. In the explicit computations, we take the onset scale as the point where $\alpha_{A \bar \psi \psi} = 1.5$ is satisfied, although no significant dependence was found when varying the onset within $\alpha_{A \bar \psi \psi} \approx(1, 2)$.

The properties of this single-avatar flow can be summarised as follows. In the UV, the gauge dynamics is weak and there are no presence of thresholds. The flow \labelcref{eq:gsImproved} reads as the $n$-loop perturbative coupling given that the ratio encoding the non-perturbative fRG effects is smoothly cancelled. This way we enjoy sensitivity to fixed-point solutions precisely encoded in the perturbative expressions. As the gauge dynamics strengthen towards the IR, thresholds smoothly appear as shown in \Cref{sec:Scaleconfinement,sec:ScalechiralSB}. These progressively decouple the respective modes on the perturbative and fRG contributions in an organised manner. It is convenient to note here that at intermediate regimes there is a minor double counting effect caused by the overlap of both contributions. Furthermore, aside from the modification discussed on the gauge sector, the bosonised sector couplings and flows of $h$, $u(\bar \rho)$ and $\lambda_j$ are kept in full glory. 

 In conclusion, the present single-avatar approximation is simple, versatile and is semi-quantitative. This allows for an easy access to a regime in theory space which has proven to be quite challenging for first principles approaches. This setup allows to understand the interplay between the dynamics of interest as well as the self-consistent implementation of other sectors. For example, this latter point is particularly relevant for BSM investigations which can be self-consistently implemented within functional implementations of the SM see eg. \cite{Pastor-Gutierrez:2022nki,Goertz:2023pvn,Gies:2023jzd,Gies:2017zwf,Gies:2019nij,Gies:2018vwk,Gies:2015lia,Gies:2013pma}.

\subsection{Error estimate of the many-flavour truncation}
\label{app:errorestimateSingleavatar}

The single-avatar truncation discussed in the previous Appendix provides a simplified setup to describe the pure gauge and chiral dynamics in a unified manner. In this Appendix we discuss the regime of reliability of this truncation as well as provide an error estimate.

In \Cref{sec:Nf<4} we discussed how this approximation displays in the $N_f=1$ limit a dip in quantities dependent on $\kSSB$ in comparison to the multi-avatar truncation. This is a product of accounting only for a single avatar (the fermion-gauge coupling) demanded to describe both, the confining and $\dSSB$ dynamics. In the QCD limit, we have seen in \Cref{sec:ScalechiralSB} that the confinement and $\dSSB$ are rather apart and consequently distinguishing the different avatars is necessary for a quantitative determination. Moreover, it is then well understood that the leading source of error of the single-avatar approximation in the $N_f=1$ limit. This hints to the qualitative nature of this truncation in the few flavour limit. Moreover, a good estimate on the error in this limit can be obtained by comparing to the results from in multi-avatar truncation. Furthermore, we note that this error is reduced when analysing quantities as a function of the respective $\kSSB$ scale. For example, this is visible in \Cref{fig:mqmsigmasigma_diffNf,fig:mqmsigmasigma_diffNf_appendix}. Moreover, the errors displayed account for the difference between both truncations as well as an inherent source of error providing an approximate and generous  40\% uncertainty on the obtained values. We note that part of the uncertainty arises from varying the  definition of the $\kconf$ scale in each of the truncations, as it is a proxy and not uniquely defined.

\begin{figure}[t]
	\centering
	\includegraphics[width=1\columnwidth]{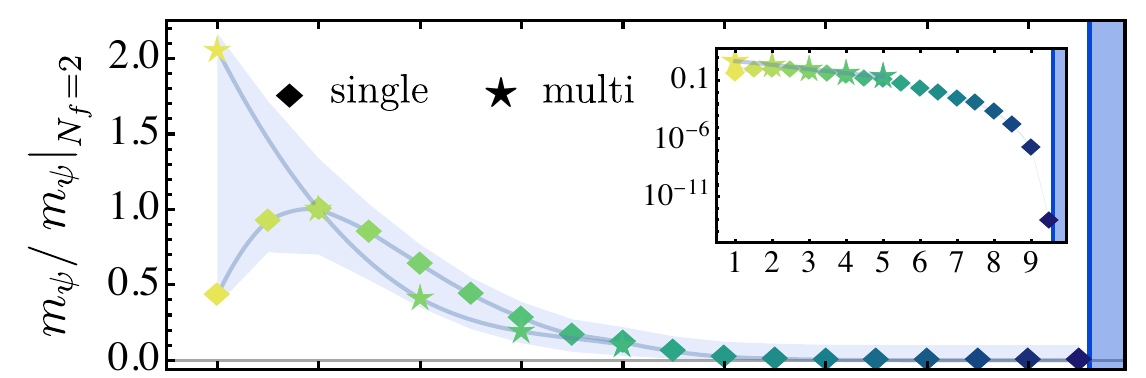}
	\includegraphics[width=1\columnwidth]{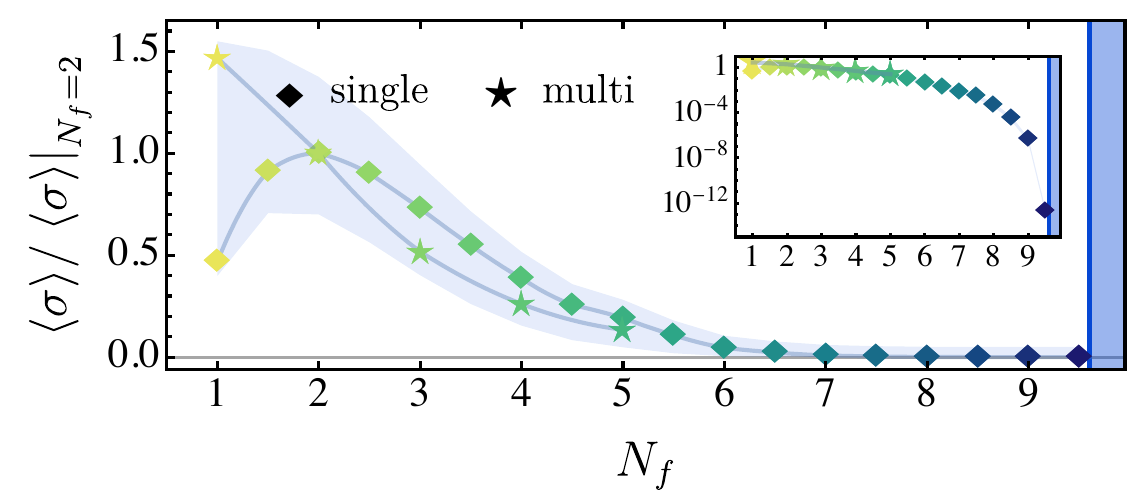}
	\caption{The constituent fermion masses $m_\psi$ (top panel) and the chiral condensate $\langle \sigma\rangle$ (bottom panel) in units of an absolute UV scale (where all couplings are initiated with $\alpha_g\approx0.08$) for a SU(3) gauge theory with $N_f$ fundamental  flavours. Both quantities are shown normalised to the respective $N_f=2$ values. We show results in the single (diamonds) and multi-avatar (stars) approximations detailed in \Cref{sec:InterplayLowNf} and \Cref{app:largeNfapproximation}, respectively. The shaded areas depict the estimated errors discussed in \Cref{app:errorestimateSingleavatar}. The inlay plots show the same quantities in a logarithmic scale. } 
\label{fig:mpsi_sigma_largeNf}
\end{figure}

In the many flavour limit, the scales of confinement and $\dSSB$ are drawn together in both truncations (see \Cref{fig: alpha_g semipert diff Nf,fig: alpha_s Nf=123456 scaling}) and this source of error is not present. This can be seen in the following \Cref{app:Nfscaling} and \Cref{fig:mpsi_sigma_largeNf,fig:mqmsigmasigma_diffNf_appendix} where fundamental quantities are displayed in both approximations.

%

\subsection{Many-flavour scaling of fundamental quantities}
\label{app:Nfscaling}

In this Appendix we provide more results on the dependence of  fundamental quantities and parameters as a function of  the number of flavours.

In \Cref{fig:mpsi_sigma_largeNf}, we show the equivalent to \Cref{fig:kSSBkconf_largeNf} for the constituent fermion masses $m_\psi$ (top panel) and the chiral condensate $\langle \sigma\rangle$ (bottom panel) in units of an absolute UV scale (where all couplings are initiated with $\alpha_g\approx0.08$) for a SU(3) gauge theory with $N_f$ fundamental  flavours. Both of these quantities are tightly related to $\kSSB$ and hence in the few flavour limit we observe a difference between the multi- and single-avatar truncations.

\begin{figure}[t!]
	\centering
	\includegraphics[width=1\columnwidth]{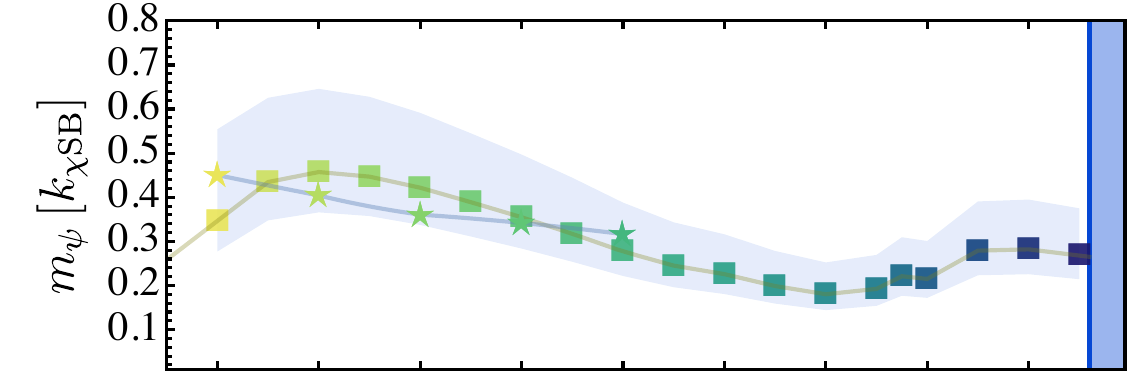}
	\includegraphics[width=1\columnwidth]{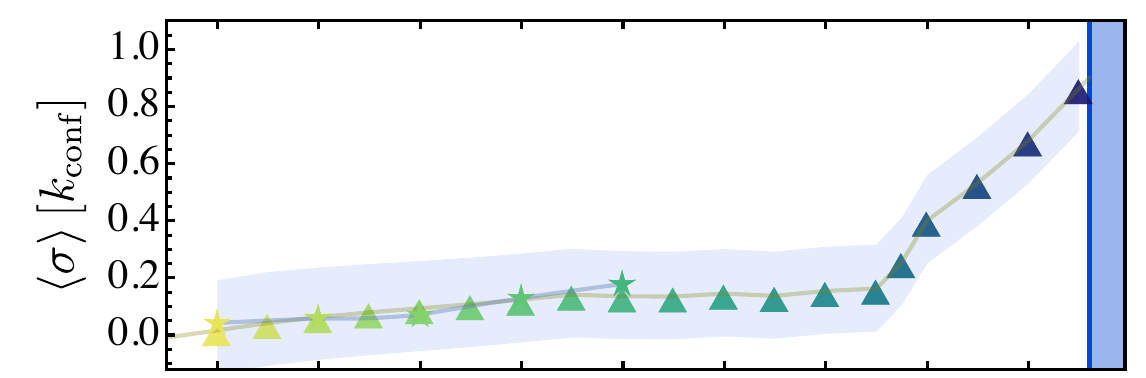}
	\includegraphics[width=1\columnwidth]{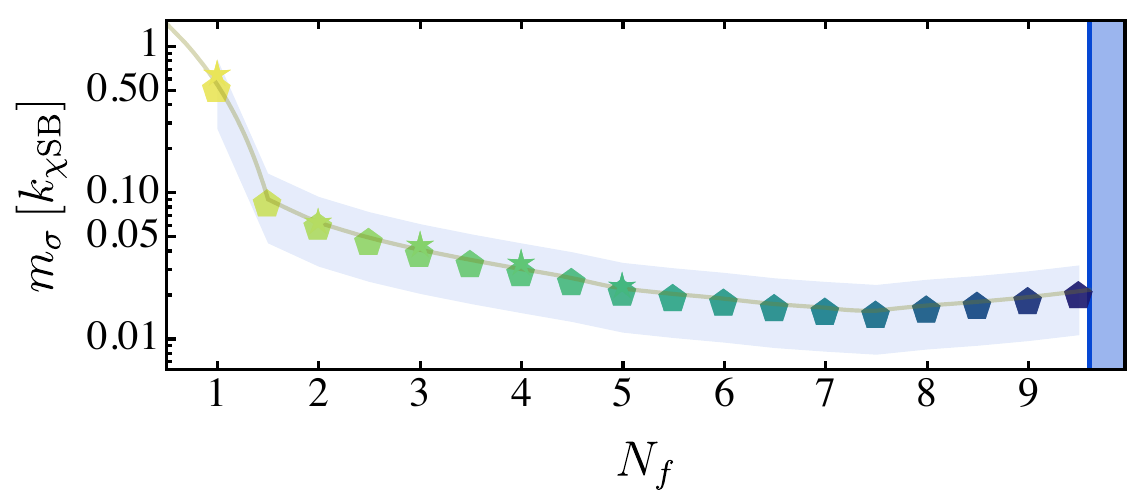}
	\caption{  From top to bottom, the constituent fermion masses $m_\psi$ (squares), the chiral condensate $\langle \sigma\rangle$ (triangles) and the mass $m_\sigma$ of the $\sigma$-mode (polygons) in a SU(3) gauge theory with $N_f$ fundamental flavours. The fermion and sigma masses are shown in units of the respective $\kSSB$ scale and the chiral condensate, in units of $\kconf$. The shaded areas depict the estimated errors. }
	\label{fig:mqmsigmasigma_diffNf_appendix}
\end{figure}

In \Cref{fig:mqmsigmasigma_diffNf_appendix} we show the constituent fermion mass in units of the $\kSSB$ scale and the chiral condensate in units of $\kconf$, similarly to \Cref{fig:mqmsigmasigma_diffNf}.  Additionally, we depict the mass of the scalar $\sigma$-mode in units of the $\kSSB$ scale. We note that the mass of the $\sigma$-mode stays finite for all $N_f$ considered, for a more quantitative evaluation in physical QCD with a full effective potential see \cite{Ihssen:2024miv}. We note in passing that this mode has been conjectured to play the role of the dilaton, namely the Goldstone boson of conformal symmetry, see e.g.~\cite{Zwicky:2023krx,Ingoldby:2023mtf,Appelquist:2019lgk,LSD:2018inr,Appelquist:2017wcg,Appelquist:2010gy}. A respective discussion based on the present results and \cite{Ihssen:2024miv} will be done elsewhere. 

\begin{figure}[t]
	\centering
	\includegraphics[width=.975\columnwidth]{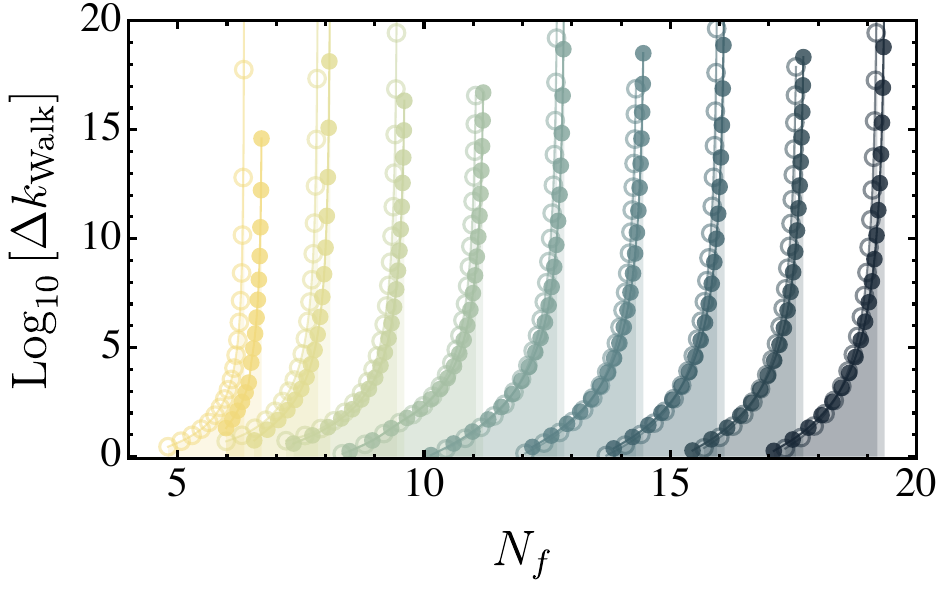}
	\caption{Size of walking regimes ($\Delta k_{\rm walk}$) as a function of $N_f$ for  SU($N_c$) gauge theories with different numbers of colours from $N_c=2$ (left-most light orange points) to $N_c=6$ (right-most and dark blue points) in half integer steps. We show estimates employing three (circles) and four (filled circles) loop $\overline{\rm MS}$ beta functions. The shaded bands indicate 40\% variation purely for indicative purposes.}
	\label{fig:walking regimes all Nc}
\end{figure}
\begin{figure*}[th]
	\centering
	\includegraphics[width=1.6\columnwidth]{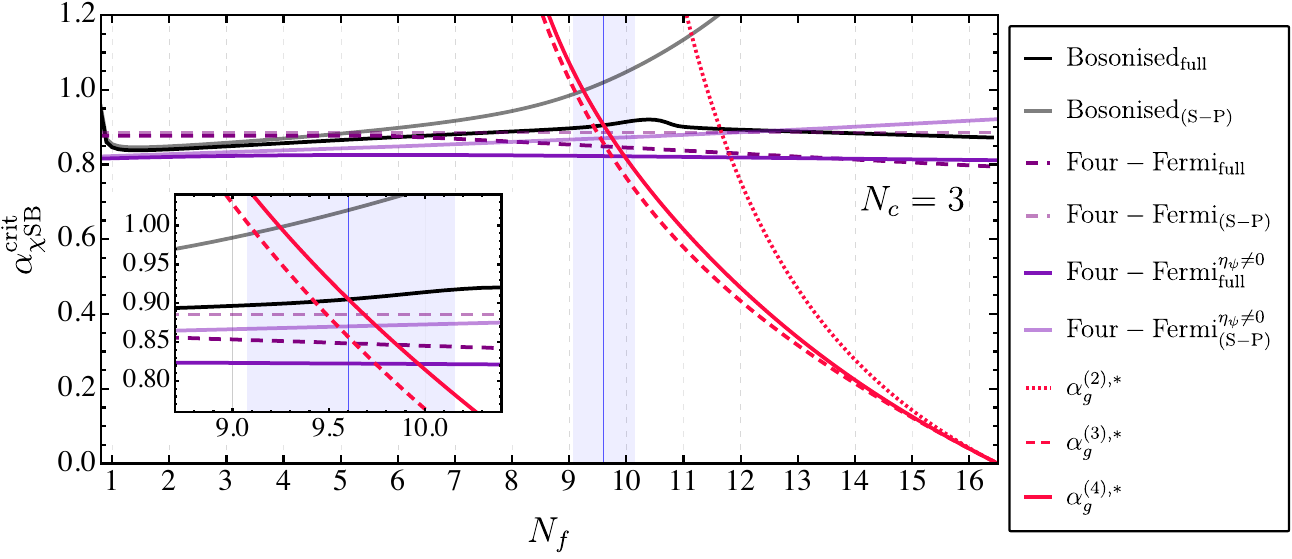}
	\caption{ The critical gauge coupling ($\acrit$) for $\dSSB$ in a SU(3) gauge theory with $N_f$ fundamental flavours and in different truncations. In black, the result obtained in the best approximation with the bosonised setup novel to this work with all tensor structures (black, labelled as ${\rm Bosonised}_{\rm full}$) and only the dominant scalar-pseudoscalar channels (grey, labelled as ${\rm Bosonised}_{\rm (S-P)}$). We compare to the results in \cite{Gies:2005as} with all four-Fermi tensor structures (dashed dark purple, ${\rm Four-Fermi}_{\rm full}$) and only the dominant (S-P) (dashed light purple,  ${\rm Four-Fermi}_{\rm (S-P)}$). Additionally, we show the approximation in \cite{Gies:2005as} but including the effect of gauge and fermion anomalous dimensions with all (plain dark purple, ${\rm Four-Fermi}^{\eta_\psi\neq0}_{\rm full}$) and only the dominant (plain light purple, ${\rm Four-Fermi}^{\eta_\psi\neq0}_{\rm (S-P)}$) tensor structures. The vertical red lines show the two (dotted), three (dashed) and four-loop (plain) $\overline{\rm MS}$-scheme fixed-point values. The vertical blue line marks \labelcref{eq:Nfcritbest} and the respective shaded area the estimated error on its determination. In the inlay, we zoom into the regime of interest to determine the boundary of the CBZ window where $\acrit=\alpha^{*}_{g}$.}
	\label{fig:alphas_crit Comparisons}
\end{figure*}
%

\section{Boundary of the conformal window}
\label{app:comparisonBoundaryCBZ}

In this Appendix we provide the details on the computation of the lower boundary of the CBZ conformal window. In \Cref{app:convergenceCBZ}, we determine $\alpha^{\rm crit}_{\chi{\rm SB}}$ in different truncations and provide a detailed comparison with previous computations. In \Cref{app:systematicsCBZboundary}, we discuss the error estimate of the critical number of flavours delimiting the onset of conformality \labelcref{eq:Nfcritbest}. In \Cref{app:gammam} we compute the fermion mass anomalous dimensions which is usually employed to estimate the emergence of $\dSSB$. We compare the values obtained perturbatively and non-perturbatively offering an estimate on the validity of this approach.

\subsection{Convergence of the truncation}
\label{app:convergenceCBZ}

The lower boundary of the conformal window was previously estimated using the fRG in \cite{Gies:2005as}, employing the four-Fermi language introduced in \Cref{sec:fermiongaugedynamics}. In contrast, in this work, we progress by bosonising the ($\sigma-\pi$)) channel while keeping the remaining tensor structures in their original four-Fermi form. The  (S--P) channel is the most relevant for $\dSSB$, as it manifests the appearance of a condensate and carries the information of the lightest resonant channels, which become critical. Technically, the currently employed bosonised formulation provides a qualitative improvement, as it allows for the inclusion of higher-dimensional fermionic operators via a higher-order meson potential and incorporates momentum dependencies for free. This is necessary to account for non-perturbative effects and to obtain reliable and quantitative estimates for the appearance of $\dSSB$.

In this Appendix, we begin with the approximation in \cite{Gies:2005as} and systematically improve upon it. Using the four-Fermi formalism, we observe a rather constant $\alpha^{\rm crit}_{\chi{\rm SB}}$ when considering only the (S-P) mode and the subleading tensor structures, as shown by the dashed horizontal and plain purple lines in \Cref{fig:alphas_crit Comparisons}. This indicates that, as discussed in \cite{Mitter:2014wpa}, the dominant channel for $\chi{\rm SB}$ at small $N_f$ in vacuum is the (S-P) mode. 

Moreover, we improve upon the computation in \cite{Gies:2005as} by incorporating the gauge and fermion anomalous dimensions derived in this work into the flow of the four-Fermi couplings. These findings are summarised in \Cref{fig:alphas_crit Comparisons}, where we show how this improvement slightly decreases the required critical coupling in the few $N_f$ limit towards the bosonised result. These are depicted with plain purple lines. 

In the present approach, bosonising the ($\sigma-\pi$) channel at the lowest order already includes higher four-Fermi operator effects, leading to a straightforward improvement over the point-like limit. However, we observe that neglecting the other tensor channels, in particular in the bosonised computation, leads to sizeable modifications for $N_f \geq 5$. This highlights the  relevance of accounting for the complete basis in the many flavour regime, see the grey and black lines in \Cref{fig:alphas_crit Comparisons}. 

In \Cref{fig:alphas_crit Comparisons Nmax}, we study $\alpha^{\rm crit}_{\chi{\rm SB}}$ for different expansion orders ($N_{\rm max}$) of the mesonic potential \labelcref{eq:Veff}. Higher orders in the Taylor-expanded potential correspond to considering fermionic operators up to $\left(\bar \psi {\cal T} \psi\right)^{2 N_{\rm max}}$. We see that the effect of higher-order operators modify $\alpha^{\rm crit}_{\chi{\rm SB}}$ for $N_f \gtrsim 8$ and quickly converges at the $N_{\rm max}=4$ order, see inlay in the right plot of \Cref{fig:alphas_crit}.  However, it is important to stress that $\acrit$ is affected by high-order fermionic interactions in the regime of interest for the determination of the boundary CBZ window. 

An interesting future improvement of the present computation consists of bosonising the remaining channels \labelcref{eq:4FermiTensors}, and in particular the ($\eta-a$) channel. This accounts for the effect of higher order operators in an unbiased way.

\subsection{Systematics of the boundary}
\label{app:systematicsCBZboundary}

For the determination of the boundary of the conformal window we have employed the relation \labelcref{eq:boundarycondition} which requires of the determination of the critical coupling for $\dSSB$, $\acrit$, and the CBZ fixed point, $\alpha^*_g$
as a function of $N_c$ and $N_f$. Moreover, its determination leads to two sources of error in the determination of the critical number of flavours in \labelcref{eq:Nfcritbest}.

The first arises from the obtention of $\alpha^*_g$ for which we have employed  results from perturbation theory including high-order loop effects. For this task, perturbative computations have proven to be very powerful as they account in a systematic manner for the higher-order in the gauge beta functions. However, these are known to be scheme dependent beyond two loops and hence subject to an error. In this work we employed four-loop results in the $\overline{\rm MS}$ scheme \cite{Herzog:2017ohr,Vermaseren:1997fq,Ritbergen1997,Ruijl:2017eht} without further improvement (eg. resummation). Moreover, we estimate a modest $10\%$ error in the $N_f$ dependence of the fixed-point values. We find this to be a sufficient estimate given that the relative difference of the four- and  three-loop ones is less than $3\%$. 
 
\begin{figure}[t]
	\centering
	\includegraphics[width=1.\columnwidth]{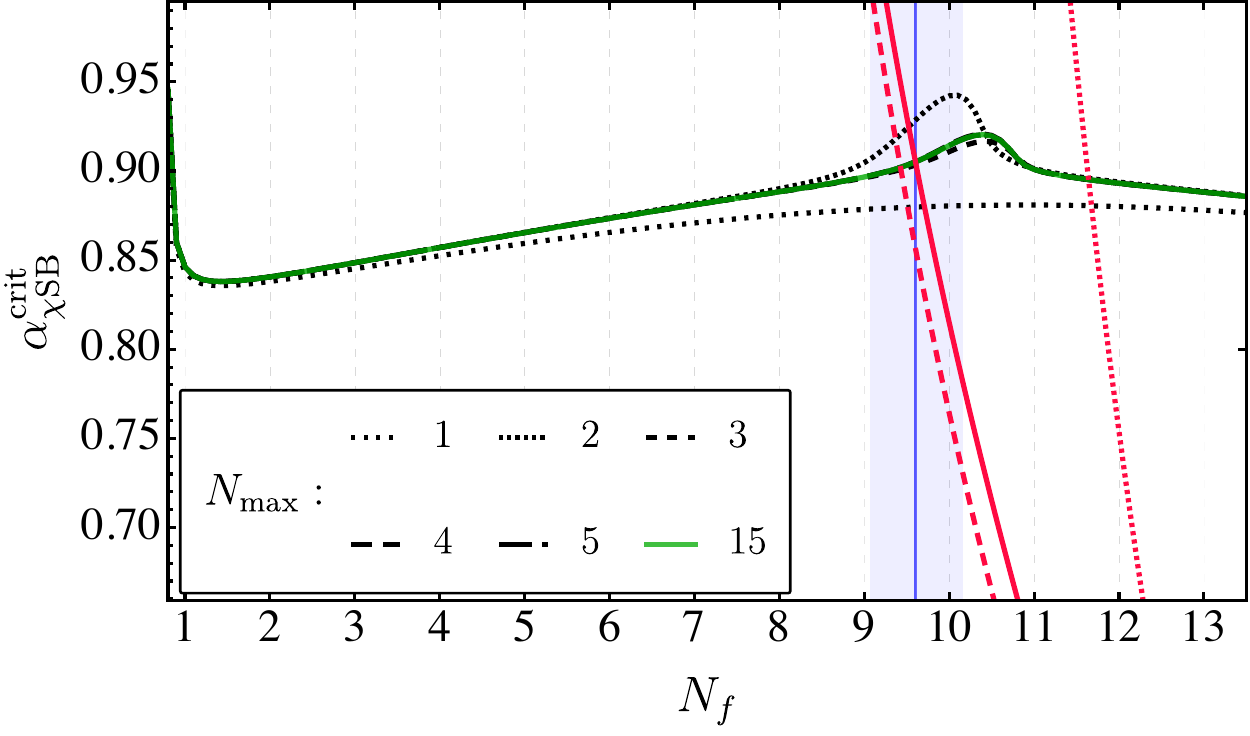}
	\caption{ Comparison of $\alpha^{\rm crit}_{\chi{\rm SB}}$ for different expansion orders of the Taylor chiral potential defined in \labelcref{eq:Veff}. By light red lines we display the fixed-point values derived from the two (dotted), three (dashed) and four (plain) loop $\overline{\rm MS}$-scheme beta functions. We find the dependence on the  that the dependence is considerable only at large numbers of flavours $N_f \gtrsim 9$. For expansion orders $N_{\rm max}\geq3$ we find well converged results. The vertical blue line marks \labelcref{eq:Nfcritbest} and the respective shaded area the estimated error on its determination.}
	\label{fig:alphas_crit Comparisons Nmax}
\end{figure}

The second source of error arises from truncating the effective action for the determination of $\acrit$. An analysis of its solid stability has been done in \Cref{app:comparisonBoundaryCBZ} where we have systematically improved different aspects of the computation. Considering higher orders in the Taylor expanded potential in \labelcref{eq:Veff} is equivalent to considering up to  fermionic operators $\left(\bar \psi {\cal T} \psi\right)^{2 N_{\rm max}}$ in the bosonised scalar-pseudo scalar channel. We find that these operators are relevant up to $ N_{\rm max}=4$ where very good convergence is found, see \Cref{fig:alphas_crit Comparisons Nmax}. Worth noting is the necessity of high-order operators in the regime relevant for $N_f^{\rm crit}$. Furthermore, we have also checked what is the impact of the subdominant tensor structures (non (S-P)) in the determination of $\alpha^{\rm crit}_{\chi{\rm SB}}$. As shown in \Cref{fig:alphas_crit Comparisons}, the impact of 
neglecting their contribution is sizeable for $N_f>8$. 

To estimate the potential impact of higher dimensional operators in these subleading channels which are not explicitly accounted in the present computation,  
we have multiplied the flows of the respective four-Fermi couplings by an enhancement factor. Considering twice its original strength, we observe only minor modifications in the many flavour regime well accounted by the error estimate.  If applying the same technique to enhance or suppress the flows of the bosonised (S-P) channel by 30\% leads to less than 1\% change in $N_f^{\rm crit}$ showing the remarkable stability of the results. 

In the present computation we have only explicitly included the classical tensor structure (${\cal T}^{(1)}_{A \bar \psi \psi}(p_1,p_2)$ in \labelcref{eq:Abarpsipsidominant tensors}) in the fermion-gauge vertex \labelcref{eq:AbarpsipsiFull}. As discussed in detail in \Cref{app:comparisonQCD}, it is known from functional QCD computations that two additional ones are required (${\cal T}^{(4)}_{A \bar \psi \psi}(p_1,p_2)$ and ${\cal T}^{(7)}_{A \bar \psi \psi}(p_1,p_2)$ in \labelcref{eq:Abarpsipsidominant tensors}), see \cite{Mitter:2014wpa, Cyrol:2017ewj, Gao:2021wun, Gao:2020qsj}. 
Previously we discussed its impact in the determination of $m_\psi$ in the $N_f=2$ and 3 cases and considered and enhancement in the fermion-gauge vertex to reproduce for its effect. In the many-flavour limit, these subleading tensor structures are expected to become less relevant as the gauge coupling becomes more perturbative due to the fermionic effects and consequently higher dimensional operators are expected not feed significantly into the dynamics. Despite of this observation, we estimate the potential effect of the missing tensor structures in the determination of $\acrit$. For it, we enhance and suppress by 5\% the $\alpha_{A \bar \psi \psi}$ exchange coupling feeding into the bosonised and four-Fermi flows. The resulting uncertainty is depicted by a dark grey band in \Cref{fig:alphas_crit} and leads to the error presented in \labelcref{eq:Nfcritbest}.

\subsection{$\gamma_m$ in the fRG and perturbation theory}
\label{app:gammam}

A common approach employed for the analysis of $\dSSB$ is based on the magnitude of the anomalous dimension $\gamma_m$ of the fermion mass in the chiral limit. Deep in the conformal regime, $\gamma_m\ll 1$  and goes to zero at the upper boundary of the conformal window, see \Cref{fig:gammam}. In turn, it rises towards smaller number of flavours. It is suggestive that the onset of $\dSSB$ is signalled by $\gamma_m\approx 1$ and this signal has been used for its determination. Moreover, given the good convergence properties of perturbation theory deep in the conformal regime, $\gamma_m$ has been computed within high perturbative loop orders and the perturbative $\gamma_m \sim 1$  has been used for the determination of the critical flavour number, $N_f^{\rm crit}$. 

For a detailed analysis of this approach see e.g.~\cite{Kim:2020yvr}, and we find $N_f^{\rm crit}(N_c=3) \approx(9.2,9.8)$ in the $\overline{\rm MS}$ scheme. In \cite{Lee2021}, $N_f^{\rm crit}(N_c=3)\approx9.79 ^{+1.25}_{-1.18}$ is found, which is compatible with the present fRG estimate. For other computations employing the same methodology and different schemes see  e.g.~ \cite{Ryttov2011,Ryttov2016,Ryttov2016a,Ryttov2017,Ryttov2018,Lee2021,Appelquist:1998rb} and references therein.

While perturbation theory works successively well deep in the conformal regime, it lacks sensitivity to the dynamical onset of $\dSSB$. For example, pre-condensation phenomena as described in \Cref{sec:precondensation} cannot be captured by perturbation theory. Moreover, increasingly large anomalous dimensions are typically also a signal for the potential failure of perturbation theory. In summary, we expect the perturbative estimates for $\gamma_m$ to hold true qualitatively for $N_f\to N_f^{\rm crit}$ from above, but with quantitative deviations close to $N_f^{\rm crit}$. Furthermore, the present approach enables us to compute the critical $\gamma_m$. While we expect it to be of the order of unity, it may also deviate considerably. 

The anomalous dimension of the fermionic mass in the conformal window is given by 
\begin{align} 
\gamma_m = \lim_{m_\psi \to 0} 	\frac{\partial_t m_\psi}{m_\psi}  \,, 
\label{eq:gammamapp} 
\end{align}
which amounts to the computation of the anomalous dimension in the conformal window for a small explicit mass $m_\psi$, which is then taken to zero. In the present fRG approach with emergent composites, the RG-scale dependent fermionic mass is given by 
\begin{align} 
	m_{\psi,k} = \frac{h_{\phi,k}(\rho_{\textrm{\tiny{EoM}},k}) \sigma_{\textrm{\tiny{EoM}},k}}{\sqrt{2N_f}} \,, 
\end{align} 
as in \labelcref{eq:mpsi} where $	\phi_{\textrm{\tiny{EoM}},k}$ is the solution of the equation of motion 
\begin{align} 
	\left. \frac{\partial V(\rho)}{\partial \sigma }\right|_{\phi=\phi_\textrm{\tiny{EoM}} } = c_\sigma\,, \quad \textrm{with}\quad \phi_\textrm{\tiny{EoM}} =  \left( \begin{array}{c} 0 \\ \sigma_\textrm{\tiny{EoM}}\end{array}  \right)\,.
\end{align}
Here, $c_\sigma$ is  a linear shift of the effective action in $\sigma$, 
\begin{align} 
	\Gamma[\Phi]\to \Gamma[\Phi] - c_\sigma \int_x \sigma\,, \qquad c_\sigma \propto m_\psi\,,
\label{eq:FermionMassShift} 
\end{align}  
which introduces an explicit chiral symmetry breaking, see \cite{Fu:2019hdw, Ihssen:2024miv} for a detailed discussion. The vanishing mass limit is tantamount to $c_\sigma\to 0$, thus removing the explicit breaking of chiral symmetry. Then, the anomalous dimension $\gamma_m$ is given by  
\begin{align}  \nonumber 
	\gamma^{\rm fRG}_m =&\, \lim_{c_\sigma\to 0} \frac{ \partial_t \left[ h_{\phi}\sigma_{\textrm{\tiny{EoM}}} \right] }{h_{\phi} \sigma_{\textrm{\tiny{EoM}}}  } \\[1ex]\nonumber 
	=&\lim_{c_\sigma\to 0} \frac{\partial_t h_{\phi}+ 2 \partial_t \rho_{\textrm{\tiny{EoM}}} h_\phi'}{h_{\phi}} + \frac{\partial_t \sigma_{\textrm{\tiny{EoM}}} }{  \sigma_{\textrm{\tiny{EoM}}} }\\[1ex]
	=&\,  \frac{\partial_t h_\phi}{h_{\phi}} + \frac{\partial_t \sigma_{\textrm{\tiny{EoM}}} }{  \sigma_{\textrm{\tiny{EoM}}} }\,, 
\label{eq:gammamfull}
\end{align}
where we have used that the $h_\phi'$-term vanishes at $\dot \rho_\textrm{\tiny{EoM}}=0$ at $\rho_\textrm{\tiny{EoM}}=0$. Moreover, the limit $c_\sigma\to 0$ can be safely done in the $(\partial_t h_\phi)/h_\phi$ part: both the flow and in particular the Yukawa coupling in the denominator have a finite limit. This leads us to 
\begin{widetext}
	\begin{align}
		\frac{\partial_t h_\phi}{h_\phi}=&-\frac{\lambda_{\psi\bar \psi A}^2}{(4\pi)^2} C_{\rm F}\Bigg[\frac{3(1-\eta_A/6)}{ (1+\bar m_\psi^2)  (1+\bar m_{\rm gap}^2)^2 }\notag+\frac{3 (1-\eta_\psi/5) }{ (1+\bar m_\psi^2)^2 (1+\bar m_{\rm gap}^2) }\Bigg]+\frac{\lambda_{+}}{(4\pi)^2}\frac{ (1-\eta_\psi/5) }{(1+\bar m_\psi^2)^2}\nonumber\\[1ex]
		&-\frac{h_{(\sigma-\pi)}^2}{(4\pi)^2} \Bigg[\left(N_f-\frac{1}{N_f}\right)\left\{\frac{1-\eta_\pi/6 }{2\, (1+\bar m_\psi^2)  \left(1+\bar m^2_\pi\right)^2}+\frac{1-\eta_\psi/5 }{2\, (1+\bar m_\psi^2)^2  \left(1+\bar m^2_\pi\right)}\right\}\nonumber\\[1ex]
		&\hspace{1.75cm}-\frac{1}{N_f }\left\{\frac{1-\eta_\sigma/6}{2\, (1+\bar m_\psi^2) \left(1+\bar m^2_\sigma\right)^2}+\frac{1-\eta_\psi/5}{2 \,(1+\bar m_\psi^2)^2  \left(1+ \bar m^2_\sigma\right)}\right\}\notag\\
		&+\frac{h_{(\eta-a)}^2}{(4\pi)^2} \Bigg[\left(N_f-\frac{1}{N_f}\right)\left\{\frac{1-\eta_\eta/6 }{2\, (1+\bar m_\psi^2)  \left(1+\bar m^2_\eta\right)^2}+\frac{1-\eta_\psi/5 }{2\, (1+\bar m_\psi^2)^2  \left(1+\bar m^2_\eta\right)}\right\}\nonumber\\[1ex]
		&\hspace{1.75cm}-\frac{1}{N_f }\left\{\frac{1-\eta_a/6}{2\, (1+\bar m_\psi^2) \left(1+\bar m^2_a\right)^2}+\frac{1-\eta_\psi/5}{2 \,(1+\bar m_\psi^2)^2  \left(1+ \bar m^2_a\right)}\right\}\,\Bigg]+\frac{1}{2}\eta_\pi+ \eta_\psi\,,	\label{eq:flowhdiagramatic}
	\end{align}
\end{widetext}
where the $\dot{\bar  A}$ term in \labelcref{eq:flowh} is absent in the broken phase and also in the limit $c_\sigma\to0$.  The first line contains the gauge diagrams and the contribution from the (V+A) four-Fermi channel. In the remaining lines, we have made explicit the bosonised channels of the ($\sigma-\pi$) and ($\eta-a$) modes. In the axially symmetric case, both contributions are equal with opposite sign and therefore cancel. However, a remaining part of the ($\sigma-\pi$) channel is contained in the flow of the minimum of the potential,
\begin{align}
	\frac{\partial_t \sigma_{\textrm{\tiny{EoM}}}}{\sigma_{\textrm{\tiny{EoM}} } }=&  -\frac{h_{(\sigma-\pi)}^2 N_c}{(4\pi)^2 \, \bar m^2_\sigma} \left(1-\frac{\eta_\psi}{5}\right)-\frac{1}{2}\eta_\pi\,,\label{eq:flowsigmadiagramatic}
\end{align}
which appears proportional to $N_c$. 

The anomalous dimension can also be determined in the conformal window purely from the four-Fermi language, leading to $\gamma^{\rm fRG\, 4F}_m$. It is determined completely from the flow of the fermion two-point function in the $T^0_f$ projection, 
\begin{align}\label{eq:gammam4f}
	\gamma^{{\rm fRG\, 4}\psi}_m =& \cdots+ \frac{(1-\eta_\psi/5)}{(4\pi)^2 N_f(1+\bar m_\psi^2)^2} \Big[ 8 \lambda_{+}\\[1ex]
	&\hspace{-.7cm}+\lambda_\textrm{\tiny{SP}}^{\left({\eta-a}\right)}(N_f^2-2)-\lambda_\textrm{\tiny{SP}}^{\left({\sigma-\pi}\right)}(N_f^2-2 + 4 N_c  N_f)\Big]\notag\,,
\end{align}
and the $\cdots$ here stand for the gauge field loops in the first line of \labelcref{eq:flowhdiagramatic}. All terms in this four-Fermi expression can be identified to those in the bosonised one finding the correct correspondence.

In \Cref{fig:gammam} we provided an explicit comparison between the anomalous dimensions computed in different orders of perturbation theory and the non-perturbative fRG ones, \labelcref{eq:gammamfull,eq:gammam4f}, evaluated on the four-loop $\overline{\rm MS}$ fixed-point values. We note that the mesonic thresholds account for higher order fermionic correlations. They are still present even in the absence of $\dSSB$ and confinement, leading to $\bar m_\psi = \bar m_{\rm gap}=0$. 

For $N_f\gtrsim 11$, the fRG result $\gamma^{\rm fRG}_m$ agrees well with the three-loop result and the four-Fermi follows closely. The latter computation displays an $N_f^{{\rm crit,\, 4}\psi}=10$ as shown in \Cref{fig:alphas_crit Comparisons} well in agreement with \cite{Gies:2005as} and with a
\begin{align}
	\left|\gamma^{{\rm fRG,\, 4}\psi,\,*}_m\right|(N_f^{{\rm crit,  \,4}\psi}) = 0.82\,.
\end{align}
On the other hand, the present computation displays the anomalous dimension in \labelcref{eq:gammamcrit}. Close to $N_f^{\rm crit}$, $	\left|\gamma^{\rm fRG,\,*}_m\right|$ grows rapidly driven by the high-order mesonic interactions, in particular displaying $\bar m^*_\sigma= \bar m^*_\pi\lesssim10$ entering in \labelcref{eq:flowhdiagramatic,eq:flowsigmadiagramatic}. This regime supports the potential presence of a small strongly coupled conformal field theories.

   \bibliography{references}

\end{document}